\documentclass[preprint2]{aastex}
\usepackage{amsmath}
\usepackage{color}
\usepackage{comment}
\usepackage{url}

\begin{document}
\shorttitle{Candidate Water Vapor Lines to Locate the $\mathrm{H_2O}$ Snowline through High-Dispersion Spectroscopic Observations I. The Case of a T Tauri Star}
\shortauthors{Notsu et al.}


\received{2015/12/30} 
\accepted{2016/6/16}


\title{Candidate Water Vapor Lines to Locate the $\mathrm{H_2O}$ Snowline through High-Dispersion Spectroscopic Observations I. The Case of a T Tauri Star}


\author{Shota Notsu\altaffilmark{1,7},  Hideko Nomura\altaffilmark{2}, Daiki Ishimoto\altaffilmark{1,2}, Catherine Walsh\altaffilmark{3}, Mitsuhiko Honda\altaffilmark{4}, Tomoya Hirota\altaffilmark{5}, and T. J. Millar\altaffilmark{6}}
\affil{\altaffilmark{1}Department of Astronomy, Graduate School of Science, Kyoto University, Kitashirakawa-Oiwake-cho, Sakyo-ku, Kyoto 606-8502, Japan}
\affil{\altaffilmark{2}Department of Earth and Planetary Science, Tokyo Institute of Technology, 2-12-1 Ookayama, Meguro-ku, Tokyo 152-8551, Japan}
\affil{\altaffilmark{3}Leiden Observatory, Leiden University, P.O. Box 9513, 2300 RA Leiden, The Netherlands}
\affil{\altaffilmark{4}Department of Physics, School of Medicine, Kurume University, 67 Asahi-machi, Kurume, Fukuoka 830-0011, Japan}
\affil{\altaffilmark{5}National Astronomical Observatory of Japan, 2-21-1 Osawa, Mitaka, Tokyo 181-8588, Japan}
\affil{\altaffilmark{6}Astrophysics Research Centre, School of Mathematics and Physics, Queen's University Belfast, University Road, Belfast, BT7 1NN, UK}
\affil{\altaffilmark{7}Research Fellow of Japan Society for the Promotion of Science (DC1)}
\email{snotsu@kusastro.kyoto-u.ac.jp}
\affil{}
\affil{This paper was received by The Astrophysical Journal (ApJ) on December 30th, 2015, \\ and was accepted on June 16th, 2016.}


\begin{abstract}
\noindent
Inside the $\mathrm{H_2O}$ snowline of protoplanetary disks, water evaporates from the dust-grain surface into the gas phase, whereas it is frozen out on to the dust in the cold region beyond the snowline.
$\mathrm{H_2O}$ ice enhances the solid material in the cold outer part of a disk, which promotes the formation of gas-giant planet cores. We can regard the $\mathrm{H_2O}$ snowline as the surface that divides the regions between rocky and gaseous giant planet formation.
Thus observationally measuring the location of the $\mathrm{H_2O}$ snowline is crucial for understanding the planetesimal and planet formation processes, and the origin of water on Earth.
In this paper, we find candidate water lines to locate the $\mathrm{H_2O}$ snowline through future high-dispersion spectroscopic observations.
First, we calculate the chemical composition of the disk and investigate the abundance distributions of $\mathrm{H_2O}$ gas and ice, and the position of the $\mathrm{H_2O}$ snowline. We confirm that the abundance of $\mathrm{H_2O}$ gas is high not only in the hot midplane region inside the $\mathrm{H_2O}$ snowline but also in the hot surface layer of the outer disk. Second, we calculate the $\mathrm{H_2O}$ line profiles and identify those $\mathrm{H_2O}$ lines which are promising for locating the $\mathrm{H_2O}$ snowline: the identified lines are those which have small Einstein $A$ coefficients and high upper state energies.
The wavelengths of the candidate $\mathrm{H_2O}$ lines range from mid-infrared to sub-millimeter, and they overlap with the regions accessible to ALMA and future mid-infrared high dispersion spectrographs (e.g., TMT/MICHI, SPICA).
\end{abstract}
%
%
%





\section{Introduction}
\noindent Protoplanetary disks are rotating accretion disks surrounding young newly formed stars (e.g., T Tauri stars, Herbig Ae/Be stars). 
They are composed of dust grains and gas, and contain all the material which will form 
planetary systems orbiting main-sequence stars (e.g., \citealt{Armitage2011}).
They are active environments for the creation of simple and complex molecules, including organic matter and $\mathrm{H_2O}$ (e.g., \citealt{CaselliCeccarelli2012, Henning2013, Pontoppidan2014PPIV}). 
The physical and chemical environments of protoplanetary disks determine the properties of various planets, including mass and chemical composition (e.g., \citealt{Oberg2011, Pontoppidan2014PPIV}). 
Among all molecules in disks, $\mathrm{H_2O}$ is one of the most important in determining physical and chemical properties.
\\
\\
$\mathrm{H_2O}$ gas and ice likely carries most of the oxygen that is available, the only competitors are CO and possibly $\mathrm{CO_2}$ \citep{Pontoppidan2014, Walsh2015}.
In the hot inner regions of protoplanetary disks, $\mathrm{H_2O}$ ice evaporates from the dust-grain surfaces into the gas phase.
On the other hand, it is frozen out on the dust-grain surfaces in the cold outer parts of the disk. 
The $\mathrm{H_2O}$ snowline is the surface that divides the disk into these two different regions \citep{Hayashi1981}. $\mathrm{H_2O}$ ice enhances the solid material in the cold region beyond the $\mathrm{H_2O}$ snowline, and $\mathrm{H_2O}$ ice mantles on dust grains beyond the $\mathrm{H_2O}$ snowline allow dust grains to stick at higher collisional velocities and promote efficient coagulation compared with collisions of refractory grains (e.g., \citealt{Wada2013}). 
As a result, the formation of the cores of gaseous planets are promoted in such regions. 
In the disk midplane, we can thus regard the $\mathrm{H_2O}$ snowline as the line that divides the 
regions of rocky planet and gas-giant planet formation (e.g., \citealt{Hayashi1981}, \citeyear{Hayashi1985}, \citealt{Oberg2011}).
In the upper layers of the disk, the surface separating water vapor from water ice is determined by 
the photodesorption of water ice by the stellar UV radiation field in competition with freeze-out of water onto dust grains (e.g., \citealt{Blevins2015}, see 
also Section 3.1).
\\
\\
Icy planetesimals, comets, and/or icy pebbles coming from outside the $\mathrm{H_2O}$ snowline may bring water to rocky planets including the Earth (e.g., \citealt{Morbidelli2000, Morbidelli2012, CaselliCeccarelli2012, vanDishoeck2014, Sato2016, Matsumura2016}). 
In the case of disks around solar-mass T Tauri stars, the $\mathrm{H_2O}$ snowline is calculated to exist at a few AU from the central T Tauri star (e.g., \citealt{Hayashi1981}). 
However, if we change the physical conditions such as the luminosity of the central star, the mass accretion rate, and the dust-grain size distribution in the disk, the location of the $\mathrm{H_2O}$ snowline will change (e.g., \citealt{Du2014}, \citealt{Piso2015}). 
Recent studies \citep{Davis2005, Garaud2007, Min2011, Oka2011, Harsono2015, Mulders2015, Piso2015} calculate the evolution of the $\mathrm{H_2O}$ snowline in optically thick disks, and show that it migrates as the mass accretion rate in the disk decreases and as the dust grains grow in size. In some cases the line may lie within the current location of Earth's orbit (1AU), meaning that the formation of water-devoid planetesimals in the terrestrial planet region becomes more difficult as the disk evolves.
\citet{Sato2016} estimated the amount of icy pebbles accreted by terrestrial embryos after the $\mathrm{H_2O}$ snowline has migrated inwards to the distance of Earth's orbit (1AU). They argued that the fractional water content of the embryos is not kept below the current Earth's water content (0.023 wt$\%$) unless the total disk size is relatively small ($<$ 100 AU) and the disk has a stronger value of turbulence than that suggested by recent work, so that the pebble flow decays at early times.
In contrast, other studies \citep{Martin2012, Martin2013} model the evolution of the $\mathrm{H_2O}$ snowline in a time-dependent disk with a dead zone and self-gravitational heating, and suggest that there is sufficient time and mass in the disk for the Earth to form from water-devoid planetesimals at the current location of Earth's orbit (1AU).
\\ \\
\citet{Ros2013} showed that around the $\mathrm{H_2O}$ snowline, dust-grain growth due to the condensation from millimeter to at least decimeter sized pebbles is possible on a timescale of only 1000 years.
The resulting particles are large enough to be sensitive to concentration by streaming instabilities, pressure bumps and vortices, which can cause further growth into planetesimals, even in young disks ($<$1 Myr, e.g., \citealt{Zhang2015}).
Moreover, \citet{Banzatti2015} recently showed that the presence of the $\mathrm{H_2O}$ snowline leads to a sharp discontinuity
in the radial profile of the dust emission spectral index, due to replenishment of small grains through fragmentation because of the change in fragmentation velocities
across the $\mathrm{H_2O}$ snowline.
Furthermore, \citet{Okuzumi2016} argued that dust aggregates collisionally disrupt and pile up at the region slightly outside the snowlines due to the effects of sintering.
These mechanisms of condensation \citep{Ros2013, Zhang2015}, fragmentation \citep{Banzatti2015}, and sintering \citep{Okuzumi2016} of dust grains have been invoked to explain the multiple bright and dark ring patterns in the dust spectral index of the young disk HL Tau \citep{ALMA2015}.
\\ \\
Therefore, observationally measuring the location of the $\mathrm{H_2O}$ snowline is vital because it will provide information on the physical and chemical conditions of protoplanetary disks, such as the temperature, and water vapor distribution in the disk midplane, and will give constraints on the current formation theories of planetesimals and planets.
It will help clarify the origin of water on rocky planets including the Earth, since icy planetesimals, comets, and/or icy pebbles coming from outside the $\mathrm{H_2O}$ snowline may bring water to rocky planets including the Earth (e.g., \citealt{Morbidelli2000, Morbidelli2012, CaselliCeccarelli2012, vanDishoeck2014, Sato2016, Matsumura2016}). 
\\
\\
Recent high spatial resolution direct imaging of protoplanetary disks at infrared wavelengths (e.g., Subaru/HiCIAO, VLT/SPHERE, Gemini South/GPI) and sub-millimeter wavelengths (e.g., the Atacama Large Millimeter/sub-millimeter Array (ALMA), the Sub-Millimeter Array (SMA)) have revealed detailed structures in disks, such as the CO snowline (e.g., \citealt{Mathews2013, Qi2013ApJ, Qi2013, Qi2015, Oberg2015}), spiral structures (e.g., \citealt{Muto2012, Benisty2015}), strong azimuthal asymmetries in the dust continuum (e.g., \citealt{Fukagawa2013, van der Marel2013, van der Marel2016}), gap structures (e.g., \citealt{Walsh2014b, Akiyama2015, Nomura2016, Rapson2015, Andrews2016, Schwarz2016}), and multiple axisymmetric bright and dark rings in the disk of HL Tau \citep{ALMA2015}. 
$\mathrm{H_2O}$ ice in disks has been detected by conducting low dispersion spectroscopic observations including the 3 $\mu$m $\mathrm{H_2O}$ ice 
absorption band \citep{Pontoppidan2005, Terada2007}, and crystalline and amorphous $\mathrm{H_2O}$ ice features at 63 $\mu$m (e.g., \citealt{McClure2012, McClure2015}).
Multi-wavelength imaging including the 3 $\mu$m $\mathrm{H_2O}$ ice absorption band \citep{Inoue2008} detected $\mathrm{H_2O}$ ice grains in the surface of the disksaround the Herbig Ae/Be star, HD142527 \citep{Honda2009}.
More recently, \citet{Honda2016} report the detection of $\mathrm{H_2O}$ ice in the HD~100546 disk, 
and postulate that photodesorption of water ice from dust grains in the disk surface can help explain the radial absorption strength at 3 $\mu$m.  
As we described previously, the $\mathrm{H_2O}$ snowline around a solar-mass T Tauri star is thought to exist at only a few AU from the central star. Therefore, the required spatial resolution to directly locate the $\mathrm{H_2O}$ snowline is on the order of 10 mas (milliarcsecond) around nearby disks ($\sim$100-200pc), which remains challenging for current facilities.
\\
\\
In contrast, $\mathrm{H_2O}$ vapor has been detected through recent space spectroscopic observations of infrared rovibrational and pure rotational lines from protoplanetary disks around T Tauri stars and Herbig Ae stars using $Spitzer$/IRS (e.g., \citealt{CarrNajita2008, CarrNajita2011, Salyk2008, Salyk2011, Pontoppidan2010a, Najita2013}), $Herschel$/PACS (e.g., \citealt{Fedele2012, Fedele2013, Dent2013, Meeus2012, Riviere-Marichalar2012, Kamp2013, Blevins2015}), and $Herschel$/HIFI \citep{Hogerheijde2011, vanDishoeck2014}.
\citet{vanDishoeck2013, vanDishoeck2014} reviewed the results of these recent space spectroscopic observations.
The observations using $Spitzer$/IRS and $Herschel$/PACS are spatially unresolved; hence, large uncertainties 
remain on the spatial distribution of $\mathrm{H_2O}$ gas in the protoplanetary disk.
Although the observations using $Herschel$/HIFI \citep{Hogerheijde2011, vanDishoeck2014} have high spectral resolution (allowing some constraints on the radial location), they mainly probe the cold water vapor in the outer disk (beyond 100 AU). 
The lines they detected correspond to the ground state rotational transitions which have low upper state energies (see Section 3.2.3). 
\citet{Zhang2013} estimated the position of the $\mathrm{H_2O}$ snowline in the transitional disk around TW Hya by using the intensity ratio of $\mathrm{H_2O}$ lines with various wavelengths and upper state energies. 
They used archival spectra obtained by $Spitzer$/IRS, $Herschel$/PACS, and $Herschel$/HIFI.
\citet{Blevins2015} investigated the surface water vapor distribution in four disks 
using data obtained by $Spitzer$/IRS and $Herschel$/PACS, and found that they have critical radii of 
$3-11$~AU, beyond which the surface gas-phase water abundance decreases by at least 5 orders of magnitude. 
The measured values for the critical radius are consistently smaller than the location of the 
surface $\mathrm{H_2O}$ snowline, as predicted by temperature profiles derived from the observed spectral 
energy distribution.
\\
\\
Studies investigating the structure of the inner disk from the analyses of velocity profiles of emission lines have been conducted using lines, such as the 4.7 $\mu$m rovibrational lines 
of CO gas (e.g., \citealt{Goto2006, Pontoppidan2008, Pontoppidan2011}).
Profiles of emission lines from protoplanetary disks are usually affected by Doppler shift due to Keplerian rotation, and thermal broadening. 
Therefore, the velocity profiles of lines are sensitive to the radial distributions of molecular tracers in disks.
Follow-up ground-based near- and mid-infrared (L, N band) spectroscopic observations of $\mathrm{H_2O}$ emission lines for some of the known brightest targets have been conducted using VLT/VISIR, VLT/CRIRES, Keck/NIRSPEC, and TEXES (a visitor instrument on Gemini North), and the velocity profiles of the lines have been obtained (e.g., \citealt{Salyk2008, Salyk2015, Pontoppidan2010b, Fedele2011, Mandell2012}).
These observations suggested that the water vapor resides in the inner disk, but the spatial and spectroscopic resolution is not sufficient to investigate the detailed structure, such as the position of the $\mathrm{H_2O}$ snowline. In addition, the lines they observed are sensitive to the disk surface temperature and are potentially polluted by slow disk winds, and they do not probe the midplane where planet formation occurs (e.g., \citealt{Salyk2008, Pontoppidan2010b, Mandell2012, vanDishoeck2014}).
This is because these lines have large Einstein $A$ coefficients and very high upper state energies ($>$ 3000K), and exist in the near- to mid-infrared wavelength region where dust emission becomes optically thick in the surface regions (see also Section 3.2). 
\\
\\
In this work, 
we seek candidate $\mathrm{H_2O}$ lines for locating the $\mathrm{H_2O}$ snowline through future high-dispersion spectroscopic observations.
The outline of the paper is as follows.
First, we calculate the chemical composition of a protoplanetary disk using a self-consistently derived physical model of a T Tauri disk to investigate the abundance and distribution of $\mathrm{H_2O}$ gas and ice, as opposed to assuming the position of the $\mathrm{H_2O}$ snowline. 
Second, we use the model results to calculate the velocity profiles of $\mathrm{H_2O}$ emission lines ranging in wavelength from near-infrared to sub-millimeter, and investigate the properties of 
$\mathrm{H_2O}$ lines which trace the emission from the hot water vapor within the $\mathrm{H_2O}$ snowline and are promising for locating the $\mathrm{H_2O}$ snowline.
These calculations are explained in Section 2. The results and discussion are described in Section 3 and conclusions listed in Section 4.
\section{Methods}
\subsection{The physical model of the protoplanetary disk}
\begin{figure*}[htbp]
\begin{center}
\includegraphics[scale=0.6]{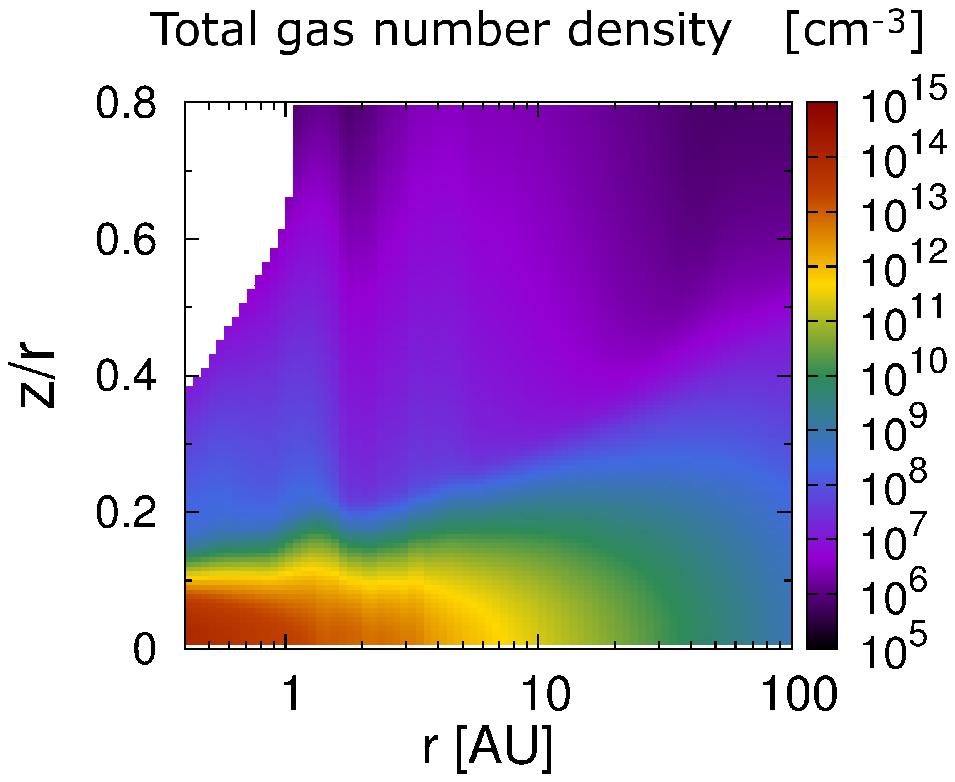}
\includegraphics[scale=0.6]{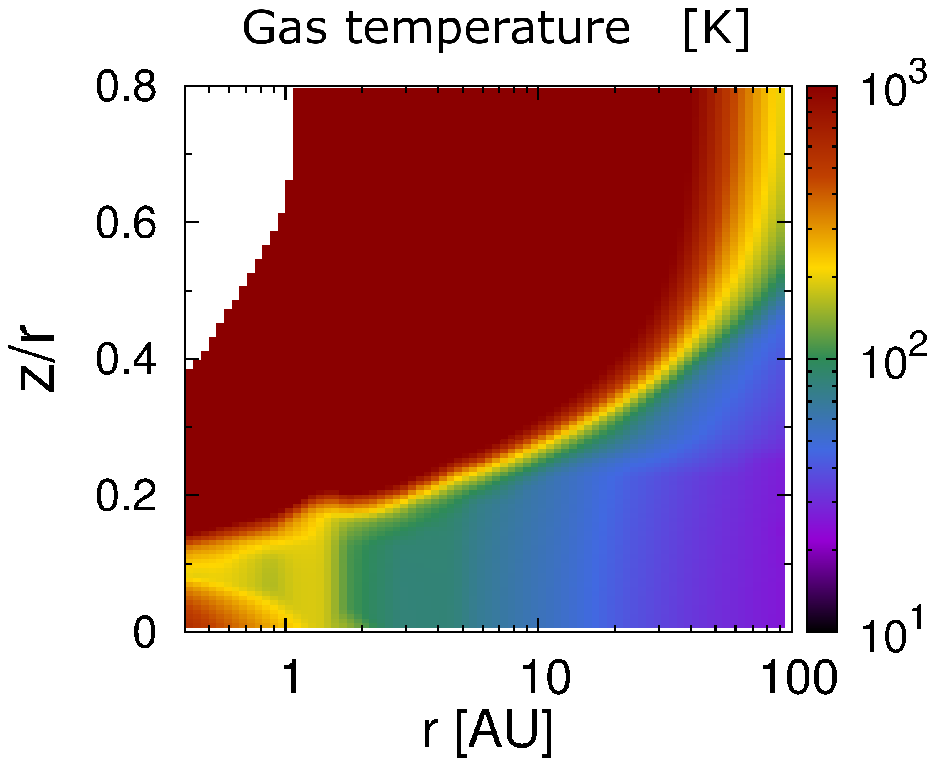}
\includegraphics[scale=0.6]{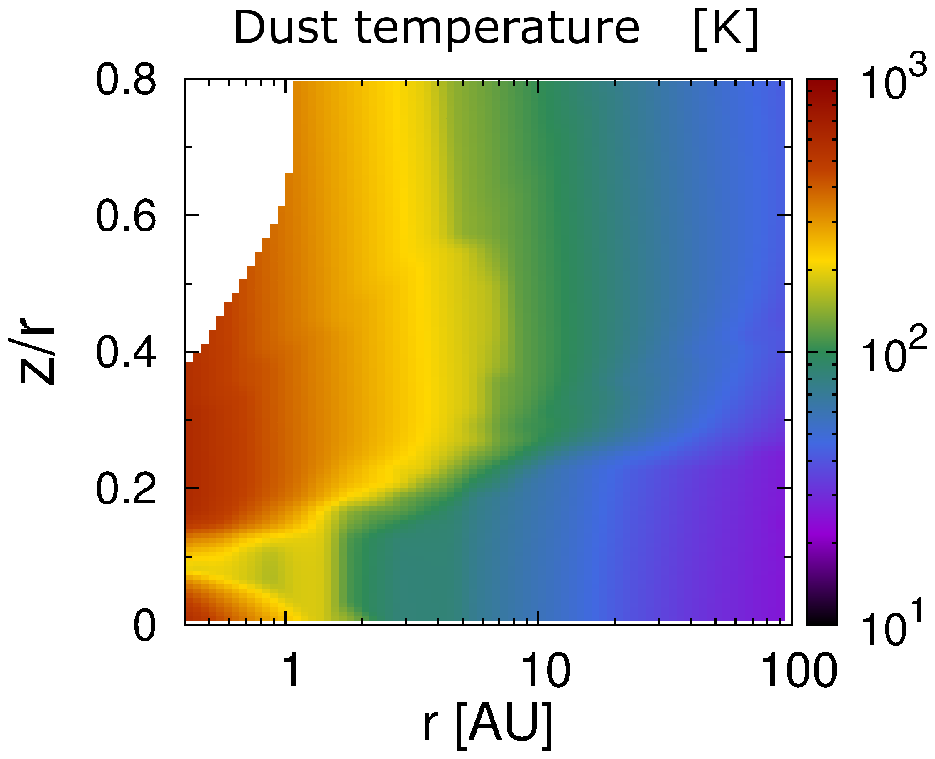}
\includegraphics[scale=0.6]{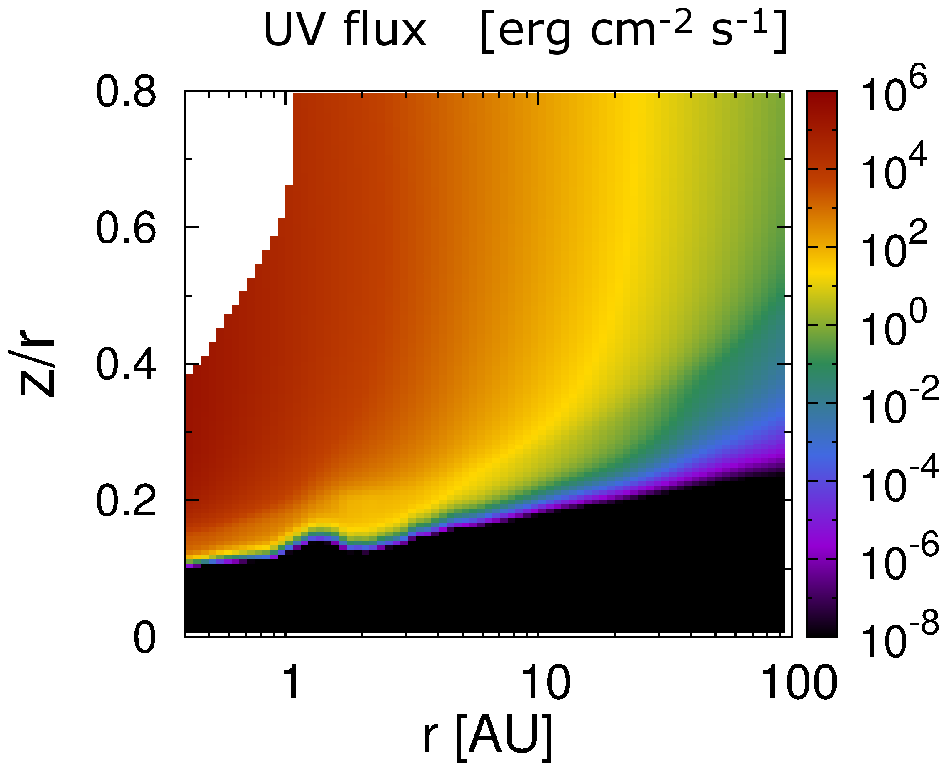}
\end{center}
\caption{The total gas number density in $\mathrm{cm}^{-3}$ (top left), the gas temperature in K (top right), the dust temperature in K (bottom left), and the UV flux in erg $\mathrm{cm}^{-2}$ s$^{-1}$ (bottom right) of a disk around a T Tauri star as a function of the disk radius in AU and height (scaled by the radius, $z/r$) up to maximum radii of $r=$100AU.}\label{Figure1_original}
\end{figure*}

\noindent 
The physical structure of a protoplanetary disk model is calculated using the methods outlined in 
\citet{NomuraMillar2005} including X-ray heating as described in \citet{Nomura2007}. 
In this subsection, we provide a brief overview of the physical model we adopt. 
A more detailed description of the background theory and computation of this physical model 
is described in the original papers \citep{NomuraMillar2005, Nomura2007}.
\citet{Walsh2010, Walsh2012, Walsh2014a, Walsh2015}, \citet{Heinzeller2011}, and \citet{Furuya2013} 
used the same physical model to study various chemical and physical effects, 
and they explain the treatment of the physical structure in detail.
\\
\\
We adopt the physical model of a steady, axisymmetric Keplarian disk surrounding a 
T Tauri star with mass $M_{\mathrm{*}}$=0.5$M_{\bigodot}$, radius $R_{\mathrm{*}}$=2.0$R_{\bigodot}$, 
and effective temperature $T_{\mathrm{*}}$=4000K \citep{KenyonHartmann1995}.
The $\alpha$-disk model \citep{ShakuraSunyaev1973} is adopted to obtain the radial surface density, 
assuming a viscous parameter $\alpha$=$10^{-2}$ and an accretion rate 
$\dot{M}$=$10^{-8}M_{\bigodot}$ yr$^{-1}$. 
The steady gas temperature and density distributions of the disk are computed self-consistently 
by solving the equations of hydrostatic equilibrium in the vertical direction and the local 
thermal balance between gas heating and cooling. 
The heating sources of the gas are grain photoelectric heating by UV photons and heating 
due to hydrogen ionization by X-rays. 
The cooling mechanisms are gas-grain collisions and line transitions 
(for details, see \citealt{NomuraMillar2005} and \citealt{Nomura2007}).
The dust temperature distribution is obtained by assuming radiative equilibrium between absorption 
and reemission of radiation by dust grains. 
The dust heating sources adopted are the radiative flux produced by viscous dissipation 
($\alpha$-disk model) at the midplane of the disk, and the irradiation from the central star.
The radial range for which the calculations are conducted is $r\sim$0.04~AU to 305~AU.
\\
\\
The dust properties are important because they affect the physical and chemical structure of protoplanetary disks in several ways  (for details, see, e.g., \citealt{NomuraMillar2005}).
Since dust grains are the dominant opacity source, they determine the dust temperature profile and the UV radiation field throughout the disk.
Photodesorption, photodissociation, and photoionization processes are affected by the UV radiation field.
The dust properties affect the gas temperature distribution, because photoelectric heating by UV photons is the dominant source of gas heating at the disk surface.
The total surface area of dust grains has an influence on the chemical abundances of molecules through determining the gas and ice balance. 
In Appendix A, we describe a brief overview of the X-ray and UV radiation fields and the dust-grain models we adopt.
\\
\\
In Figure \ref{Figure1_original}, we display the gas number density in $\mathrm{cm}^{-3}$ (top left), the gas temperature in K (top right, $T_{g}$), the dust-grain temperature in K (bottom left, $T_{d}$), 
and the wavelength-integrated UV flux in erg $\mathrm{cm}^{-2}$ s$^{-1}$ (bottom right),
as a function of disk radius in AU and height (scaled by the radius, $z/r$).
The density decreases as a function of disk radius and disk height with the densest region of the disk found in the disk midplane
close to the star ($\sim10^{14}$ $\mathrm{cm}^{-3}$) and the most diffuse, in the disk surface at large radii ($\sim10^{5}$ $\mathrm{cm}^{-3}$), so that the density range in
our adopted disk model covers almost 10 orders of magnitude. The gas temperature increases as a function of disk height and decreases as a function of disk radius with the hottest region found in the disk surface ($> 10^{3}$K), and the coldest region found in the outer disk ($\sim$10K). 
In addition, due to the influence of viscous heating at the disk midplane, the temperature increases within several AU from the central T Tauri star.
In the disk surface, the dust-grain temperature is more than ten times lower than the gas temperature.
At low densities, gas-grain collisions become ineffective so the gas cools via radiative line transitions. 
In contrast, the gas and dust-grain temperatures are similar in the midplane region with high densities.
Moreover, the disk surface closest to the parent star is subjected to the largest flux of both UV and
X-ray photons, although the disk midplane is effectively shielded from UV and X-ray photons over the radial extent of our disk model. 
 \subsection{Chemical structure of the protoplanetary disk}
\noindent
In order to investigate the chemical structure of the protoplanetary disk, we use a large chemical network which includes gas-phase reactions and gas-grain
 interactions (freezeout of gas molecules on dust grains, and thermal and non-thermal desorption from dust grains). The non-thermal desorption mechanisms we adopt include cosmic-ray-induced desorption, and photodesorption by UV photons. \citet{Walsh2010, Walsh2012, Walsh2014a, Walsh2015}, \citet{Heinzeller2011}, \citet{Furuya2013}, \citet{Furuya2014}, \citet{Ishimoto2013}, and \citet{Du2014} used similar chemical networks to calculate chemical structure of a protoplanetary disk, and they, and the reviews of \citet{Henning2013} and \citet{Dutrey2014}, explain the background theories and procedures in detail. In this subsection, we outline the key points of the chemical network we use.
\\
\\
The addition of grain-surface chemistry (e.g., \citealt{Hasegawa1992}) is expected to aid the synthesis of complex organic molecules in the outer disk where significant freezeout has occurred.
Some previous works on chemical modeling of disks (e.g., \citealt{Willacy2007, Semenov2011, Walsh2012, Walsh2014a, Walsh2015, Furuya2013, Furuya2014, Drozdovskaya2014}) have contained grain-surface reactions.
However, the chemical network we adopt in this work does not contain such grain-surface reactions, and is equivalent to one of models in \citet{Walsh2010}, which includes the same freezeout and desorption processes as considered here.
This is because we are primarily interested in the hot inner disk region where molecular line emission originates from the thermally desorbed gas reservoir. 
\subsubsection{Gas-phase reactions}
\noindent
Our gas-phase chemistry is extracted from the UMIST Database for Astrochemistry (UDfA), 
henceforth referred to as ``{\sc Rate}06" \citep{Woodall2007}. 
\citet{Walsh2010, Walsh2012}, and \citet{Heinzeller2011} used {\sc Rate}06 to calculate the 
chemical structure of a protoplanetary disk.
We include almost the entire {\sc Rate}06 gas-phase network removing only those species 
(and thus reactions) which contain fluorine, F, and phosphorus, P, in order to reduce computation time. 
We have confirmed that the loss of F- and P-containing species have a minimal impact on the 
remaining chemistry \citep{Walsh2010, Heinzeller2011}. 
Our gas-phase network thus contains 375 atomic, molecular, and ionic species composed of the elements
H, He, C, N, O, Na, Mg, Si, S, Cl, and Fe. 
Table~1 in the online material from \citet{Woodall2007} shows the list of these 375 species.
The initial elemental fractional abundances (relative to total hydrogen nuclei density) we
use are the set of oxygen-rich low-metallicity abundances from \citet{Graedel1982}, 
listed in Table 8 of \citet{Woodall2007}.
The chemical evolution is run for $10^{6}$ years.
By this time, the chemistry in the inner regions of the disk midplane inside the $\mathrm{H_2O}$ snowline is close to 
steady state, at which time the chemistry has forgotten its origins, justifying our use of initial 
elemental abundances, instead of ambient cloud abundances. 
\\ \\
Our adopted reaction network consists of 4336 reactions including 3957 two-body reactions, 214 photoreactions, 154 X-ray/cosmic-ray-induced photoreactions, and 11 reactions of direct
X-ray/cosmic-ray ionization. The adopted equations that give the reaction rates of two-body reactions, X-ray/cosmic-ray-induced photoreactions, and reactions of X-ray/cosmic-ray ionization are described in Section 2.1 of \citet{Woodall2007}.
\\ \\
Here we mention that we use the previous version of the UDfA {\sc Rate}06 \citep{Woodall2007}, 
instead of the latest version of UDfA, ``{\sc Rate}12" \citep{McElroy2013}.
There are some updates in {\sc Rate}12 such as reactions related to some complex molecules, 
and \citet{McElroy2013} described that the major difference between {\sc Rate}12 and 
{\sc Rate}06 is the inclusion of anion reactions.  
Although this has an influence on the abundances of carbon-chain molecules \citep{Walsh2009, McElroy2013}, 
it has little effect on the chemistry of main simple molecules, such as $\mathrm{H_2O}$.
\\ 
\\
In our calculations of the chemistry, we have approximated our photoreaction rates at each point in the disk, $k^{\mathrm{ph}}(r, z)$, by scaling the rates of {\sc Rate}06 which assume the interstellar UV field, $k_{0}$, using the wavelength integrated UV flux calculated at each point (see also Figure \ref{Figure1_original}), 
\begin{equation}
G_{\mathrm{FUV}}(r, z)=\int_{912\mathrm{\mathring{A}}(13.6\mathrm{eV})}^{2068\mathrm{\mathring{A}}(6\mathrm{eV})} G_{\mathrm{FUV}}(\lambda, r, z) \mathrm{d}\lambda.
\end{equation}
Using this value of $G_{\mathrm{FUV}}(r, z)$, the rate for a particular photoreaction at each $(r, z)$ is given by
\begin{equation}
k^{\mathrm{ph}}(r, z) = \frac{G_{\mathrm{FUV}}(r, z)}{G_{0}}k_{0} \ \mathrm{s}^{-1},
\end{equation}
where $G_{0}$ is the interstellar UV flux (2.67$\times 10^{-3}$ erg $\mathrm{cm}^{-2}$ s$^{-1}$, \citealt{vanDishoeck2006}).
\subsubsection{Gas-grain Interactions}
\noindent
In our calculations, we consider the freezeout of gas-phase molecules on dust grains, and the thermal and non-thermal desorption of molecules from dust grains.
For the thermal desorption of a molecule to occur, the dust-grain temperature must exceed the freezeout (sublimation) temperature of each molecule.
Non-thermal desorption requires an input of energy from an external source and is thus independent of dust-grain temperature. 
The non-thermal desorption mechanisms we investigate are cosmic-ray-induced desorption \citep{Leger1985, Hasegawa1993} and photodesorption from UV photons \citep{Westley1995, Willacy2000, Oberg2007}, as adopted in some previous studies (e.g., \citealt{Walsh2010, Walsh2012}).
In this subsection, we explain the mechanisms of freezeout and thermal desorption we use in detail. 
In Appendix C, we introduce the detailed mechanisms of non-thermal desportion we adopt (cosmic-ray-induced desorption and photodesorption from UV photons).
\\
\\
The freezeout (accretion) rate, $k_{i}^{a}$ [s$^{-1}$], of species $i$ onto the dust-grain surface is treated using the standard equation (e.g., \citealt{Hasegawa1992, Woitke2009a, Walsh2010}),
\begin{equation}
k_{i}^{a} = \alpha \sigma_{d} \langle v_{i}^{th}\rangle n_{d} \ \mathrm{s}^{-1},
\end{equation}
where $\alpha$ is the sticking coefficient, here assumed to 0.4 for all species, which is in the range of high gas temperature cases ($T_{g} \sim 100-200$K) reported in \citet{Veeraghattam2014}.
Previous theoretical and experimental studies suggested that the sticking coefficient tends to be lower as the gas and dust-grain temperature become higher (e.g., \citealt{Masuda1998, Veeraghattam2014}).
$\sigma_{d}=\pi a^{2}$ is the geometrical cross section of a dust grain with radius, $a$,
$\langle v_{i}^{th}\rangle$ is the thermal velocity of species $i$ with mass $m_{i}$ at gas temperature $T_{g}$, $k_{B}$ is the Boltzmann's constant, and $n_{d}$ is the number density of dust grains. 
We adopt the value of $\langle v_{i}^{th}\rangle=(k_{B}T_{g}/m_{i})^{1/2}$ as \citet{Walsh2010} adopted.
In this work, for our gas-grain interactions, we assume a constant grain radius $a=0.1\mu m$ and a fixed dust-grain fractional abundance ($x_{d}=n_{d}/n_{\mathrm{H}}$\footnote[1]{$n_{\mathrm{H}}$ is the total gas atomic hydrogen number density.}) of 2.2$\times 10^{-12}$, as previous studies adopted (e.g., \citealt{Walsh2012}). 
From the viewpoint of dust-grain surface area per unit volume, the adopted value of a constant grain radius $a$ is consistent with the value from the dust-grain size distributions in the disk physical model adopted in this work (see also Appendix A). This adopted value of $x_{d}$ is consistent with a gas-to-dust ratio of 100 by mass.
\\
\\
The thermal desorption rate, $k_{i}^{d}$ [s$^{-1}$], of species $i$ from the dust-grain surface is given by (e.g., \citealt{Hasegawa1992, Woitke2009a, Walsh2010}),
\begin{equation}
k_{i}^{d}=\nu_{0}(i) \exp\left(\frac{-E_{d}^{\mathrm{K}}(i)}{T_{d}} \right) \ \mathrm{s}^{-1},
\end{equation}
where $E_{d}^{\mathrm{K}}(i)$ is the binding energy of species $i$ to the dust-grain surface in units of K. 
The values of $E_{d}^{\mathrm{K}}(i)$ for several important molecules are listed in Table 2 of Appendix B. Most of these values are adopted in \citet{Walsh2010} or \citet{Walsh2012}.
$T_{d}$ is the dust-grain temperature in units of K.
The characteristic vibrational frequency of each adsorbed species $i$ in its surface potential well, $\nu_{0}(i)$, is represented by a
harmonic oscillator relation \citep{Hasegawa1992},
\begin{equation}
\nu_{0}(i) = \sqrt{\frac{2n_{surf} E_{d}^{\mathrm{erg}}(i)}{\pi^{2} m_{i}}} \ \mathrm{s}^{-1},
\end{equation}
where, $E_{d}^{\mathrm{erg}}(i)$ is in units of erg here, $m_{i}$ is the mass of each absorbed species $i$, and $n_{surf}=1.5\times 10^{15}$ $\mathrm{cm}^{-2}$ is the surface density of absorption sites on each dust grain.
\\
\\
Considering these processes of freezeout, thermal desorption, cosmic-ray-induced desorption, and photodesorption, the total formation rate of ice species $i$ is
\begin{equation}
\dot n_{i, \mathrm{ice}}=n_{i}k_{i}^{a}-n_{i, \mathrm{ice}}^{\mathrm{desorb}}(k_{i}^{d}+k_{i}^{\mathrm{crd}}+k_{i}^{\mathrm{pd}}).
\end{equation}
where  $k_{i}^{\mathrm{crd}}$ is the cosmic-ray-induced thermal desorption rate for each species $i$, $k_{i}^{\mathrm{pd}}$ is the photodesorption rate for a specific species $i$, 
$n_{i, \mathrm{ice}}$ denotes the number density of ice species $i$, and $n_{i, \mathrm{ice}}^{\mathrm{desorb}}$ is the fraction of $n_{i, \mathrm{ice}}$
located in the uppermost active surface layers of the ice mantles. The value of $n_{i, \mathrm{ice}}^{\mathrm{desorb}}$ is given by \citep{Aikawa1996, Woitke2009a} 
\begin{equation}
n_{i, \mathrm{ice}}^{\mathrm{desorb}}= \begin{cases}
      n_{i, \mathrm{ice}} & (n_{\mathrm{ice}}< n_{act}), \\
      n_{act}\frac{n_{i, \mathrm{ice}}}{n_{\mathrm{ice}}} & (n_{\mathrm{ice}} \geq n_{act}),
\end{cases}
\end{equation}
where $n_{\mathrm{ice}}$ is the total number density of all ice species, 
$n_{act} = 4\pi a^{2}n_{d}n_{surf}N_{Lay}$ is the number of active surface sites in the ice mantle per volume. $N_{Lay}$ is the number of
surface layers to be considered as ``active", and we adopt the value from \citet{Aikawa1996}, $N_{Lay}=2$.
\subsection{Profiles of $\mathrm{H_2O}$ emission lines from protoplanetary disks}
\noindent Using the $\mathrm{H_2O}$ gas abundance distribution obtained from our chemical calculations described in Section 2.2, we
calculate the profiles of $\mathrm{H_2O}$ emission lines ranging from near-infrared to sub-millimeter wavelengths, 
and investigate which lines are the best candidates for probing emission from the inner thermally desorbed water reservoir, i.e., within the $\mathrm{H_2O}$ snow line. 
We also study how the line flux and profile shape depends upon the location of the $\mathrm{H_2O}$ snowline.
In the following paragraphs, we outline the calculation methods used to determine the $\mathrm{H_2O}$ emission line profiles (based on \citet{Rybicki1986}, \citet{Hogerheijde2000}, and \citet{NomuraMillar2005}).
\\ \\
Here we define the transition frequency of each line as $\nu_{ul}$, where the subscript $ul$ means the transition from the upper level ($u$) to the lower level ($l$).
The intensity of each line profile at the frequency $\nu$, $I_{ul}(\nu)$, is obtained by solving the radiative transfer equation in the line-of-sight direction of the disk,
\begin{equation}
\frac{\mathrm{d}I_{ul}(\nu)}{\mathrm{d}s}=-\chi_{ul}(\nu)(I_{ul}(\nu)-S_{ul}(\nu)).
\end{equation}
The source function, $S_{ul}(\nu)$, and the total extinction coefficient, $\chi_{ul}(\nu)$, are given by
\begin{equation}
S_{ul}(\nu)=\frac{1}{\chi_{ul}(\nu)}n_{u}A_{ul}\Phi_{ul}(\nu)\frac{h{\nu}_{ul}}{4\pi},
\end{equation}
and
\begin{equation}
\begin{split}
\chi_{ul}(\nu)=&\rho_{d}\kappa_{ul} \\
                       &+(n_{l}B_{lu}-n_{u}B_{ul})\Phi_{ul}(\nu)\frac{h{\nu}_{ul}}{4\pi},
\end{split}
\end{equation}
where the symbols $A_{ul}$ and $B_{ul}$ are the Einstein $A$ and $B$ coefficients for
the transition $u \rightarrow l$, the symbol $B_{lu}$ is the Einstein $B$ coefficient for
the transition $l \rightarrow u$, $h$ is the Planck constant, and $n_{u}$ and $n_{l}$ are the number densities of
the upper and lower levels, respectively. The energy difference between the levels $u$ and $l$ corresponds to $h{\nu}_{ul}$.
$\rho_{d}$ is the mass density of dust grains which we calculate from the values of total gas mass density $\rho_{g}$ and
gas-to-dust mass ratio ($\rho_{g}/\rho_{d}=100$). $\kappa_{ul}$ is dust absorption coefficient at the frequency $\nu_{ul}$ as described in Section 2.1.
\\ \\
The symbol $\Phi_{ul}(\nu)$ is the line profile function at the frequency $\nu$, 
and we consider the Doppler shift due to Keplerian rotation,
and thermal broadening, in calculating the emission line profiles. 
This function is given by,
\begin{equation}
\Phi_{ul}(\nu)=\frac{1}{\Delta\nu_D\sqrt{\pi}}\exp\biggl[-\frac{(\nu+\nu_K-\nu_{ul})^2}{\Delta\nu_D^2}\biggr],
\end{equation}
where $\Delta\nu_D=(\nu_{ul}/c)(\sqrt{2kT_g/m})$ is the Doppler width,
$c$ is the speed of light, $T_{g}$ is the gas temperature,
$k$ is the Boltzmann constant, $m$ is the mass of a water molecule, and
$\nu_{K}$ is the Doppler-shift due to projected Keplerian velocity for
the line-of-sight direction and is given by,
\begin{equation}
\nu_{K}=\frac{\nu_{ul}}{c}\sqrt{\frac{GM_{\mathrm{*}}}{r}}\sin\phi\sin i,
\end{equation}
where $G$ is the gravitational constant, $M_{\mathrm{*}}$ is the mass of central star, 
$r$ is the distance from the central star, 
$\phi$ is the azimuthal angle between the semimajor axis and the line which links the point in the
disk along the line-of-sight and the center of the disk.
\\
\\
The observable profiles of flux density are obtained by integrating Eq.~(8) in the 
line-of-sight direction and summing up the integrals in the 
plane of the projected disk, $(x,y)$, as,
\begin{equation}
\begin{split}
F_{ul}(\nu) =&\frac{1}{4\pi d^2}\int d\Omega \int\int \mathrm{d}x \mathrm{d}y \\
                   &\times \int_{-s_{\mathrm{\infty}}}^{s_{\mathrm{\infty}}} j_{ul}(s,x,y,\nu)  \mathrm{d}s,
\end{split}
\end{equation}
where $d$ is the distance of the observed disk from the Earth.
$j_{ul}(s,x,y,\nu)$ is the emissivity at $(s,x,y)$ and the frequency $\nu$ considering the effect of absorption in the upper disk layer and it is given by the following equation,
\begin{equation}
\begin{split}
j_{ul}(s,x,y,\nu)=&n_{u}(s,x,y)A_{ul}\frac{h\nu_{ul}}{4\pi}\Phi_{ul}(s,x,y,\nu)\\
                         &\times \mathrm{exp}(-\tau_{ul}(s,x,y,\nu)),
\end{split}
\end{equation}
and $\tau_{ul}(s,x,y,\nu)$ is the optical depth from $s$ to the disk surface $s_{\mathrm{\infty}}$ at the frequency $\nu$ given by,
\begin{equation}
\tau_{ul}(s,x,y,\nu)=\int_{s}^{s_{\mathrm{\infty}}} \chi_{ul}(s',x,y,\nu) \mathrm{d}s'.
\end{equation}
Hence, the observable total flux of the lines, $F_{ul}$, are given by the following equation,
\begin{equation}
F_{ul}=\int F_{ul}(\nu) \mathrm{d}\nu
\end{equation}
Here, we use a distance $d=140$ pc for calculating the line profiles since this is the distance to the Taurus molecular cloud, one of the nearest star formation regions with observable protoplanetary disks.
\\ \\
The code for ray tracing which we have built for calculating emission line profiles from the protoplanetary disk
is a modification of the original 1D code called RATRAN\footnote[2]{\url{http://home.strw.leidenuniv.nl/~michiel/ratran/}} \citep{Hogerheijde2000}.
We adopt the data for the ortho- and para-$\mathrm{H_2O}$ energy levels from \citet{Tennyson2001}, the radiative rates (Einstein $A$ coefficients $A_{ul}$) from the BT2 water line list \citep{Barber2006}, and the collisional rates, $<\sigma v>$, for the excitation of $\mathrm{H_2O}$ by H$_{\mathrm{2}}$ and by electrons from \citet{Faure2008}.
We use the collisional rates to determine the critical densities of transitions of interest.
These data are part of Leiden Atomic and Molecular Database called LAMDA\footnote[3]{\url{http://home.strw.leidenuniv.nl/~moldata/}} \citep{Schoier2005}.
The level populations of the water molecule ($n_{u}$ and $n_{l}$) are calculated under the assumption of local thermal equilibrium (LTE).
In Section 3.2.5, we discuss the validity of the assumption of LTE in our work.
We do not include dust-grain emission nor emission from disk winds and jet components in calculating the emission line profiles.
However, we do include the effects of the absorption of line emission by dust grains (as described above).
 \\ \\
The nuclear spins of the two hydrogen atoms in each water molecule can be either parallel or anti-parallel,
and this results in a grouping of the $\mathrm{H_2O}$ energy levels into ortho ($K_{a}+K_{c}=$odd) and para ($K_{a}+K_{c}=$even) ladders.
The ortho to para ratio (OPR) of water in the gas gives information on the conditions, formation, and thermal history of water in specific regions, such as comets and protoplanetary disks (e.g., \citealt{Mumma2011, vanDishoeck2013, vanDishoeck2014}).
An alternative way to describe the OPR is through the ``spin temperature", defined as the temperature that characterizes the observed OPR if it is in thermal equilibrium. 
The OPR becomes zero in the limit of low temperature and 3 in the limit of high temperature ($\gtrsim$60K). The original definition of OPR of water vapor in thermal equilibrium is described in \citet{Mumma1987}. In this paper, we set the OPR$=$3 throughout the disk to calculate values of $n_{u}$ and $n_{l}$ from the $\mathrm{H_2O}$ gas abundance distribution. The lines we calculate in order to locate the position of the $\mathrm{H_2O}$ snowline mainly trace the hot water vapor for which the temperature is higher than the water sublimation temperature ($\sim 150-160$K). 
The disk physical structure of our adopted model is steady, and thermal and chemical equilibrium is mostly achieved throughout the disk. In addition, previous observational data on warm water detected at mid-infrared wavelengths in the inner regions of protoplanetary disks are consistent with OPR$=3$ (e.g., \citealt{Pontoppidan2010a}). 
\\ \\
Here we also mention that \citet{Hama2016} reported from their experiments that water desorbed from the icy dust-grain surface at 10K shows the OPR$=$3, which invalidates the assumed relation between OPR and the formation temperature of water. They argue that the role of gas-phase processes which convert the OPR to a lower value in low temperature regions is important, although the detailed mechanism is not yet understood.
\section{Results and Discussion}
\subsection{The distributions $\mathrm{H_2O}$ gas and ice}
\begin{figure}[htbp]
\begin{center}
\includegraphics[scale=0.6]{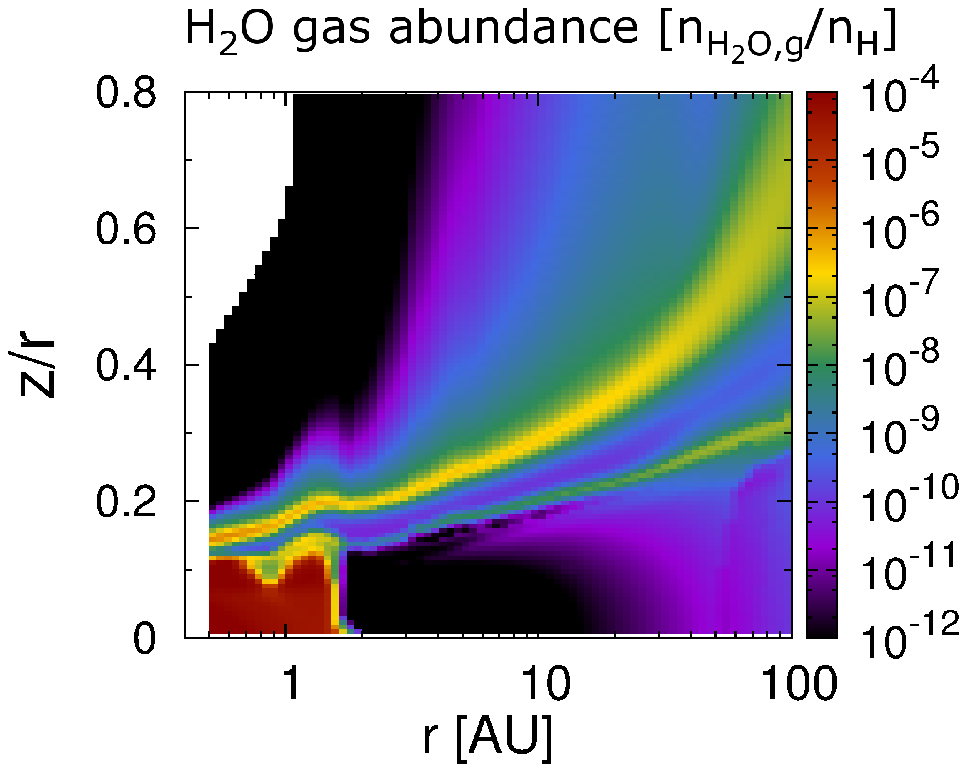}
\includegraphics[scale=0.6]{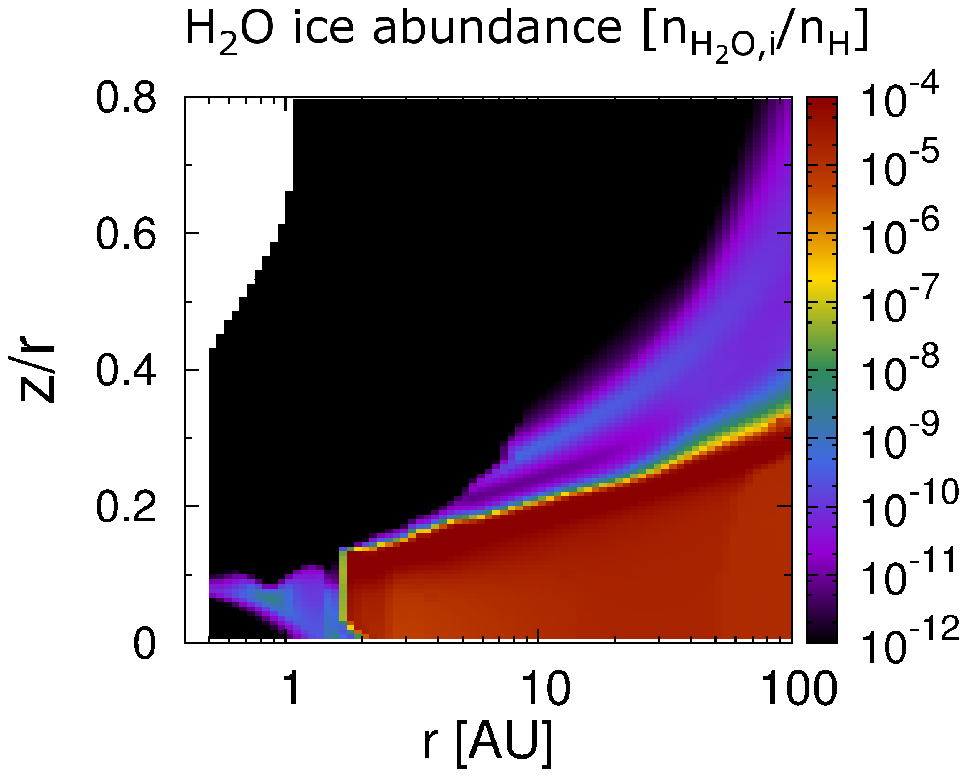}
\end{center}
\caption{\noindent The fractional abundance (relative to total hydrogen nuclei density) distributions of $\mathrm{H_2O}$ gas (top) and $\mathrm{H_2O}$ ice (bottom) of a disk around a T Tauri star as a function of disk radius and height (scaled by the radius, $z/r$) up to maximum radii of $r=$100AU.}\label{Figure2_original}
\vspace{0.2cm}
\end{figure}

\begin{figure}[htbp]
\begin{center}
\includegraphics[scale=0.5]{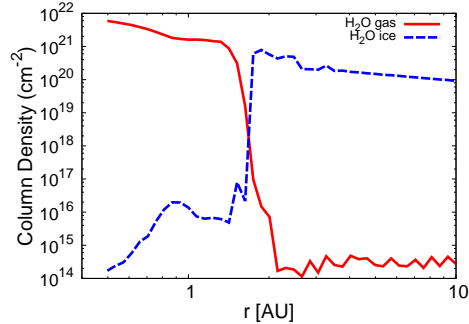}
\end{center}
\vspace{0.3cm}
\caption{\noindent The radial profile of the vertically integrated column density $\mathrm{cm}^{-2}$ of $\mathrm{H_2O}$ gas ({\it red solid line}) and ice ({\it blue dashed line}). 
}\label{Figure3_original}
\vspace{0.5cm}
\end{figure}

\noindent  
Figure \ref{Figure2_original} shows the fractional abundances (relative to total gas hydrogen nuclei density, $n_{\mathrm{H}}$) of $\mathrm{H_2O}$ gas and $\mathrm{H_2O}$ ice in a disk around a T Tauri star as a function of disk radius $r$ and height scaled by the radius ($z/r$).
The radial range over which the chemistry is computed is $r\sim$0.5AU and 100AU in order to reduce computation time.
Here we mention that at small radii, due to the high densities found in the midplane, there is a significant
column density of material shielding this region from the intense
UV and X-ray fields of the star. Therefore, molecules are expected to survive in the midplane at radii within $\sim$0.1AU, unless there are cavities in the dust and gas.
Thus the actual total amount of molecular gas in the inner disk may be larger than that of our chemical calculation results.
\\ \\
According to this figure, the fractional abundance with respect to $\mathrm{H_2}$ of $\mathrm{H_2O}$ gas is high ($\sim10^{-4}$) in the midplane region inside the $\mathrm{H_2O}$ snowline, 
and in contrast, it is low ($\lesssim10^{-12}$) in the midplane outside the $\mathrm{H_2O}$ snowline.
The fractional abundance of $\mathrm{H_2O}$ ice has the opposite distribution. It is low ($\lesssim 10^{-9}$) in the midplane region inside the $\mathrm{H_2O}$ snowline, and in contrast, it is high ($\sim10^{-5}$) in the midplane outside the $\mathrm{H_2O}$ snowline.
The $\mathrm{H_2O}$ snowline in the T Tauri disk that we adopt in this work exists at a radius of $\sim$ 1.6 AU in the midplane ($T_{g} \sim 150-160$K), consistent with the binding energy we adopt.
\\ \\
Inside the $\mathrm{H_2O}$ snowline, the temperature exceeds the sublimation temperature under the pressure conditions of the midplane ($T_{g} \sim 150-160$K) and most of 
the $\mathrm{H_2O}$ is released into the gas-phase through thermal desorption. 
In addition, this region is almost completely shielded from 
intense the UV and X-ray fields of the star and interstellar medium (\citealt{NomuraMillar2005, Nomura2007}, see also Figure 2 of \citealt{Walsh2012}), has a high temperature ($> 150$K) and large total gas particle number density ($> 10^{11}$ $\mathrm{cm}^{-3}$), and thermal equilibrium between the gas and dust is achieved ($T_{g} \sim T_{d}$).
Under these conditions, the gas-phase chemistry is close to thermochemical equilibrium and most of the oxygen atoms will be locked up into $\mathrm{H_2O}$ (and CO) molecules (e.g., \citealt{Glassgold2009, Woitke2009a, Woitke2009b, Walsh2010, Walsh2012, Walsh2015, vanDishoeck2013, vanDishoeck2014, Du2014, Antonellini2015}).
Therefore, the $\mathrm{H_2O}$ gas abundance of this region is approximately given by the elemental abundance of oxygen ($1.76\times10^{-4}$, \citealt{Woodall2007}) minus
the fraction bound in CO.
\\ \\
In addition, the fractional abundance of $\mathrm{H_2O}$ gas is relatively high in the hot surface layer of the outer disk.
First, at $z/r$ of $0.1-0.3$ between $r\sim$ 0.5$-$100 AU, the fractional abundance $\mathrm{H_2O}$ gas is $\sim10^{-8}-10^{-7}$. 
This region can be considered as the sublimation (photodesorption) front of $\mathrm{H_2O}$ molecules, driven by the relatively strong stellar UV radiation. This so-called photodesorbed layer \citep{Dominik2005} allows $\mathrm{H_2O}$ to survive in the gas phase where it would otherwise be frozen out on the dust-grain surfaces.
The abundance and extent of gas-phase $\mathrm{H_2O}$ in this layer is mediated by absorption back onto the dust grain, destruction by the stellar UV photons and by chemical reactions with other species.
\\ \\
Second, at $z/r$ of 0.15-0.7 between $r\sim$0.5$-$100 AU, the $\mathrm{H_2O}$ abundance is relatively high ($\sim10^{-7}$) compared with the cold midplane region of the outer disk ($\lesssim10^{-12}-10^{-10}$).
Since the gas temperature is significantly higher than the dust temperature (typically $T_{g} \sim 200-2000$K) and the gas density is low compared to the disk midplane, the water chemistry is controlled by chemical kinetics as opposed to thermodynamic (or chemical) equilibrium.
Due to the very high gas temperature ($>$200K), the energy barriers for the dominant neutral-neutral reactions of O$+$H$_{2}$$\rightarrow$OH$+$H and OH$+$H$_{2}$$\rightarrow$$\mathrm{H_2O}$$+$H are readily surpassed and gaseous $\mathrm{H_2O}$ is produced rapidly.
This route will drive all the available gas-phase oxygen into $\mathrm{H_2O}$, unless strong UV or a high atomic hydrogen abundance is able to convert some water back to OH and O 
(e.g., \citealt{Glassgold2009, Woitke2009b, Meijerink2012, vanDishoeck2013, vanDishoeck2014, Walsh2015}).
In the uppermost surface layers, $\mathrm{H_2O}$ is even more rapidly destroyed by photodissociation and reactions with atomic hydrogen than it is produced, so there is little water at the very top of the disk.
The OH gas abundance in our calculations and others (e.g., \citealt{Walsh2012}) is high in this hot surface region.
It is consistent with the above discussions that neutral-neutral reactions including OH and $\mathrm{H_2O}$ are dominant and strong UV or a high atomic hydrogen abundance converts some water back to OH and O \citep{Walsh2012, Walsh2015}.
\\ \\
%
%
Figure \ref{Figure3_original} shows the radial column density profile of $\mathrm{H_2O}$ gas ({\it red solid line}) and ice ({\it blue dashed line}). 
The column density of $\mathrm{H_2O}$ gas and ice in the disk midplane flips across the $\mathrm{H_2O}$ snowline ($\sim$ 1.6AU).
The column density of $\mathrm{H_2O}$ gas is high ($\sim10^{21}$ $\mathrm{cm}^{-2}$) inside the $\mathrm{H_2O}$ snowline, 
and, in contrast, is low outside the $\mathrm{H_2O}$ snowline ($\sim 10^{14}-10^{15}$ $\mathrm{cm}^{-2}$). 
The column density profile of $\mathrm{H_2O}$ ice is roughly opposite. The column density of $\mathrm{H_2O}$ ice in the outer disk is $\sim10^{20}-10^{21}$ $\mathrm{cm}^{-2}$.
Previous chemical modeling calculations (e.g., \citealt{Walsh2012, Walsh2015, Du2014}) gave a column density of $\mathrm{H_2O}$ gas inside the $\mathrm{H_2O}$ snowline of around $10^{21}-10^{22}$ $\mathrm{cm}^{-2}$.
This value is slightly higher than in our calculations, possibly due to the inclusion of
grain surface reactions.
However, since gas-phase $\mathrm{H_2O}$ in the disk midplane is likely obscured by dust grains at near- to mid-infrared wavelengths \citep{Walsh2015}, the ``visible" $\mathrm{H_2O}$ gas column density at these wavelength is much smaller than the actual amount. For example in \citet{Walsh2015}, the visible value is on the order of a few times $10^{19}$ $\mathrm{cm}^{-2}$ within the $\mathrm{H_2O}$ snowline.
Previous infrared low dispersion spectroscopic observations using $Spitzer$/IRS for classical
T Tauri stars derive the $\mathrm{H_2O}$ gas column densities ranging from 4$\times10^{17}$ to 7.9$\times10^{20}$ $\mathrm{cm}^{-2}$ \citep{CarrNajita2011, Salyk2011}.
Despite the model T Tauri disk being a generic model which is not representative of any particular source, there is significant overlap between the calculated
``visible" column densities and these observed values, although it should be acknowledged that there is a three orders-of-magnitude spread in the observed values.
\\ \\
Previous analytical models and numerical simulations derived the position of the $\mathrm{H_2O}$ snowline of an optically thick disk for given parameters, such as mass ($M_{*}$) and temperature ($T_{*}$) of the central star, a viscous parameter $\alpha$, an accretion rate $\dot{M}$, a gas-to-dust mass ratio $g/d$, and the average dust grain size $a$ and opacity (e.g., \citealt{Davis2005, Garaud2007, Min2011, Oka2011, Du2014, Harsono2015, Mulders2015, Piso2015}), and suggested that the position of the $\mathrm{H_2O}$ snowline changes, as these parameters change.
In the case of T Tauri disks with $M_{\mathrm{*}} \sim 0.5-1M_{\bigodot}$, $\dot{M} \sim 10^{-8} M_{\bigodot}$ yr$^{-1}$, and $a \sim 0.1 \mu$m, the position of the $\mathrm{H_2O}$ snowline is $\sim 1.5-2$ AU. 
In our calculations, we use similar parameters for $M_{\mathrm{*}}$,
$\dot{M}$ and $a$, and the $\mathrm{H_2O}$ snowline appears at a radius
of around 1.6AU in the midplane ($T_g\sim 150-160$K), which is
in the range of previous works.
\\ \\
\citet{Heinzeller2011} investigated the effects of physical mass transport phenomena in the radial direction by viscous accretion and in the vertical direction by diffusive turbulent mixing and disk winds.
They showed that the gas-phase $\mathrm{H_2O}$ abundance is enhanced in the warm surface layer due to the effects of vertical mixing. 
In contrast, they mentioned that the gas-phase $\mathrm{H_2O}$ abundance in the midplane inside the $\mathrm{H_2O}$ snowline is not affected by the accretion flow, since the chemical reactions are considered to be fast enough in this region to compensate for the effects of the accretion flow.
\subsection{$\mathrm{H_2O}$ emission lines from protoplanetary disks}
\noindent We perform ray-tracing calculations and investigate the profiles of $\mathrm{H_2O}$ emission lines for a protoplanetary disk in Keplerian rotation, using the methods described in Section 2.3 and next paragraph. We include rovibrational and pure rotational ortho- and para- $\mathrm{H_2O}$ lines at near-, mid-, and far-infrared and sub-millimeter wavelengths,
and find that $\mathrm{H_2O}$ lines which have small Einstein $A$ coefficients ($A_{ul}\sim 10^{-3}-10^{-6} \mathrm{s}^{-1}$) and relatively high upper energy levels ($E_{up} \sim$1000K) are most promising for tracing emission from the innermost hot water reservoir within the $\mathrm{H_2O}$ snowline.
\\ \\
Here we describe how we find 50 candidate lines which are selected from the LAMDA database of $\mathrm{H_2O}$ transition lines.
First of all, we proceeded by selecting about 20 $\mathrm{H_2O}$ lines from the LAMDA database which have various wavelengths (from near-infrared to sub-millimeter), Einstein $A$ coefficients ($A_{ul}\sim 10^{-1}-10^{-7} \mathrm{s}^{-1}$), and upper state energies ($E_{up} <$ 3000K). 
In making this initial selection we ignored lines with very small Einstein $A$ coefficients and very high upper state energies, since the emission fluxes of these lines are likely to weak to detect.
When we calculated the profiles of these lines, 
we noticed that 
$\mathrm{H_2O}$ lines with small Einstein $A$ coefficients ($A_{ul}\sim 10^{-3}-10^{-6} \mathrm{s}^{-1}$) and relatively large upper state energies ($E_{up}\sim 700-2100$K) are the best candidates to trace emission from the hot water reservoir within the $\mathrm{H_2O}$ snowline.
The number of these candidate lines is 10 lines within originally selected 20 lines.
Then we searched all other ortho-$\mathrm{H_2O}$ transition lines which satisfy these conditions, and found an additional 40 ortho-$\mathrm{H_2O}$ candidate water lines.
 \begin{table*}
  \caption{{Calculated ortho-$\mathrm{H_2O}$ line parameters and total line fluxes}}\label{tab:T1}
 \begin{center}
    \begin{tabular}{lllllll}
     \hline 
\vspace{-0.2cm}
     $J_{K_{a}K_{c}}$& \ \ $\lambda$&\ Freq.&\ $A_{ul}$&$E_{up}$&$n_{\mathrm{cr}}$& total flux$^{1}$\\
      &&&&&& \\
      &[$\mu$m]&[GHz]&[s$^{-1}$]&[K]&[$\mathrm{cm}^{-3}$]&[W $\mathrm{m}^{-2}$]\\
     \hline
      6$_{43}$-5$_{50}$ & 682.926 & 439.286 & 2.816$\times 10^{-5}$ &1088.7 & $1.0\times10^{6}$ &$3.12\times10^{-22}$ \\
      8$_{18}$-7$_{07}$ & 63.371 & 4733.995 & 1.772 & 1070.6 & $1.5\times10^{10}$ & $5.66\times10^{-18}$ \\
      1$_{10}$-1$_{01}$ & 538.664 & 556.933 & 3.497$\times 10^{-3}$& 61.0 & $2.9\times10^{7}$ & $1.13\times10^{-20}$ \\
\hline
  \multicolumn{3}{l}{\hbox to 0pt{\parbox{120mm}{
  \footnotesize
     \footnotemark[1] In calculating total flux of these $\mathrm{H_2O}$ lines, we use a distance $d=140$pc and the inclination angle of the disk $i=$30 deg.}}}
   \end{tabular}
\end{center}
    \end{table*}
\\ \\
In the remaining part of this Section, we describe the detailed properties of three characteristic pure rotational ortho-$\mathrm{H_2O}$ lines ($\lambda$=682.93, 63.37, 538.66$\mu$m). These three lines have different values of $A_{ul}$ and $E_{up}$.
We find that the $\mathrm{H_2O}$ 682.93$\mu$m line, which falls in ALMA band 8 (see Section 3.2.6), 
is a candidate for tracing emission from the innermost hot water reservoir within the $\mathrm{H_2O}$ snowline.
The 63.37 and 538.66$\mu$m lines are examples of lines which are less suited to trace emission from water vapour within the $\mathrm{H_2O}$ snowline.
We consider these two particular lines to test the validity of our model calculations, since the fluxes of these two lines from protoplanetary disks are observed with $Herschel$ (see Section 3.2.2, 3.2.3).
The list of suitable lines from mid-infrared (Q band) to sub-millimeter, and their properties, especially the variation in line fluxes with wavelength, are described in detail in our companion paper (paper II, \citealt{Notsu2016b}).
\begin{figure*}[htbp]
\begin{center}
\includegraphics[scale=0.42]{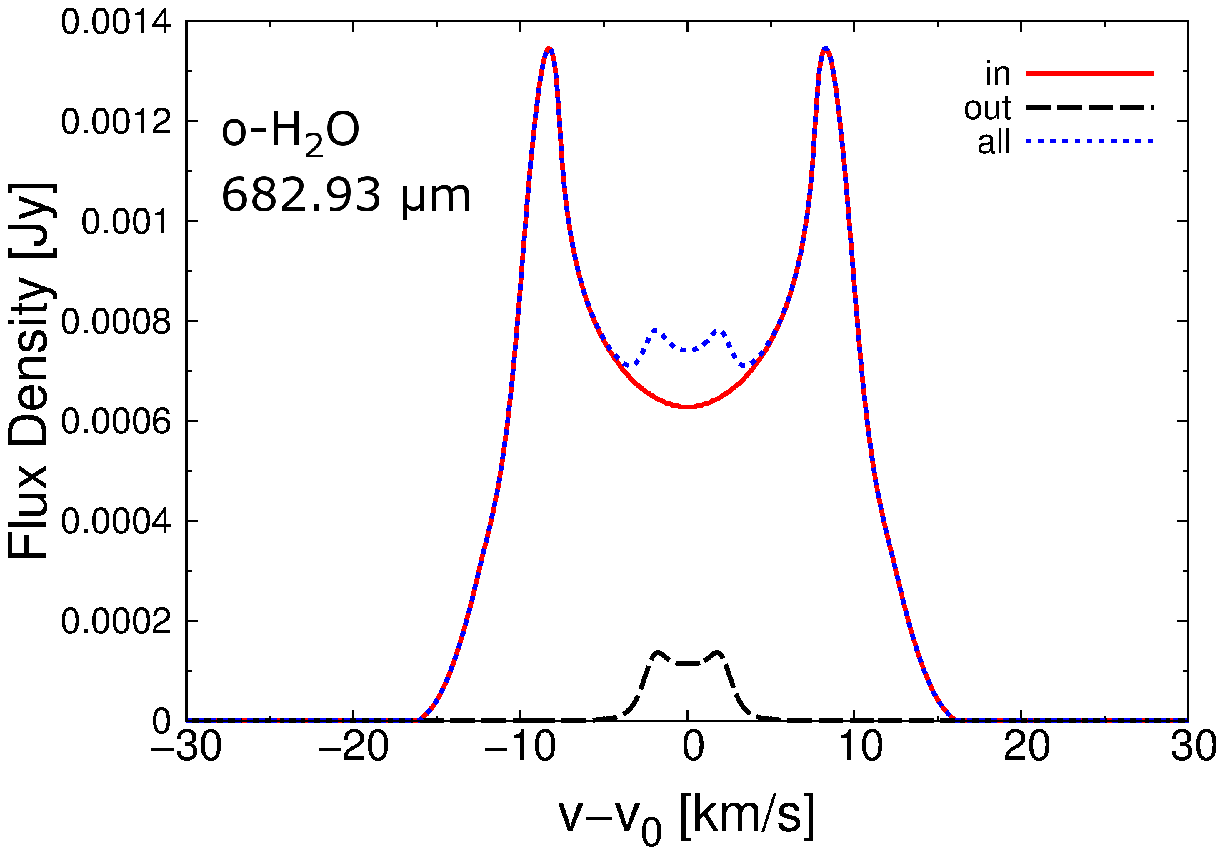}
\includegraphics[scale=0.42]{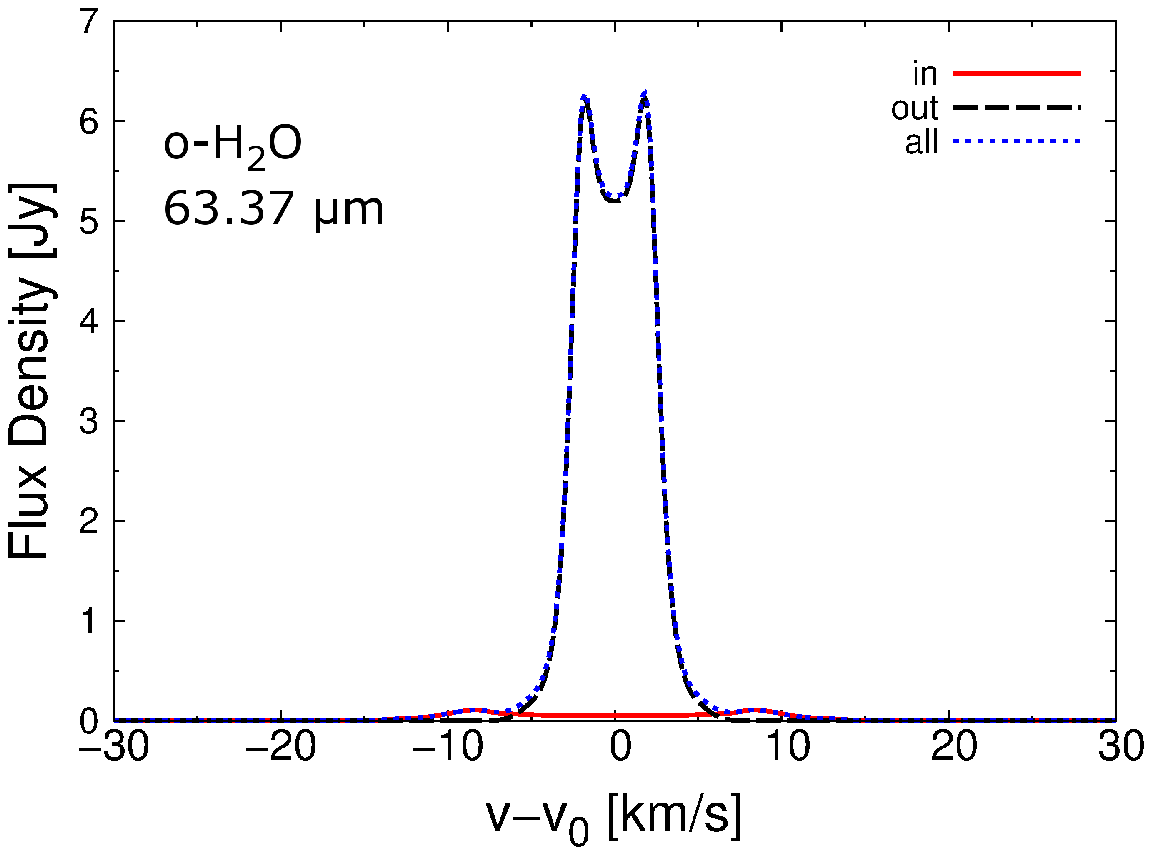}
\includegraphics[scale=0.42]{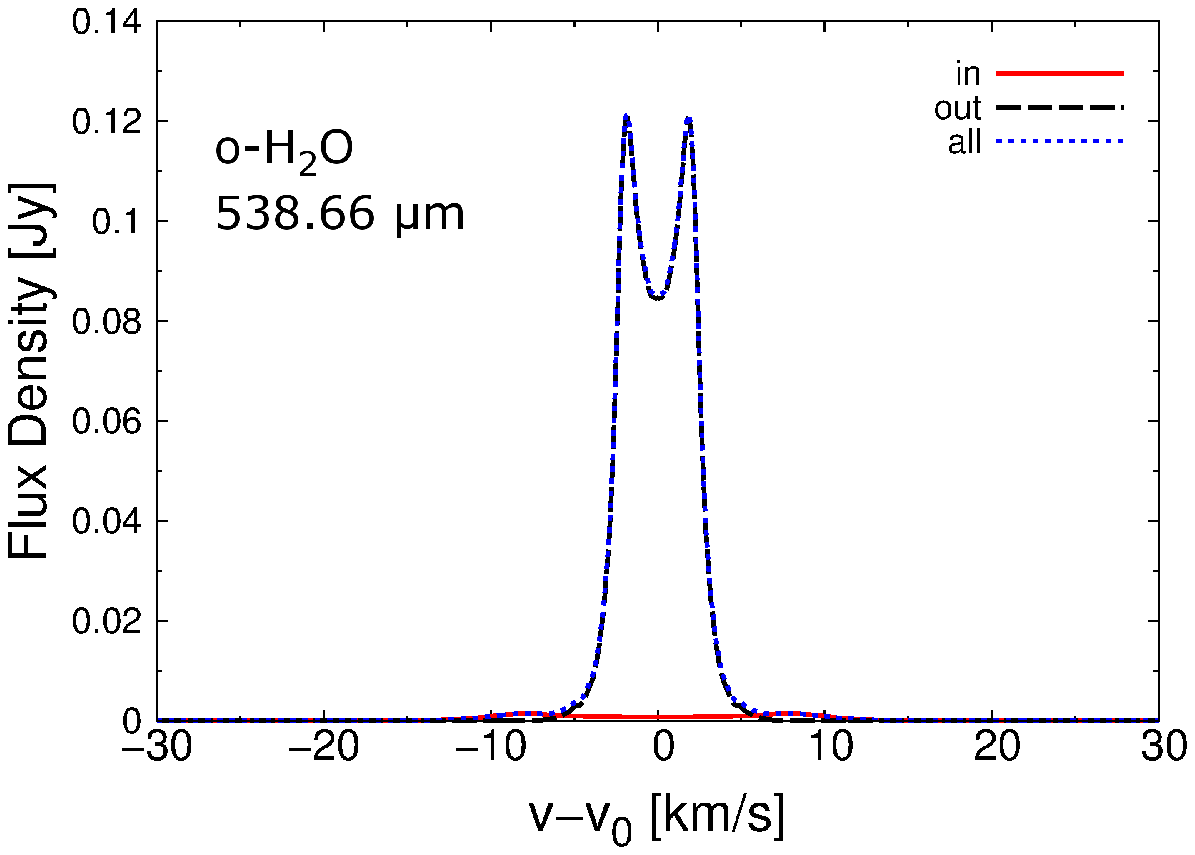}
\includegraphics[scale=0.42]{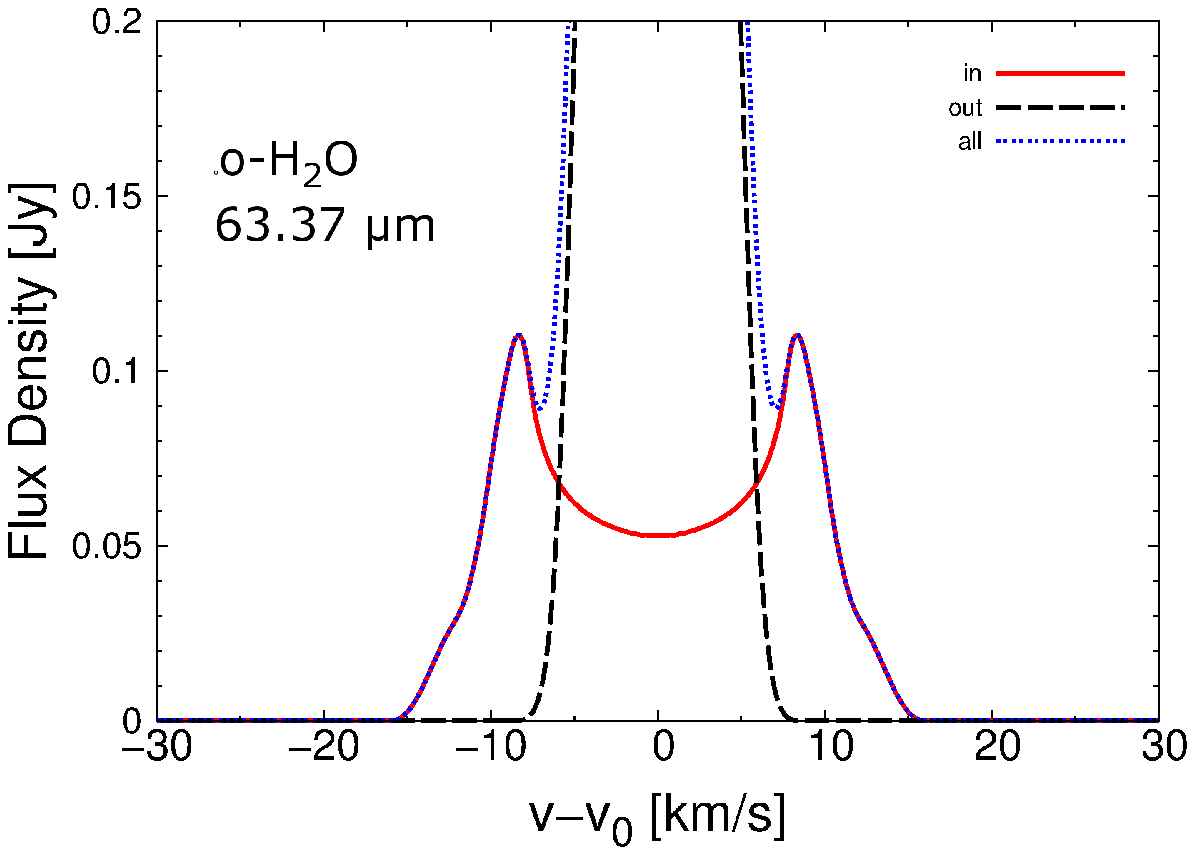}
\includegraphics[scale=0.42]{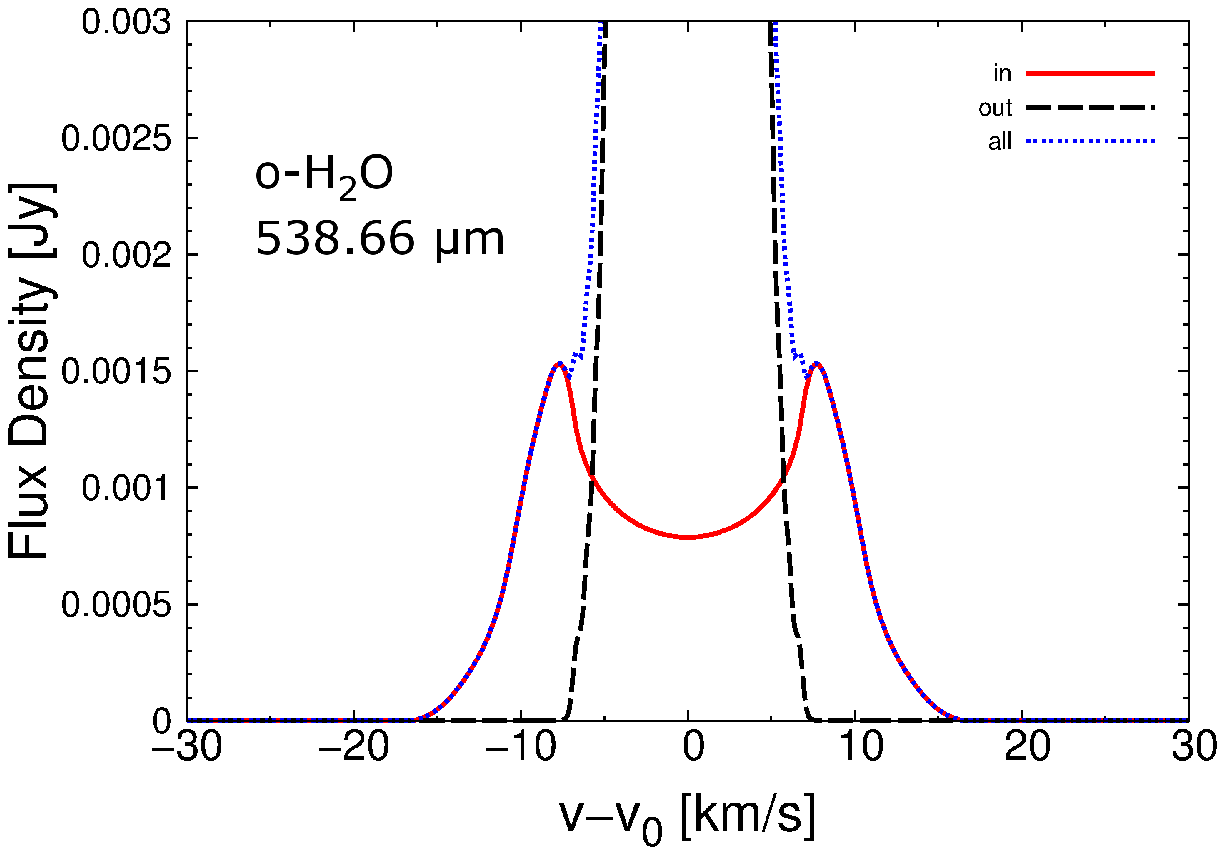}
\end{center}
\caption{\noindent Top: the velocity profiles of three characteristic pure rotational ortho-$\mathrm{H_2O}$ lines at $\lambda$=682.93$\mu$m ($J_{K_{a}K_{c}}$=6$_{43}$-5$_{50}$, top left), 63.37$\mu$m ($J_{K_{a}K_{c}}$=8$_{18}$-7$_{07}$, top middle), and 538.66$\mu$m ($J_{K_{a}K_{c}}$=1$_{10}$-1$_{01}$, top right), which have various Einstein $A$ coefficients $A_{ul}$ and upper state energies $E_{up}$.
Bottom: the velocity profiles of the ortho-$\mathrm{H_2O}$ 63.37$\mu$m line (bottom left) and 538.66$\mu$m line (bottom right), which enlarge the inner components.
{\it Red solid lines} are the emission line profiles from inside 2AU ($\sim$inside the $\mathrm{H_2O}$ snowline), {\it black dashed lines} are those from 2-30AU ($\sim$outside the $\mathrm{H_2O}$ snowline), and {\it blue dotted lines} are those from the total area inside 30AU. 
In calculating these profiles, we assume that the distance to the object $d$ is 140pc ($\sim$ the distance of Taurus molecular cloud), and the inclination angle of the disk $i$ is 30 deg.}\label{Figure4_original}
\end{figure*}  
\subsubsection{The case of a suitable $\mathrm{H_2O}$ emission line to trace emission from the hot water reservoir within the $\mathrm{H_2O}$ snowline}
\begin{figure}[htbp]
\begin{center}
\includegraphics[scale=0.65]{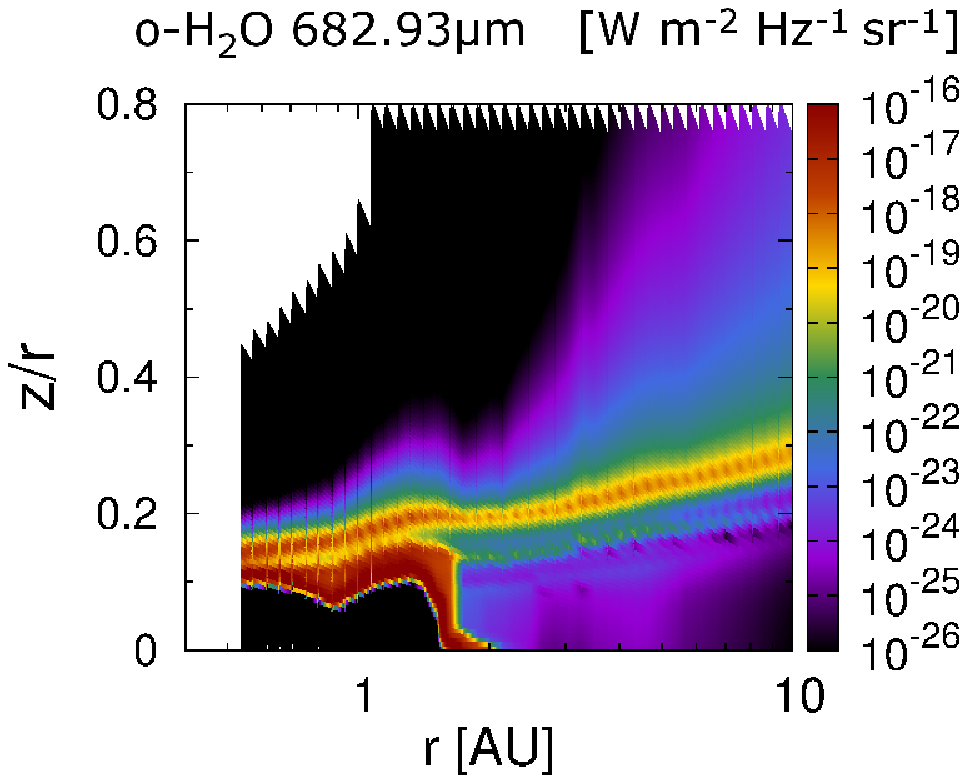}
\includegraphics[scale=0.65]{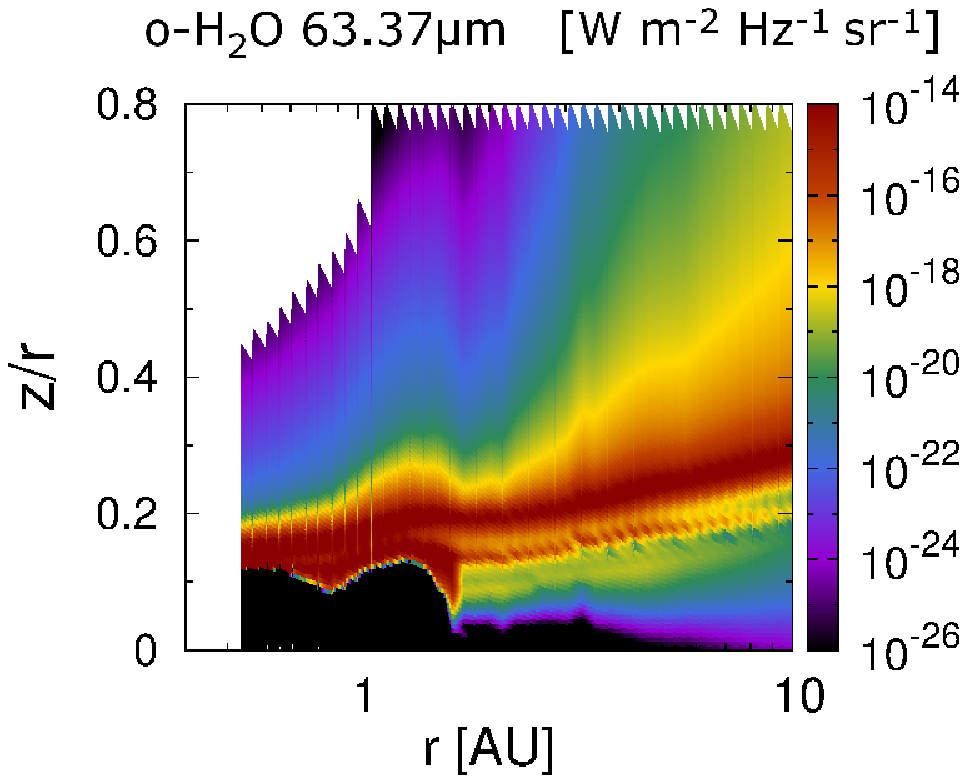}
\includegraphics[scale=0.65]{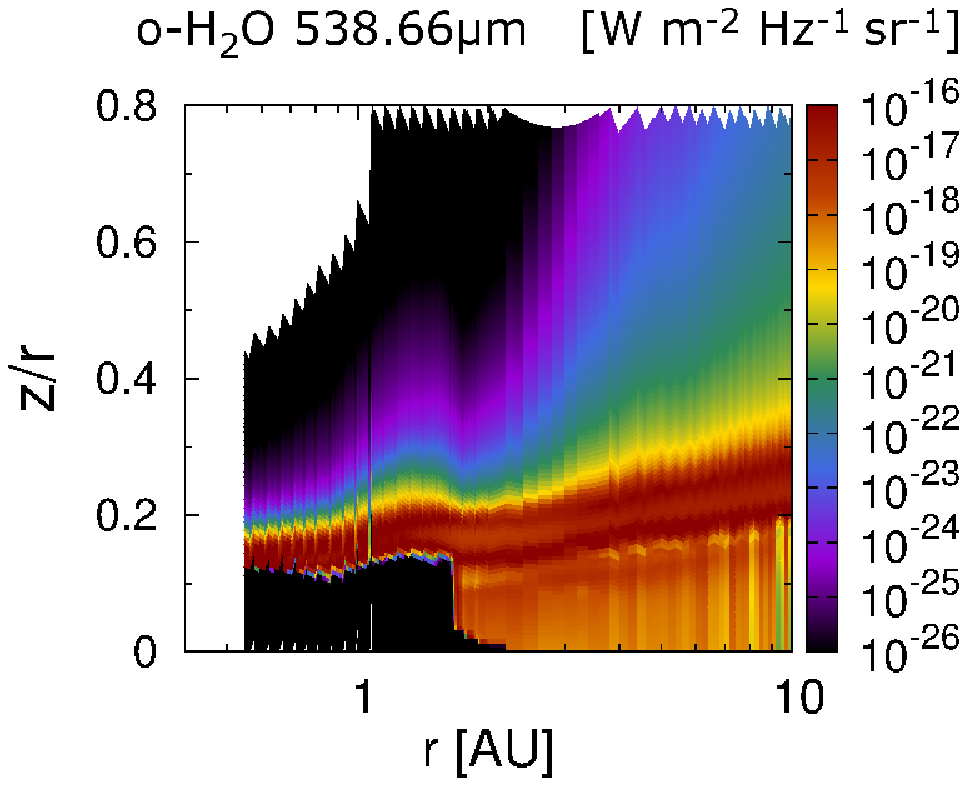}
\end{center}
\caption{\noindent The line of sight emissivity distributions of the three characteristic pure rotational ortho-$\mathrm{H_2O}$ lines with $\lambda$=682.93$\mu$m (top), 63.37$\mu$m (middle), and 538.66$\mu$m (bottom). 
The dimension is W $\mathrm{m}^{-2}$ $\mathrm{Hz}^{-1}$ ${\mathrm{sr}}^{-1}$.
We assume that the inclination angle of the disk $i$ is 0 degree in making these figures, and thus the direction of line-of-sight is from z=+$\infty$ to -$\infty$ at each disk radius.}\label{Figure5_original}\end{figure}

\noindent 
The top panels in Figure \ref{Figure4_original} show the emission profiles of three pure rotational ortho-$\mathrm{H_2O}$ lines at $\lambda$=682.93$\mu$m ($J_{K_{a}K_{c}}$=6$_{43}$-5$_{50}$, top left), 63.37$\mu$m ($J_{K_{a}K_{c}}$=8$_{18}$-7$_{07}$, top middle), and 538.66$\mu$m ($J_{K_{a}K_{c}}$=1$_{10}$-1$_{01}$, top right), 
which have various Einstein $A$ coefficients ($A_{ul}$) and upper state energies ($E_{up}$).
The detailed parameters, such as transitions ($J_{K_{a}K_{c}}$), wavelength, frequency, $A_{ul}$, $E_{up}$, critical density $n_{\mathrm{cr}}$, and total line fluxes of these three $\mathrm{H_2O}$ lines are listed in Table \ref{tab:T1}.
In calculating these profiles, we assume that the distance $d$ to the object is 140pc ($\sim$ the distance of Taurus molecular cloud), and the inclination angle $i$ of the disk is 30 degs.
The total fluxes of these three lines ($\lambda$=682.93, 63.37, 538.66$\mu$m) are $3.12\times 10^{-22}$, $5.66\times 10^{-18}$, $1.13\times 10^{-20}$ W $\mathrm{m}^{-2}$, respectively.
The bottom panels in Figure \ref{Figure4_original} show the velocity profiles of the $\mathrm{H_2O}$ 63.37$\mu$m line (bottom left) and the 538.66$\mu$m line (bottom right), which enlarge the inner components.
\\ \\
Since the $\mathrm{H_2O}$ lines at $\lambda$=682.93 and 63.37$\mu$m have large upper state energies ($E_{up}$=1088.7K and 1070.6K), these lines trace the hot water vapor ($T_{g} \gtrsim$ a few hundred K). 
On the basis of the results of our chemical calculations, the abundance of $\mathrm{H_2O}$ gas is high in the optically thick hot inner region within the $\mathrm{H_2O}$ snowline near the equatorial plane ($T_{g}>$ 150K) and in the hot optically thin surface layer of the outer disk.
\\ \\
In the top left panel of Figure \ref{Figure4_original}, we show the $\mathrm{H_2O}$ line emission at 682.93$\mu$m. The contribution from the optically thin surface layer of the outer disk ({\it black dashed line}, 2-30AU, ``out" component) is very small compared with that from the optically thick region near the midplane of the inner disk ({\it red solid line}, 0-2AU, ``in" component).
This is because this $\mathrm{H_2O}$ 682.93$\mu$m line has a small $A_{ul}$ (=$2.816\times10^{-5}$ s$^{-1}$).
On the basis of Eqs. (13)-(16) in Section 2.3, the observable flux density is calculated by summing up the emissivity at each point ($j_{ul}(s,x,y,\nu)$) in the line-of-sight
direction. In the optically thin ($\tau_{ul}$$<<$ 1) region (e.g., the disk surface layer), the flux density is roughly characterized by integrating the values of $n_{u}(s,x,y)A_{ul}$ at each point. On the other hand, in the optically thick ($\tau_{ul}$$\geq$ 1) region (e.g., the disk midplane of the inner disk), the flux density is independent of $n_{u}(s,x,y)$ and $A_{ul}$ at each point, and it becomes similar to the value of the Planck function at $T_{g}$ around the region of $\tau_{ul}\sim1$. Therefore, the emission profile of the $\mathrm{H_2O}$ 682.93$\mu$m line which has a small $A_{ul}$ and a relatively high $E_{up}$ mainly traces the hot $\mathrm{H_2O}$ gas inside the $\mathrm{H_2O}$ snowline, and shows the characteristic double-peaked profile due to Keplerian rotation. 
In this profile, the position of the two peaks and the rapid drop in flux density between the peaks gives information on the distribution of hot $\mathrm{H_2O}$ gas
within the $\mathrm{H_2O}$ snowline.
This profile potentially contain information which can be used to determine the $\mathrm{H_2O}$ snowline position.
The spread in the wings of the emission profile (high velocity regions) represents the inner edge of the $\mathrm{H_2O}$ gas distribution in the disk.
This is because emission from each radial region in the disk is Doppler-shifted due to the Keplerian rotation. Because the area near the outer emitting region is larger than that of the inner region ($\propto r^{2}$), the contribution to the emission from the region near the outer edge is larger if the emissivity at each radial point is similar.
\\ \\ 
Figure \ref{Figure5_original} shows the line-of-sight emissivity distributions of these three pure rotational ortho-$\mathrm{H_2O}$ lines. 
Figure \ref{Figure6_original} shows the total optical depth (gas emission and dust) distributions for the same transitions. 
We assume that the inclination angle, $i$, of the disk is 0 deg in making these figures, and thus the line-of-sight direction is from z=+$\infty$ to -$\infty$ at each disk radius.
According to the top panels of Figures \ref{Figure5_original} and \ref{Figure6_original}, the values of the emissivity at $r<$1.6AU (= the position of the $\mathrm{H_2O}$ snowline) and $z/r \sim 0.1$ are stronger than that of the other regions including the optically thin hot surface layer of the outer disk and the photodesorbed layer.
Although we cannot detect the emission from $z \sim 0$ because of the high optical depth of the inner disk midplane due to the absorption by dust grains and excited $\mathrm{H_2O}$ molecules, we can get information about the distribution of hot $\mathrm{H_2O}$ gas within the $\mathrm{H_2O}$ snowline.
This is because the $\mathrm{H_2O}$ gas fractional abundance is close to constant within $r<$1.6AU (= the position of the $\mathrm{H_2O}$ snowline) and $z/r \sim$ 0-0.1 (see also Section 3.1 and Figure \ref{Figure2_original}).
\subsubsection{The case of a $\mathrm{H_2O}$ emission line that traces the hot surface layer}
\noindent The top middle panel of Figure \ref{Figure4_original} where we show the line profile for the $\mathrm{H_2O}$ 63.37$\mu$m line, the contribution from the optically thin surface layer of the outer disk ({\it black dashed line}, 2-30AU, ``out" component) is large compared with that of the optically thick region near the midplane of the inner disk ({\it red solid line}, 0-2AU, ``in" component), and the shape of the line profile is a much narrower double peaked profile.
This is because this $\mathrm{H_2O}$ 63.37$\mu$m line has a large $A_{ul}$ (=1.772 s$^{-1}$), although $E_{up}$ (=1070.6K) is similar to that of the $\mathrm{H_2O}$ 682.93$\mu$m line (=1088.7K), and thus the flux density from the hot surface layer of the outer disk becomes strong.
%
Here we note that since the peak velocities of the ``in" and ``out" components are different, water lines with large $A_{ul}$ at infrared wavelengths, such as the $\mathrm{H_2O}$ 63.37$\mu$m line, can in principal trace the hot $\mathrm{H_2O}$ gas within the $\mathrm{H_2O}$ snowline.
However, there is no current or future instrument with enough sensitivity and spectral resolution to distinguish the peaks of the ``in" component from the ``out" component in these lines.
For example, SPICA/SAFARI is a future instrument with far-infrared spectrograph, but its spectral resolution is low ($R\sim$3000) and is not enough to distinguish the peaks of these line profiles.
The difference in the peak flux density is very large ($\gtrsim$ several tens) and the wings of both components are blended (see also the bottom left panel of Figure \ref{Figure4_original}).
\begin{figure}[htbp]
\begin{center}
\includegraphics[scale=0.65]{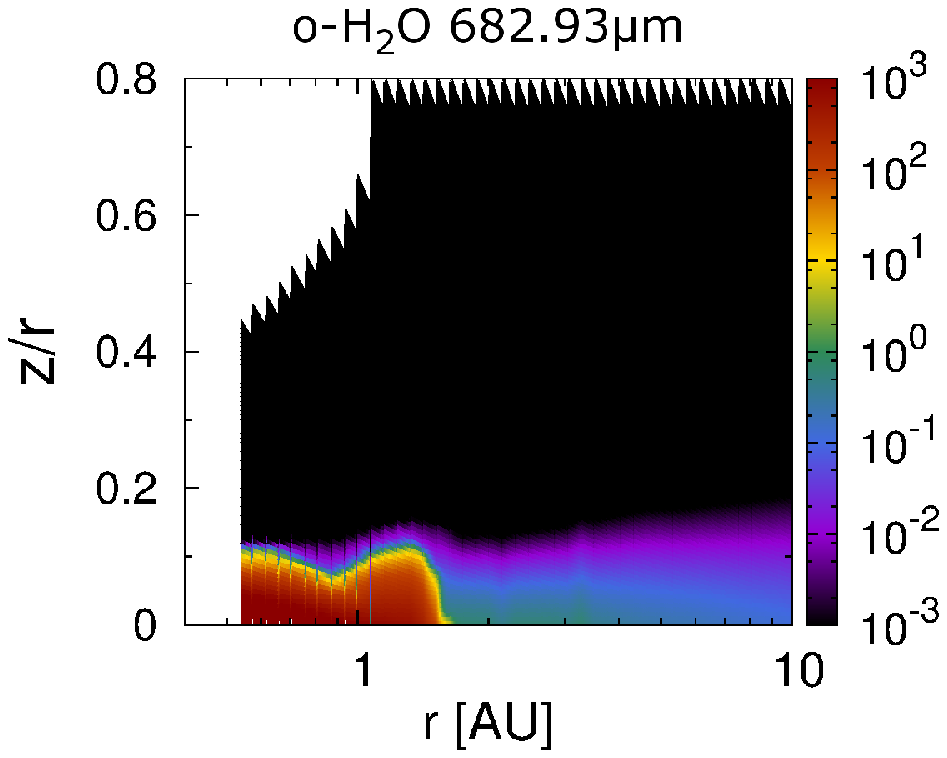}
\includegraphics[scale=0.65]{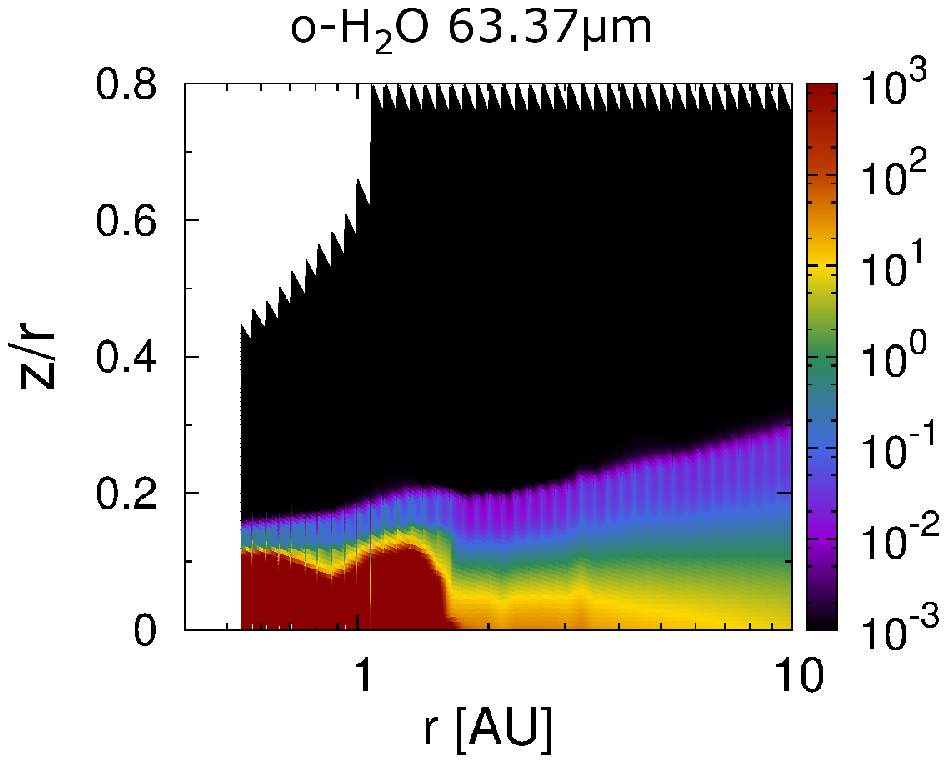}
\includegraphics[scale=0.65]{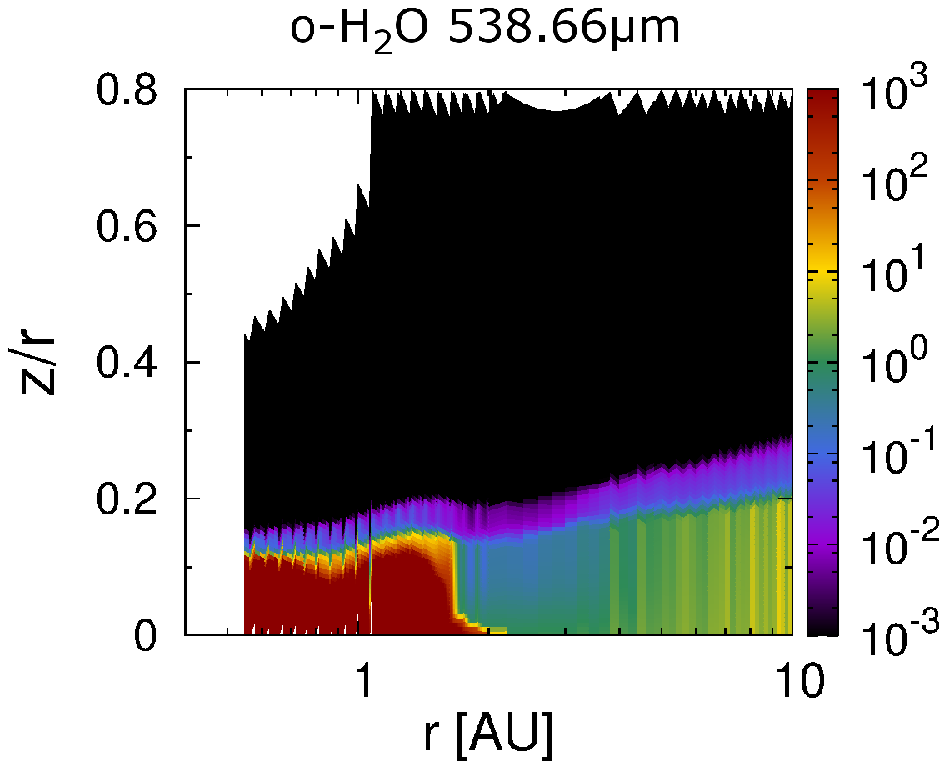}
\end{center}
\caption{\noindent The line of sight optical depth $\tau_{ul}(s,x,y,\nu)$ distributions of three characteristic pure rotational ortho-$\mathrm{H_2O}$ lines at $\lambda$=682.93$\mu$m (top), 63.37$\mu$m (middle), and 538.66$\mu$m (bottom). 
We assume that the inclination angle of the disk $i$ is 0 degree in making these figures, and thus the direction of line-of-sight is from z=+$\infty$ to -$\infty$ at each disk radius.}\label{Figure6_original}\end{figure}
\\ \\
According to the middle panels of Figures \ref{Figure5_original} and \ref{Figure6_original}, the values of the emissivity at each $(r,z)$ point in the optically thin hot surface layer of the outer disk and the photodesorbed layer 
are as strong as that of the optically thick region inside the $\mathrm{H_2O}$ snowline. 
For similar reasons as the case for the 682.93$\mu$m line, emission from the outer disk dominates.
In addition, the outer disk midplane opacity of this line is larger than that of the $\mathrm{H_2O}$ 682.93$\mu$m line, because the dust opacity becomes large at shorter wavelengths (e.g, \citealt{NomuraMillar2005}).
\\ \\
We mention that previous space far-infrared low dispersion spectroscopic observations with $Herschel$/PACS ($R\sim$1500) detected this line from some T Tauri disks and Herbig Ae disks (e.g., \citealt{Fedele2012, Fedele2013, Dent2013, Meeus2012, Riviere-Marichalar2012}).
Although the profiles of these lines are unresolved, comparison with models indicates that the emitting regions of these observations are thought to originate in the hot surface layer  (e.g., \citealt{Fedele2012, Riviere-Marichalar2012}). 
In addition, the total integrated line flux of classical T Tauri objects in the Taurus molecular cloud are observed to be $\sim 6\times 10^{-18} - 3\times 10^{-16}$ W $\mathrm{m}^{-2}$ (e.g., \citealt{Riviere-Marichalar2012}). These values have a dispersion factor of 50. \citet{Riviere-Marichalar2012} suggested that the objects with higher values of line flux have extended emission from outflows, in contrast to those with lower values which have no extended emissions (e.g., AA Tau, DL Tau, and RY Tau). The latter lower values are of the same order as the value we calculate here assuming a T Tauri disk model with no outflow and envelope.
\subsubsection{The case of a $\mathrm{H_2O}$ emission line that traces the cold water}
\noindent In the top right panel of Figure \ref{Figure6_original} where we show the line profile for the $\mathrm{H_2O}$ 538.66$\mu$m line, the contribution from the outer disk ({\it black dashed line}, 2-30AU, ``out" component) is large compared with that of the optically thick region near the midplane of the inner disk ({\it red solid line}, 0-2AU, ``in" component) and the shape of the profile is a much narrower double peaked profile (closer to a single peaked profile), although the $A_{ul}$ is not so high (=3.497$\times10^{-3}$s$^{-1}$). This is because this $\mathrm{H_2O}$ 538.66$\mu$m line is the ground-state rotational transition and has low $E_{up}$ (=61.0K) compared with the other lines.
The flux of this line comes mainly from the outer cold water reservoir in the photodesorbed layer (see also Section 3.1).
We propose that this line is not optimal to detect emission from the innermost water reservoir within the $\mathrm{H_2O}$ snowline 
for the same reasons explained in Section 3.2.2 for the 63.37$\mu$m line (see also the bottom right panel of Figure \ref{Figure4_original}).
\\ \\
According to the bottom panels of Figure \ref{Figure5_original} and \ref{Figure6_original}, the value of the emissivity at each $(r,z)$ point in the photodesorbed layer is comparable to that of the optically thick region inside the $\mathrm{H_2O}$ snowline.
The larger surface area of the outer disk, however, means that most disk-integrated emission arises from this region.
In addition, the outer disk midplane opacity of this line is larger than that of the $\mathrm{H_2O}$ 682.93$\mu$m line, although the wavelength and thus the dust opacity is similar. 
This is because the abundance of cold $\mathrm{H_2O}$ is relatively high, and because this line has low $E_{up}$.
\\ \\
We mention that previous space high dispersion spectroscopic observations with $Herschel$/HIFI detected the profiles of this line from disks around one Herbig Ae star (HD100546) and TW Hya (e.g., \citealt{Hogerheijde2011, vanDishoeck2014}). The number of detections is small since the line flux is low compared with the sensitivity of that instrument \citep{Antonellini2015}.
The detected line profiles and other line modeling work (e.g., \citealt{Meijerink2008, Woitke2009b, Antonellini2015, Du2015}) suggested that the emitting region arises in the cold outer disk, consistent with the results of our model calculations.
In addition, the total integrated line flux of TW Hya is observed to be $(1.7\pm1.1)\times 10^{-19}$ W $\mathrm{m}^{-2}$ \citep{Hogerheijde2011, Du2015}.
Considering the difference in distance between TW Hya ($\sim$51pc, e.g., \citealt{Zhang2013, Du2015}) 
and our assumed value, 140 pc, the observed flux is within about a factor $\approx 2$ of our estimated value (see also Table 1).
\\ \\
We note that previous observations suggested that the OPR of the emitting region is 0.77 for TW Hya \citep{Hogerheijde2011} derived using the observed para-$\mathrm{H_2O}$ ground state 1$_{11}$-$0_{00}$ 269.47$\mu$m line ($A_{ul}$=1.86$\times 10^{-2}$ and $E_{up}$=53.4K) and the observed ortho-$\mathrm{H_2O}$ ground state 538.66$\mu$m line.
Since we define OPR as 3 (=the value in the high temperature region) throughout the disk (see also Section 2.3), we likely overestimate the line flux of the ortho-$\mathrm{H_2O}$ 538.66$\mu$m line.
In addition, since the flux of this line is controlled by the outer cold $\mathrm{H_2O}$ gas which is desorbed from the cold dust-grain surfaces, it is necessary to include grain-surface reactions (e.g., \citealt{Hasegawa1992}) to calculate the $\mathrm{H_2O}$ gas and ice abundance to more accurately model this region.
\begin{figure}[htbp]
\begin{center}
\includegraphics[scale=0.53]{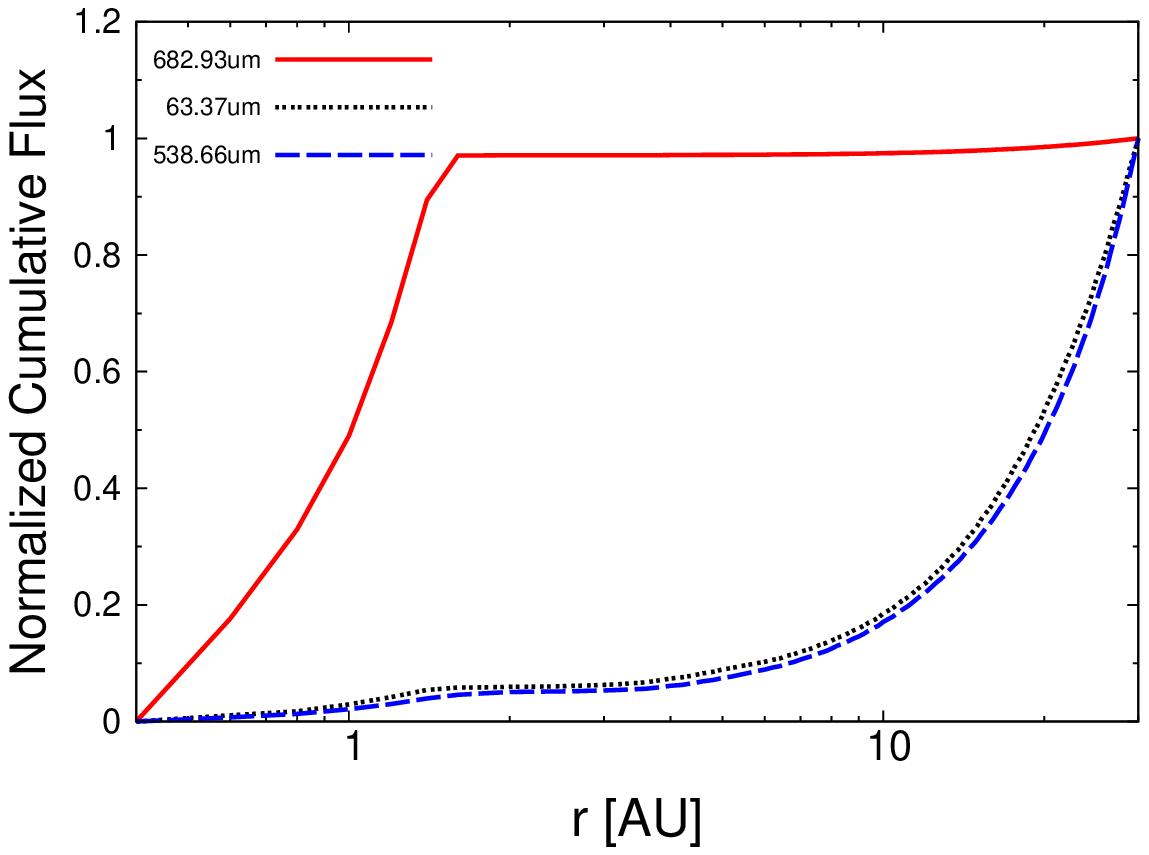}
\includegraphics[scale=0.53]{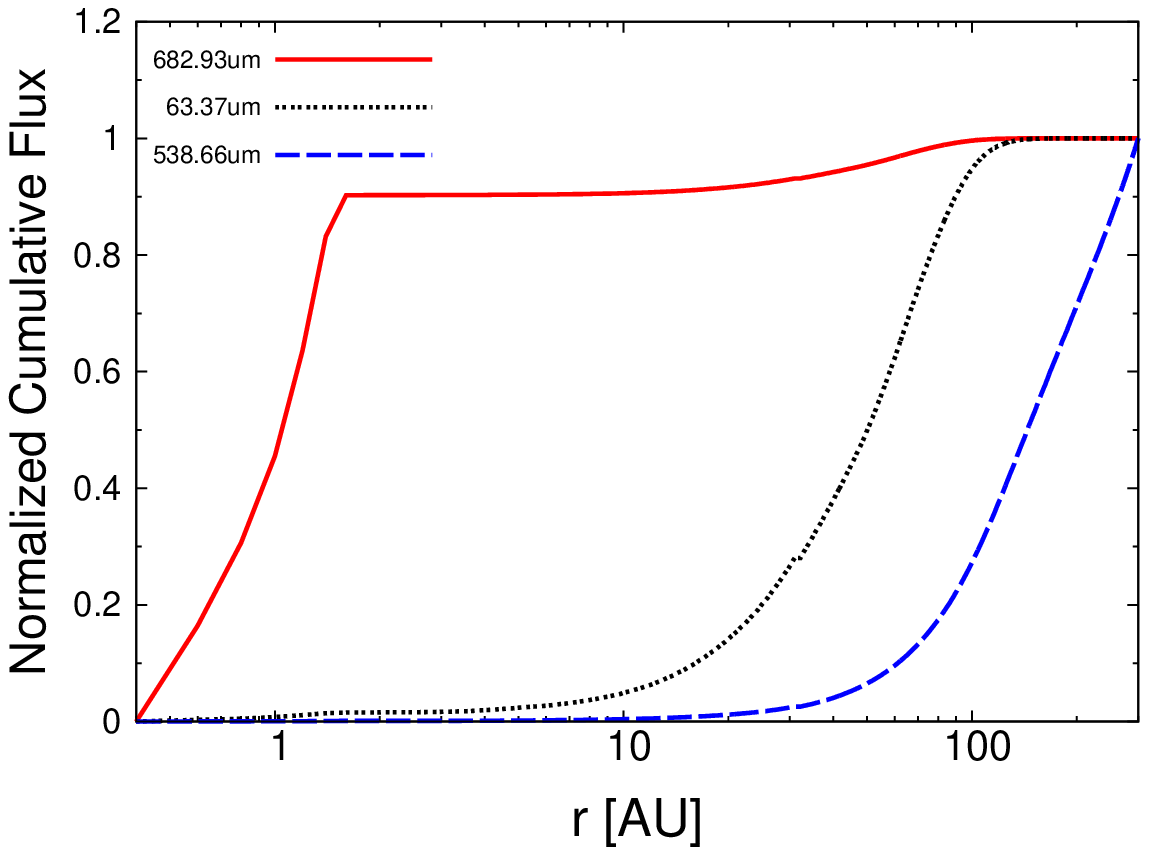}
\end{center}
\vspace{0.3cm}
\caption{\noindent The radial distributions of the normalized cumulative flux for three pure rotational ortho-$\mathrm{H_2O}$ lines at $\lambda$=682.93$\mu$m ({\it red solid line}), 63.37$\mu$m ({\it black dotted line}), and 538.66$\mu$m ({\it blue dashed line}). We normalized the cumulative flux of each line using the values at $r=30$AU (top panel) and at $r=300$AU (bottom panel).
We assume that the inclination angle of the disk $i$ is 0 degree in making these figures.}\label{FigureA2_add}
\end{figure}
\subsubsection{Influence of model assumptions}
\noindent Figure \ref{FigureA2_add} shows the radial distributions of normalized cumulative fluxes for these three pure rotational ortho-$\mathrm{H_2O}$ lines. 
We normalized the values of cumulative fluxes of these lines using the values at $r=30$AU (top panel) and at $r=300$AU (bottom panel).
According to these panels, the 682.93$\mu$m line is emitted mostly from the region inside the $\mathrm{H_2O}$ snowline.
In contrast, the 63.37$\mu$m line and the 538.66$\mu$m line are emitted mostly from the region outside the $\mathrm{H_2O}$ snowline.
In addition, although the 63.37$\mu$m line is mainly emitted from the region between $r\sim10-100$AU, the 538.66$\mu$m line is mainly emitted from a region much further out ($r\sim 50-300$AU).
This is because the 682.93$\mu$m line has a small $A_{ul}$ and a relatively high $E_{up}$, and thus it mainly emits from the hot $\mathrm{H_2O}$ gas inside the $\mathrm{H_2O}$ snowline. In contrast, the 63.37$\mu$m line has a large $A_{ul}$, although $E_{up}$ is similar to that of the 682.93$\mu$m line, and thus the flux density from the hot surface layer of the outer disk is strong (see also Section 3.13 and 3.2.2).
Moreover, the flux density of the 538.66$\mu$m line from the outer cold water reservoir in the photodesorbed layer is strong, since this line is the ground-state rotational transition and has low $E_{up}$ compared with the other lines (see also Section 3.1 and 3.2.3).
These results suggest that the total fluxes of the 538.66$\mu$m line (and partly the 63.37$\mu$m line) will be influenced by the size of the disk which is included in the calculation of the  line profiles, although the 682.93$\mu$m line does not have this problem because the line emitting region is sufficiently small.
\\ \\
\noindent Although we adopt a dust-grain size distribution with a maximum radius of $a_{\mathrm{max}}\sim$10$\mu$m throughout the disk (see Appendix A), dust grains are expected to grow in size due to settling and coagulation as the disk evolves and planet formation proceeds.
\citet{Aikawa2006} calculated disk physical structures with various dust-grain size distributions.
In addition, \citet{Vasyunin2011} and \citet{Akimkin2013} calculated the chemical structure of the outer disk ($\gtrsim$10AU) with grain evolution and discuss its features.
They showed that the dust-grain settling and growth reduce the total
dust-grain surface area and lead to higher UV irradiation rates in the upper disk. Therefore, the hot surface layer of the outer disk which contains abundant gas-phase molecules, including $\mathrm{H_2O}$, gets wider and shifts closer to the disk midplane, thus the abundances and column densities of species are enhanced. 
However, they did not discuss the midplane structure of the inner disk including the position of the $\mathrm{H_2O}$ snowline, since they restricted their calculations to the outer disk ($\gtrsim$10AU).
Here, we note that the position of the $\mathrm{H_2O}$ snowline in such an evolved disk is expected to be closer to the central star, since the total dust-grain surface area and thus dust opacity decreases as the size of dust grains becomes large, leading to a decrease in dust-grain and gas temperatures in the midplane of the inner disk \citep{Oka2011}.
Moreover, \citet{Ros2013}, \citet{Zhang2015}, and \citet{Banzatti2015} discussed the effects of rapid dust-grain growth that leads to pebble-sized particles near the $\mathrm{H_2O}$ snowline.
\\ \\
As we explained in Section 2.1, the dominant dust heating source in the disk midplane of the inner disk is the radiative flux produced by viscous dissipation ($\alpha$-disk model) which determines the dust and
gas temperature of the region.
Recent studies (e.g., \citealt{Davis2005, Garaud2007, Min2011, Oka2011, Harsono2015, Piso2015}) calculated the evolution of the position of the $\mathrm{H_2O}$ snowline in optically thick disks, and showed that it migrates as the disk evolves and as the mass accretion rate in the disk decreases, since the radiative flux produced by viscous dissipation becomes larger as the mass accretion rate increases.
We suggest that younger protoplanetary disks like HL Tau  \citep{ALMA2015} are expected to have a larger mass accretion rate compared with that of our reference T Tauri disk model, and the position of the $\mathrm{H_2O}$ snowline will reside further out in the disk midplane.
\citet{Zhang2015} argue that the center of the prominent innermost gap at 13 AU is coincident with the expected midplane condensation front of water ice.
Here we note that \citet{Banzatti2015} and \citet{Okuzumi2016} report the position of the $\mathrm{H_2O}$ snowline in HL Tau as $\lesssim$ 10 AU.
The difference occurs because the midplane radial temperature profile of \citet{Zhang2015} is larger than those of \citet{Banzatti2015} and \citet{Okuzumi2016}.
%
\\ \\
As we described in Section 2.2 and Appendix C, we adopt the wavelength integrated UV 
flux calculated at each point by Eqs.~(1) and (2) to approximate the photoreaction rates $k^{\mathrm{ph}}(r, z)$ 
and photodesorption rate $k_{i}^{\mathrm{pd}}$.  
This UV flux is estimated by summing up the fluxes of three components: photospheric blackbody radiation, 
optically thin hydrogenic bremsstrahlung radiation, and strong Ly$\alpha$ line (see also Appendix A). 
\citet{Walsh2012} pointed out that using Eqs.~(1) and (2), we may overestimate the strength of the UV field 
at wavelengths other than the Ly$\alpha$ ($\sim 1216\mathrm{\mathring{A}}$).  
On the basis of their calculations, if we adopt the wavelength dependent UV flux to calculate photochemical reaction 
rates, the fractional abundance of $\mathrm{H_2O}$ vapor in the outer disk surface becomes larger 
because of the combination of increased gas phase production and decreased photodestruction.  
In contrast, the fractional abundance of $\mathrm{H_2O}$ vapor in the inner disk midplane is not expected to 
change, since the UV flux plays a minor role in determining physical and chemical structures around the $\mathrm{H_2O}$ snowline (see Figure \ref{Figure1_original}).
\citet{Walsh2012} suggested that the column density of $\mathrm{H_2O}$ vapor in the outer disk can be 
enhanced by an order of magnitude depending on the method used to calculate the photodissociation rates.
\\ \\
\begin{figure*}[htbp]
\begin{center}
\includegraphics[scale=0.5]{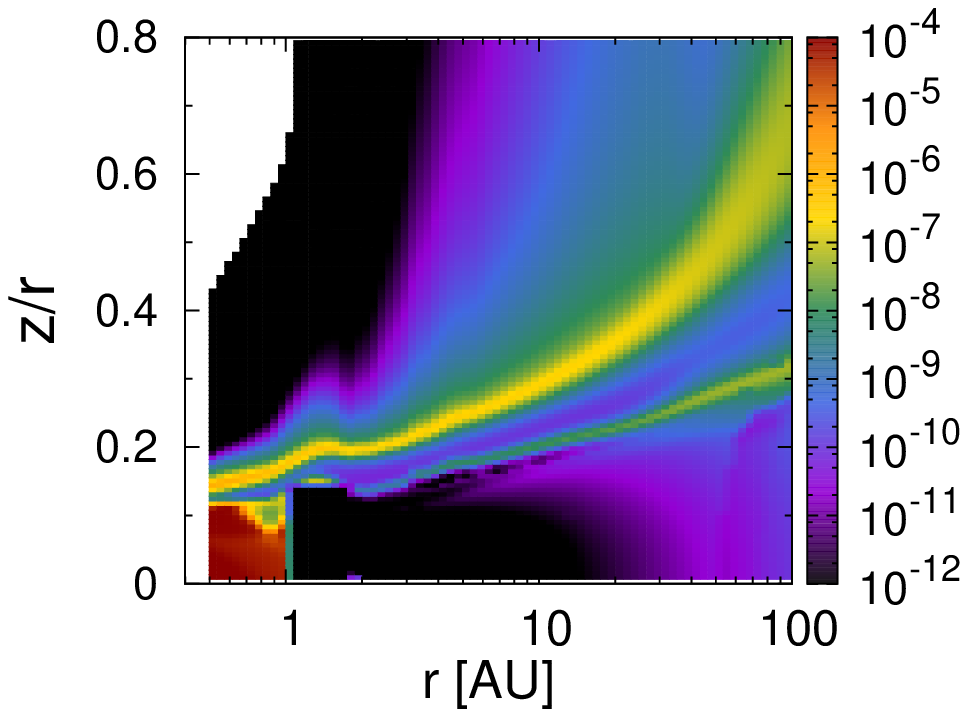}
\includegraphics[scale=0.5]{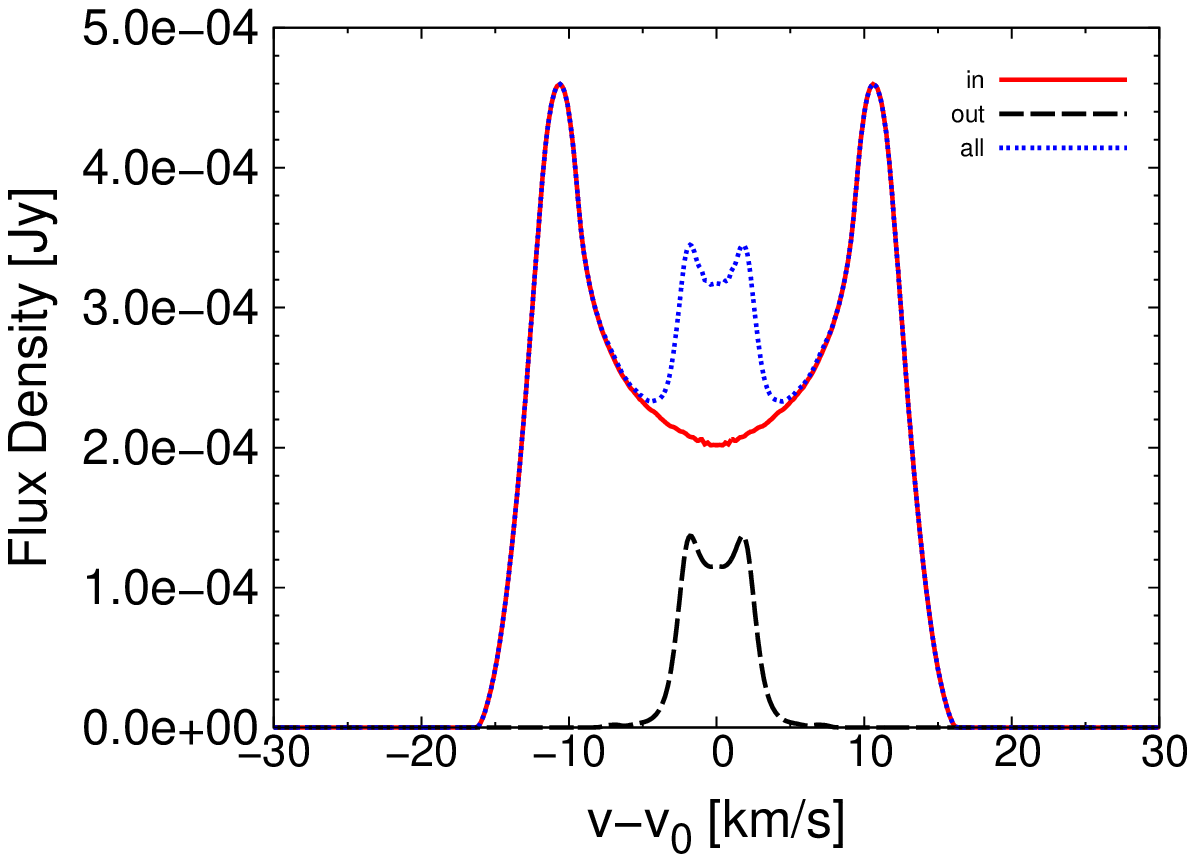}
\includegraphics[scale=0.5]{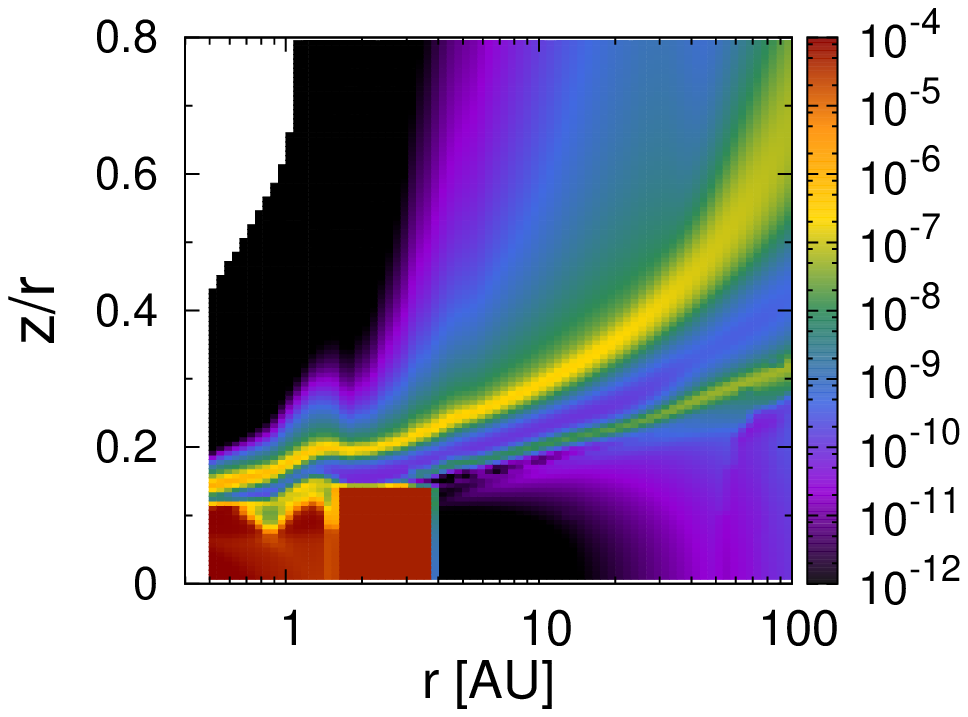}
\includegraphics[scale=0.5]{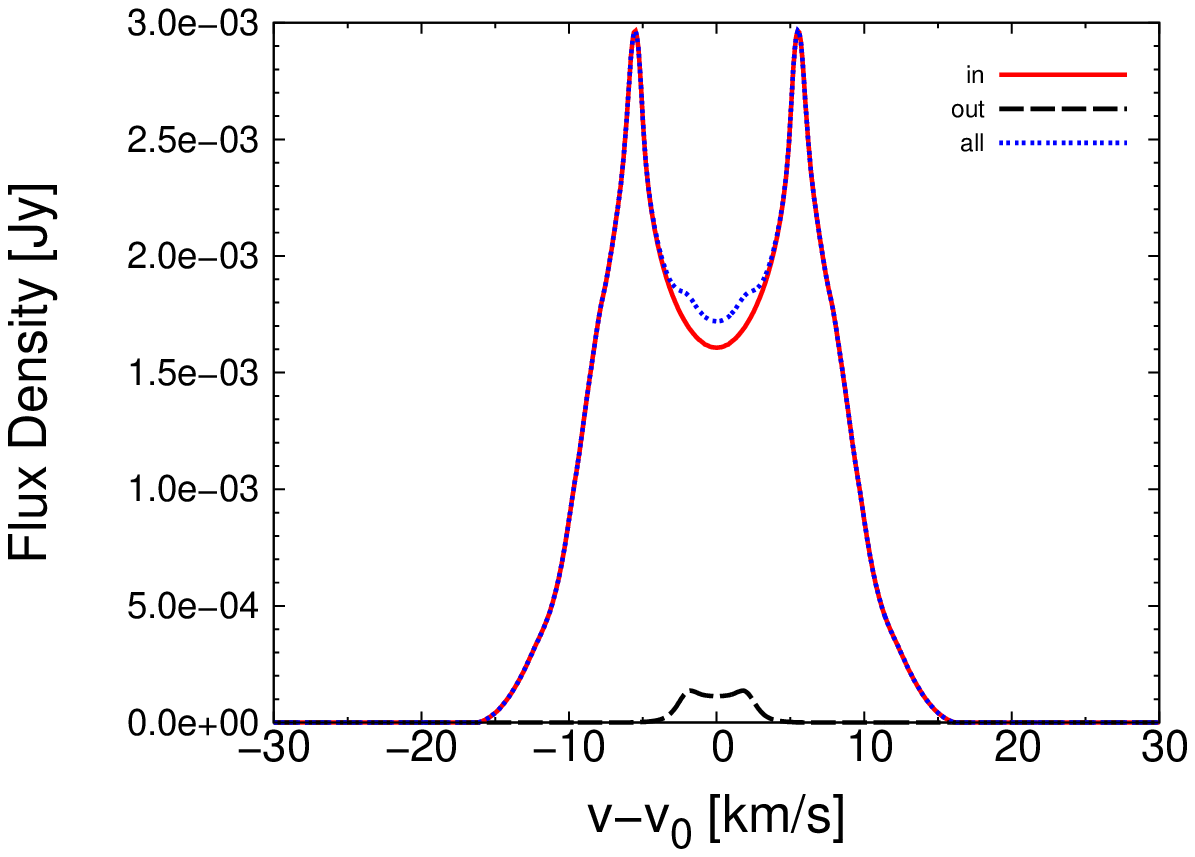}
\includegraphics[scale=0.5]{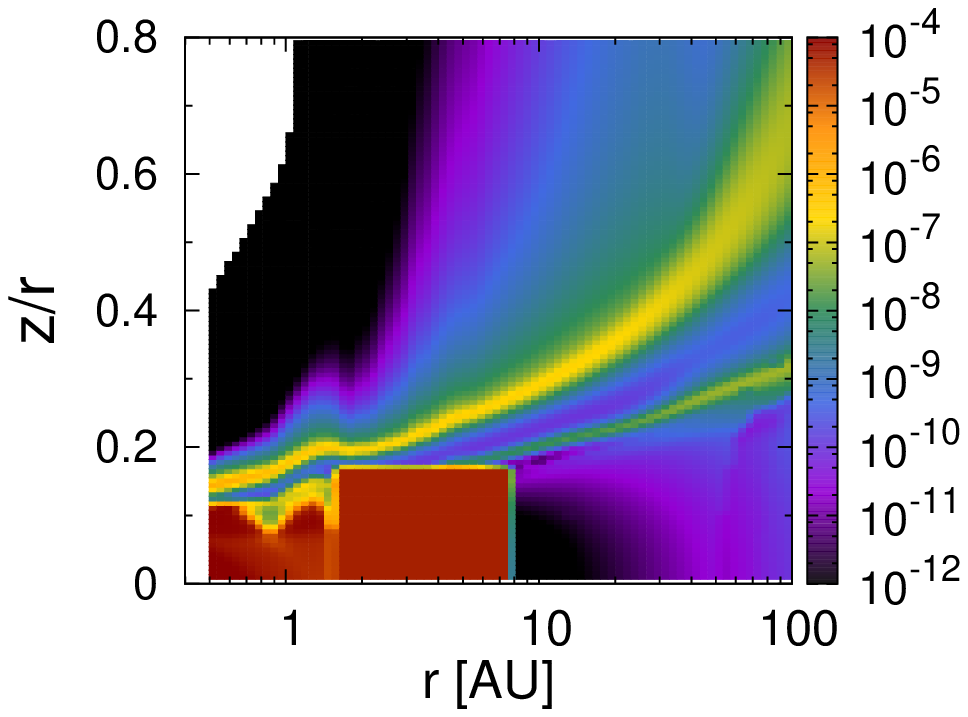}
\includegraphics[scale=0.5]{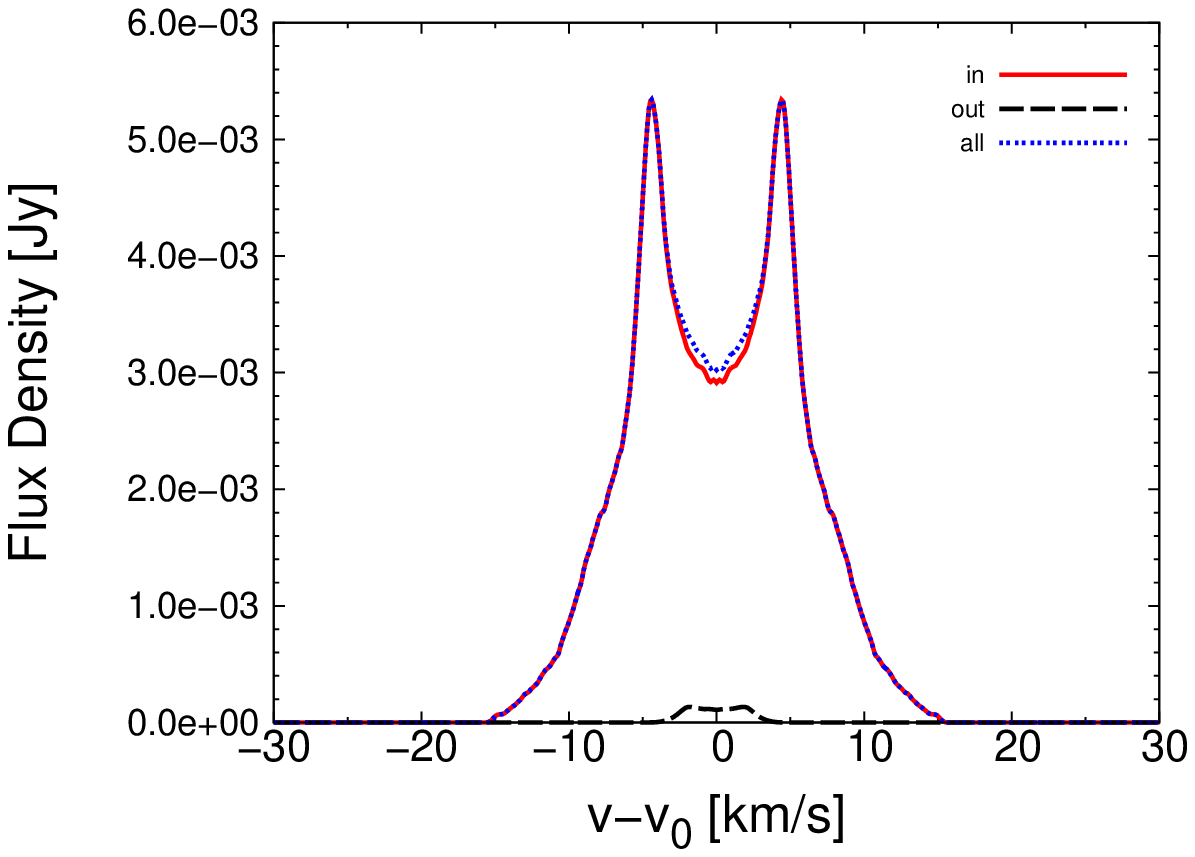}
\end{center}
\vspace{0.3cm}
\caption{\noindent The left three panels: the fractional abundance (relative to total hydrogen nuclei density) distributions of $\mathrm{H_2O}$ gas of a disk around a T Tauri star as a function of disk radius and height (scaled by the radius, $z/r$) up to maximum radii of $r=$100AU. We change the positions of the $\mathrm{H_2O}$ snowline to 1 AU (top left), 4 AU (middle left), 8 AU (bottom left) by hand, in order to test the sensitivity for the position of the $\mathrm{H_2O}$ snowline. The right three panels: The velocity profiles of the pure rotational ortho-$\mathrm{H_2O}$ lines at $\lambda$=682.93$\mu$m ($J_{K_{a}K_{c}}$=6$_{43}$-5$_{50}$). The three panels correspond to the cases that the $\mathrm{H_2O}$ snowline is assumed to be 1 AU (top right) 4AU (middle right), 8AU (bottom right). {\it Red solid lines} are the emission line profiles from inside 1, 4, 8AU ($\sim$inside the $\mathrm{H_2O}$ snowline), {\it black dashed lines} are those from 1-30, 4-30, 8-30AU ($\sim$outside the $\mathrm{H_2O}$ snowline), and {\it blue dotted lines} are those from the total area inside 30AU, respectively. In calculating these profiles, we assume that the distance to the object $d$ is 140pc ($\sim$ the distance of Taurus molecular cloud), and the inclination angle of the disk $i$ is 30 deg.}\label{FigureA3_add}
\end{figure*} 

\setcounter{figure}{7}
 \begin{figure*}[htbp]
\begin{center}
\includegraphics[scale=0.5]{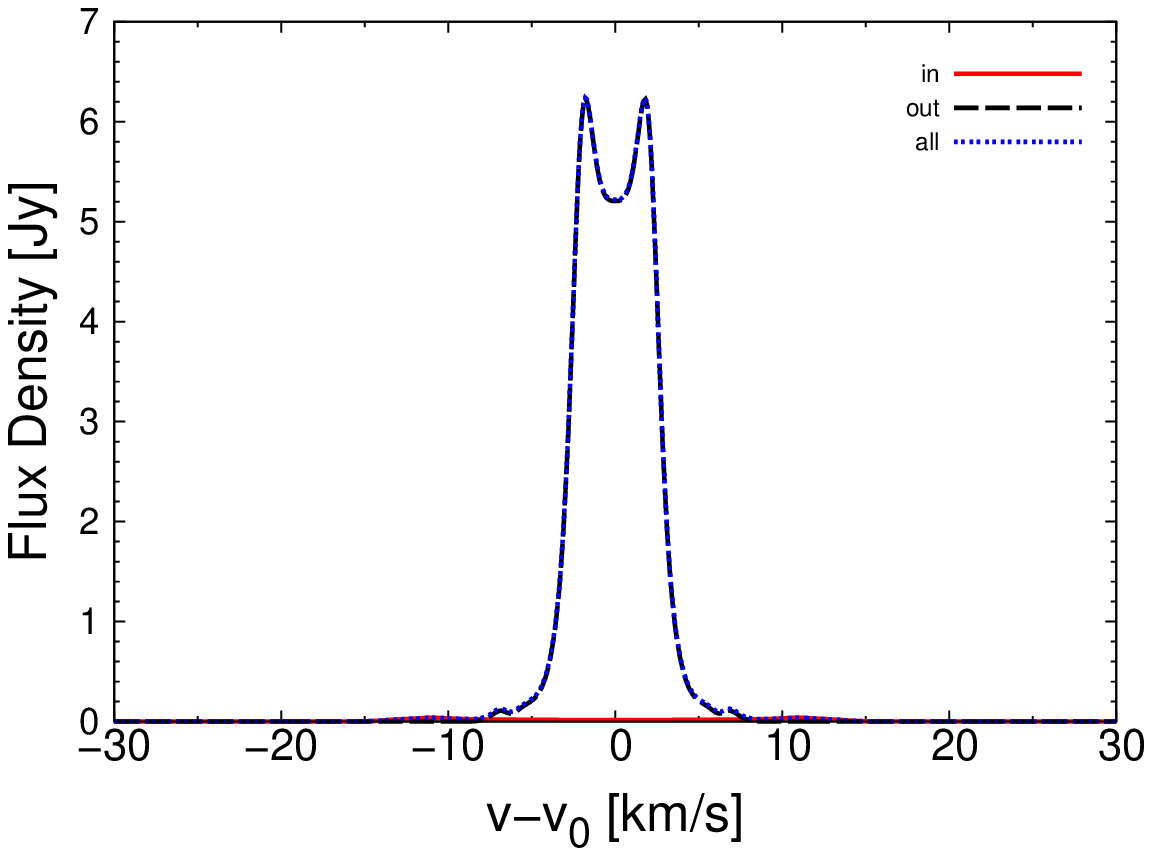}
\includegraphics[scale=0.5]{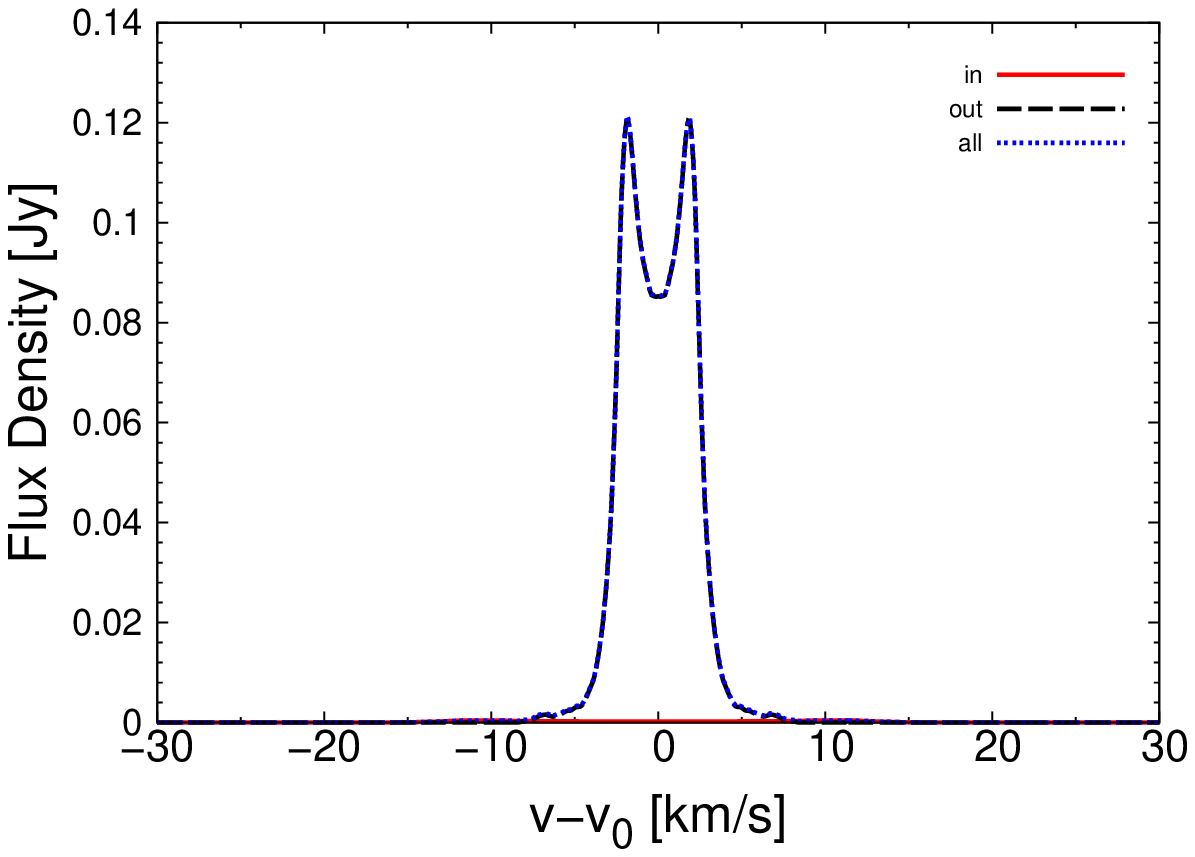}
\includegraphics[scale=0.5]{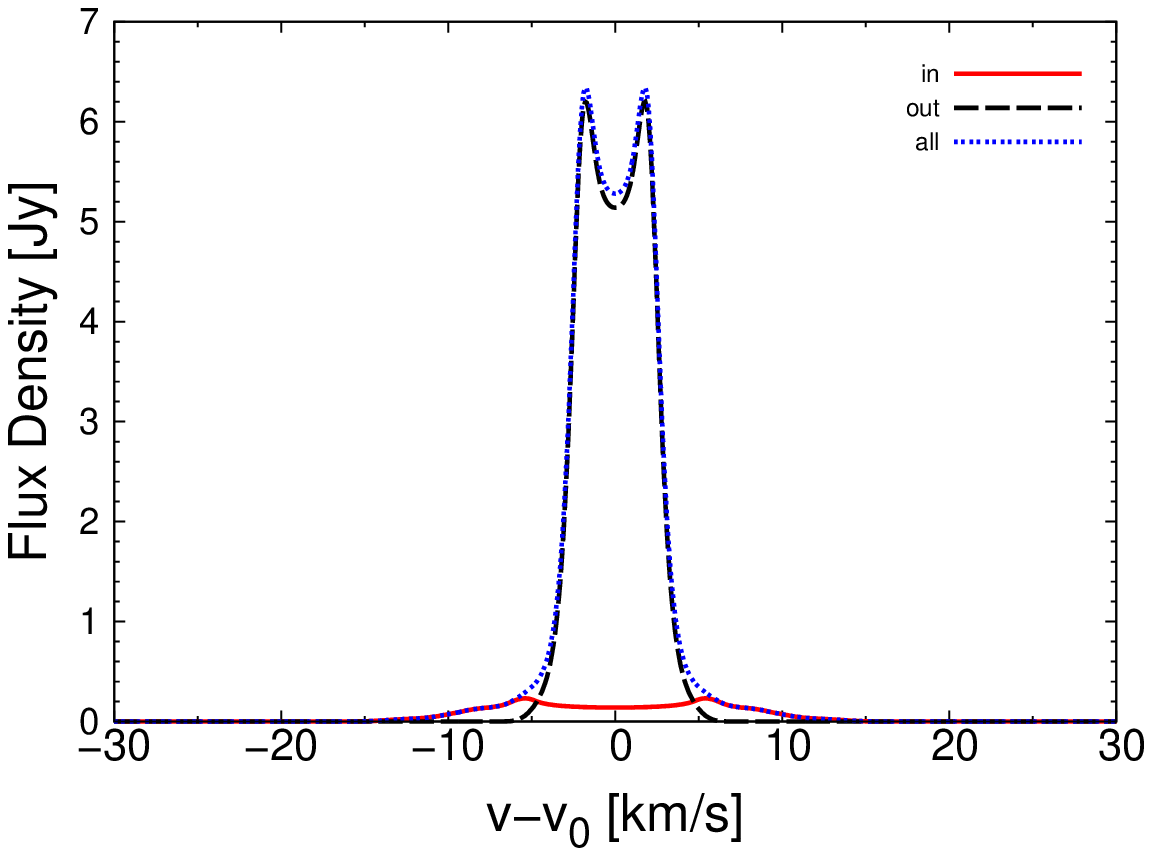}
\includegraphics[scale=0.5]{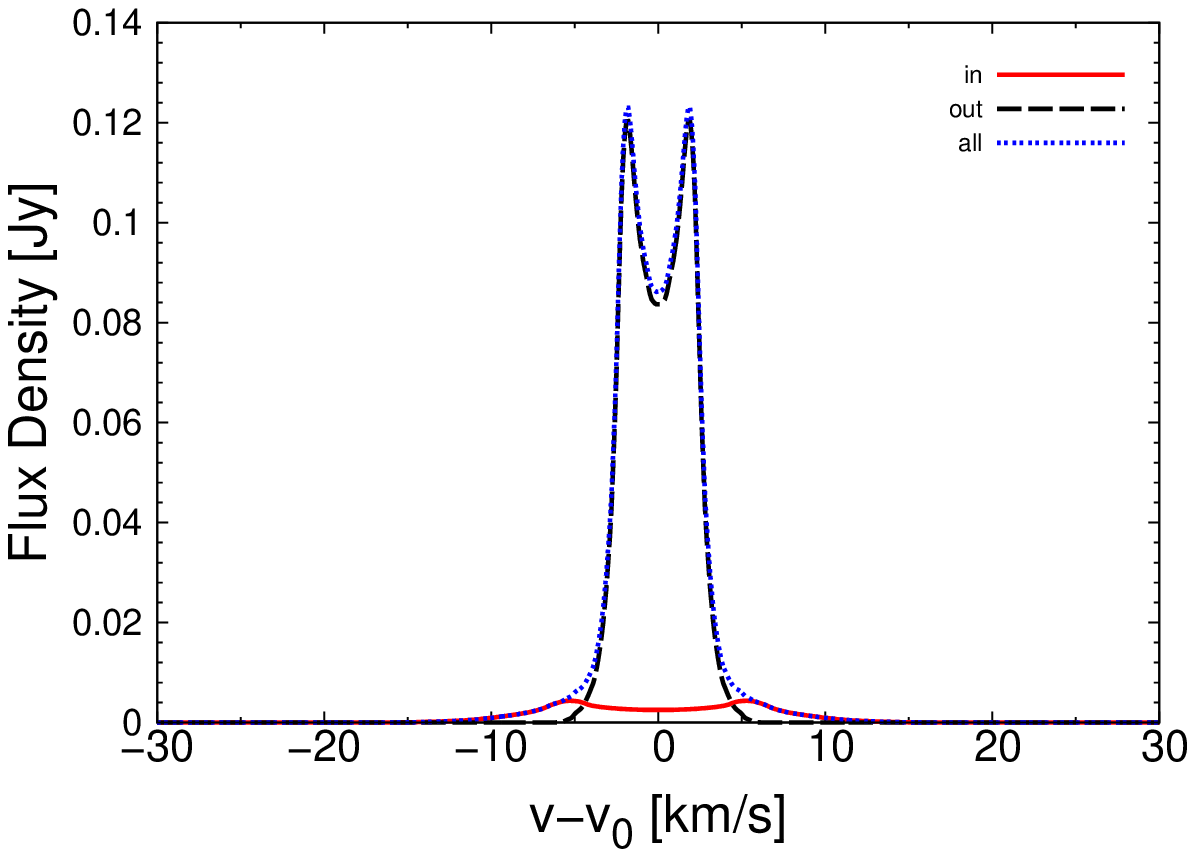}
\includegraphics[scale=0.5]{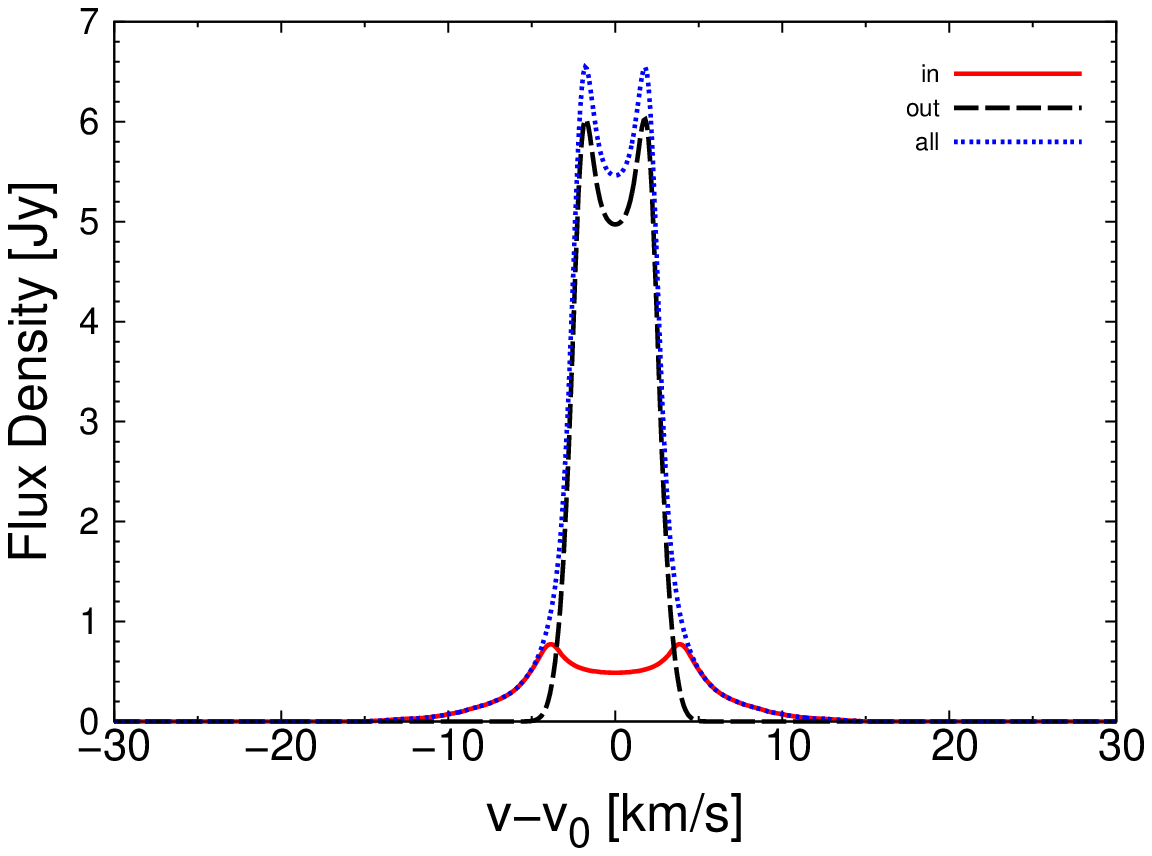}
\includegraphics[scale=0.5]{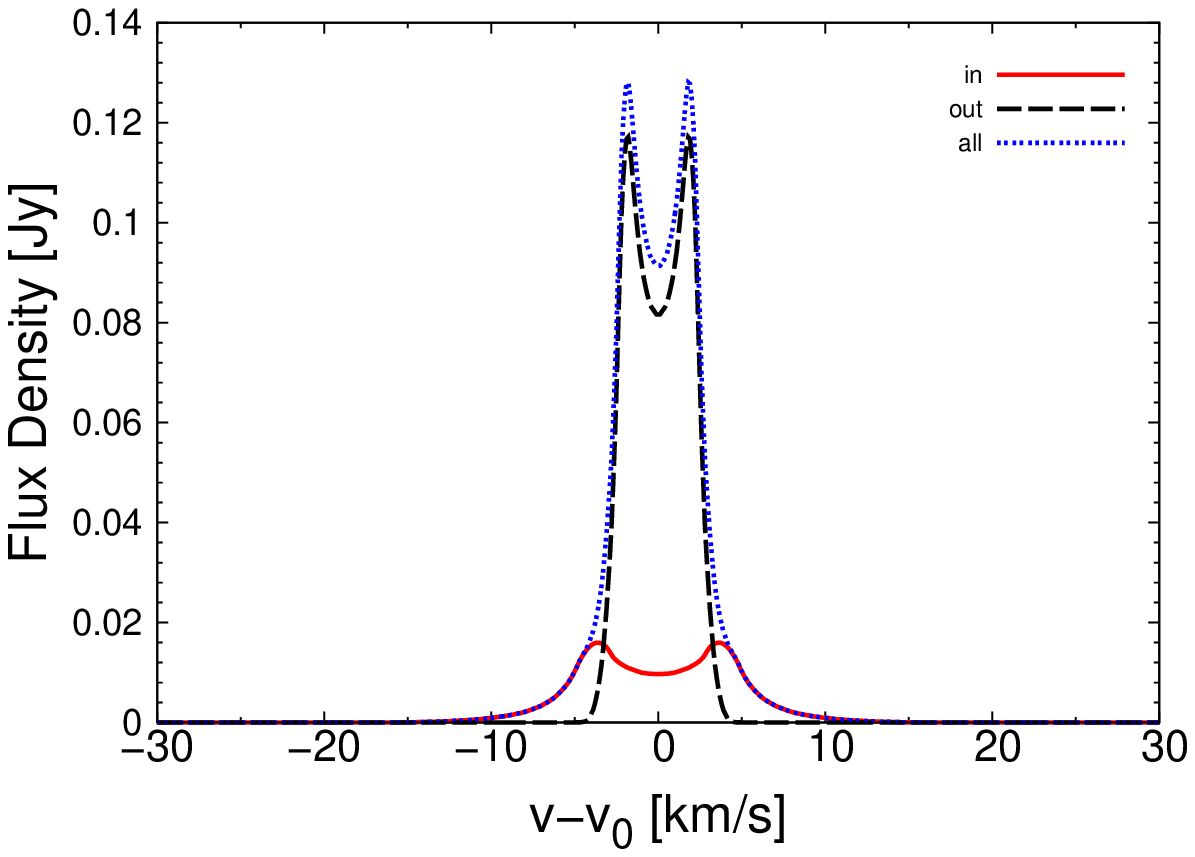}
\end{center}
\vspace{0.3cm}
\caption{
\noindent 
(Continued.) The left three panels: The velocity profiles of the pure rotational ortho-$\mathrm{H_2O}$ lines at $\lambda$=63.37$\mu$m ($J_{K_{a}K_{c}}$=8$_{18}$-7$_{07}$). 
The right three panels: The velocity profiles of the pure rotational ortho-$\mathrm{H_2O}$ lines at $\lambda$=538.66$\mu$m ($J_{K_{a}K_{c}}$=1$_{10}$-1$_{01}$). 
These panels correspond to the cases that the $\mathrm{H_2O}$ snowline is assumed to be 1 AU (top left and right) 4AU (middle left and right), 8AU (bottom left and right). 
}\label{Figure9_add}
\end{figure*} 
In the remaining part of this subsection, we discuss the behavior of the $\mathrm{H_2O}$ lines for some cases in which we artificially change the distribution of $\mathrm{H_2O}$ vapor, the position of the $\mathrm{H_2O}$ snowline and the fractional abundance of $\mathrm{H_2O}$ gas in the outer disk surface, and test the validity of our model predictions.
We explore different values of the $\mathrm{H_2O}$ snowline radius to simulate the effects of viscous heating and of different dust opacities due to dust evolution, and of the water abundance to simulate the effects of the strength of photo-reactions, as outlined above.
\\ \\
In Figure \ref{FigureA3_add}, we show the distributions of $\mathrm{H_2O}$ gas and profiles of the 682.93$\mu$m, 63.37$\mu$m, and 538.66 $\mu$m lines when we change the positions of the $\mathrm{H_2O}$ snowline ($r_{\mathrm{snowline}}$) to 1 AU (top panels), 4 AU (middle panels), 8 AU (bottom panels) by hand.
In the case of $r_{\mathrm{snowline}}=$1 AU, we change the fractional abundance of $\mathrm{H_2O}$ gas by hand to 10$^{-12}$ in the regions of $r=1-1.6$AU and $z/r\sim 0-1.5$.
In the cases of $r_{\mathrm{snowline}}=$4 AU and 8AU, we change the fractional abundance to $5\times 10^{-5}$ in the regions of $r=1.6-4$AU and $z/r\sim 0-1.5$, and $r=1.6-8$AU and $z/r\sim 0-1.7$, respectively.
In calculating these line profiles, we assume that the distance to the object $d$ is 140pc ($\sim$ the distance of Taurus molecular cloud), and the inclination angle of the disk $i$ is 30 deg. 
Here we note that the disk physical structure is the same as the original reference model (see Figure \ref{Figure1_original}). As the position of the $\mathrm{H_2O}$ snowline moves outward, the flux of these three line from the inner disk becomes larger, that from the outer disk becomes weaker, and the line width, especially the width between the two peaks becomes narrower. 
In the case of the 682.93$\mu$m line, the emission flux inside the $\mathrm{H_2O}$ snowline is still larger than that outside the $\mathrm{H_2O}$ snowline, even when the $\mathrm{H_2O}$ snowline is artificially set at 1 AU. In addition, the position of the $\mathrm{H_2O}$ snowline can be distinguished using the difference in the peak separations, although the sensitivity to its position will depend on the spectral resolution of the observations and the uncertainty of other parameters (e.g., inclination $i$).
In the cases of the 63.37$\mu$m and 538.66 $\mu$m lines, the emission fluxes inside the $\mathrm{H_2O}$ snowline are still much smaller than that outside the $\mathrm{H_2O}$ snowline, even when the $\mathrm{H_2O}$ snowline is at 8 AU. 
However, if we calculate the line fluxes using self-consistent physical models, the emission flux of the 63.37$\mu$m line inside the $\mathrm{H_2O}$ snowline is around ten times larger in the case of $r_{\mathrm{snowline}}=$ 8 AU, and its emission flux could be similar to that outside the $\mathrm{H_2O}$ snowline (see below). 
\\ \\
We use the same disk physical structure as the original reference model, because calculating several different disk physical structures 
and chemical structures self-consistently using our method (see Section 2.1 and 2.2) 
is computationally demanding and beyond the scope of this work.
Even if we adopt self-consistent models, we expect that the line widths will not be affected; however, we do expect that the line fluxes will be affected
since the temperature of line emitting regions will be different.
In our original reference model, the gas and dust temperatures around the $\mathrm{H_2O}$ snowline are about 150$-$160K.
In contrast, the temperatures of the line emitting regions
around the $\mathrm{H_2O}$ snowline for the models with a snowline radius ($r_{\mathrm{snowline}}$) of 1 AU, 4 AU, and 8 AU are 180$-$300K, 85$-$90K, and $\sim$65K, respectively. 
Therefore, estimation of blackbody intensities at $\lambda \sim$63$-$683$\mu$m suggests that the line peak flux densities could be $\sim$0.3$-$0.85 times lower for the model with $r_{\mathrm{snowline}}=$1AU, 
and $\sim$$2-4$ times and $\sim$$2.5-10$ times higher for the models with $r_{\mathrm{snowline}}$=4AU and 8AU, respectively, if we calculate the line fluxes using self-consistent physical models. These differences in the peak flux densities are larger in the lines at shorter wavelengths.
 \begin{figure*}[htbp]
\begin{center}
\includegraphics[scale=0.5]{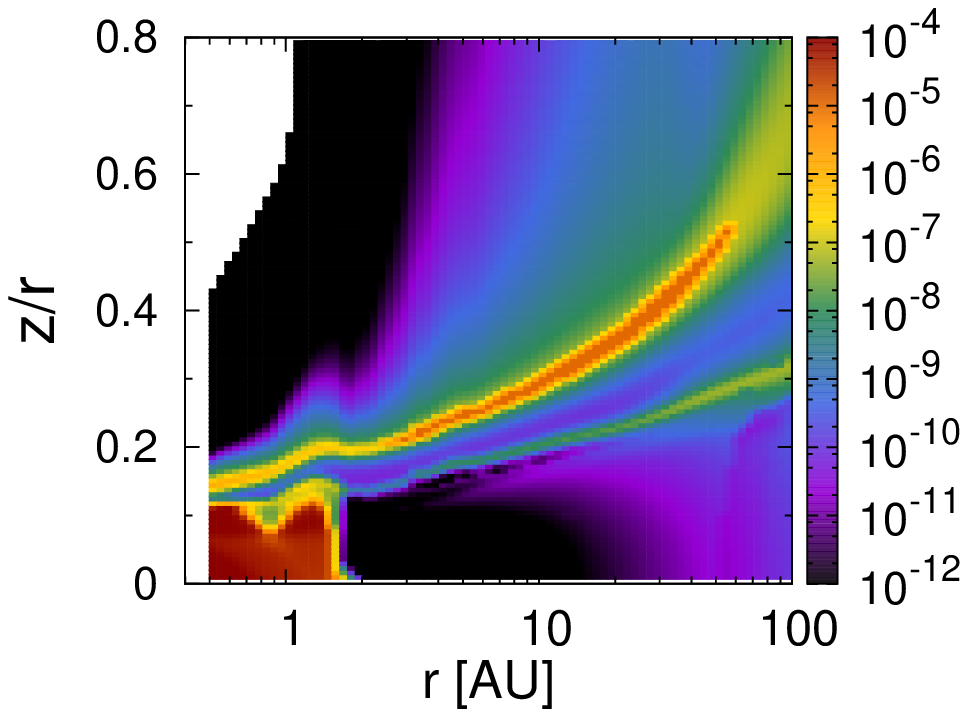}
\includegraphics[scale=0.5]{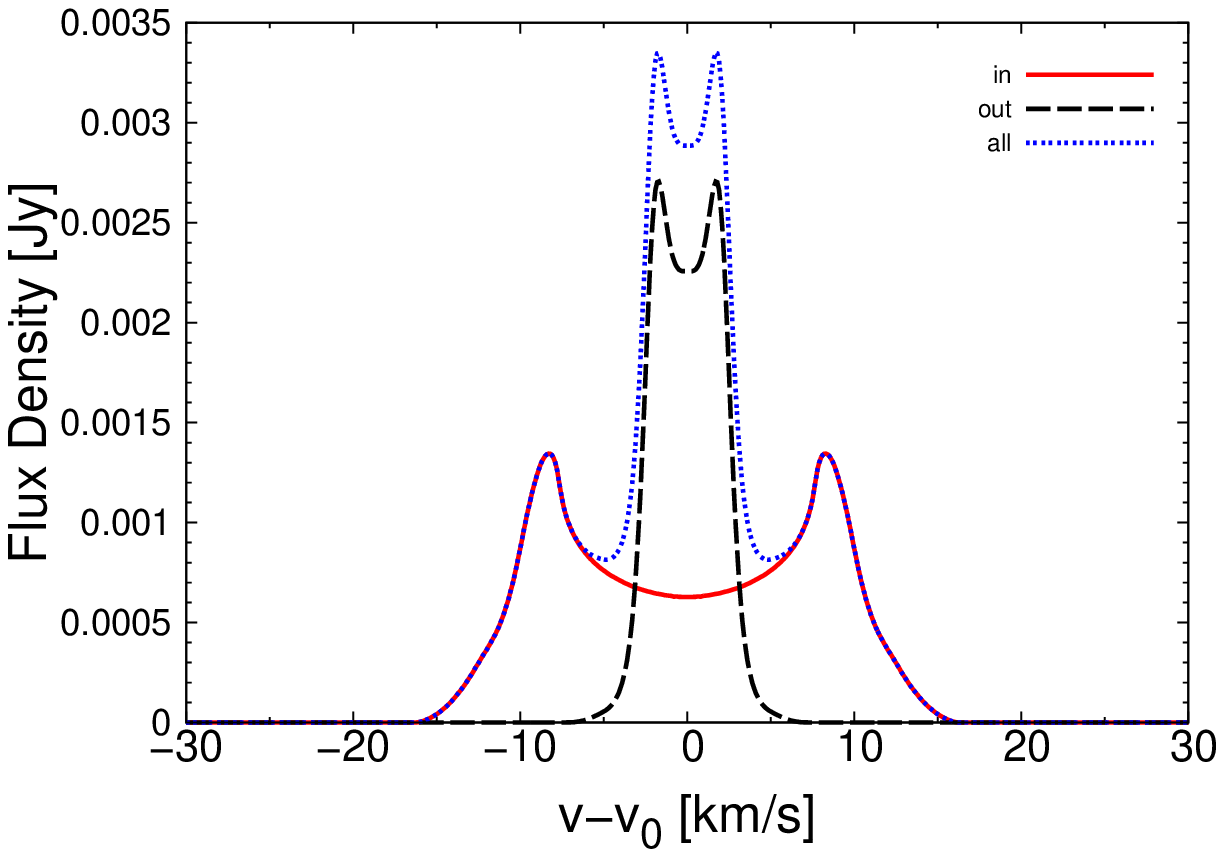}
\includegraphics[scale=0.5]{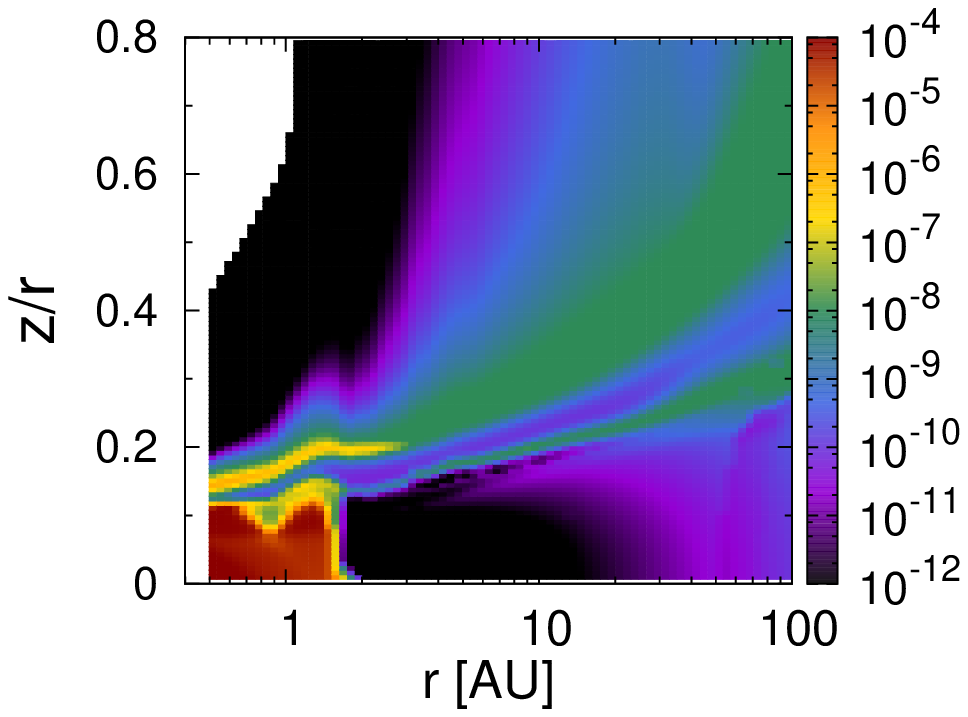}
\includegraphics[scale=0.5]{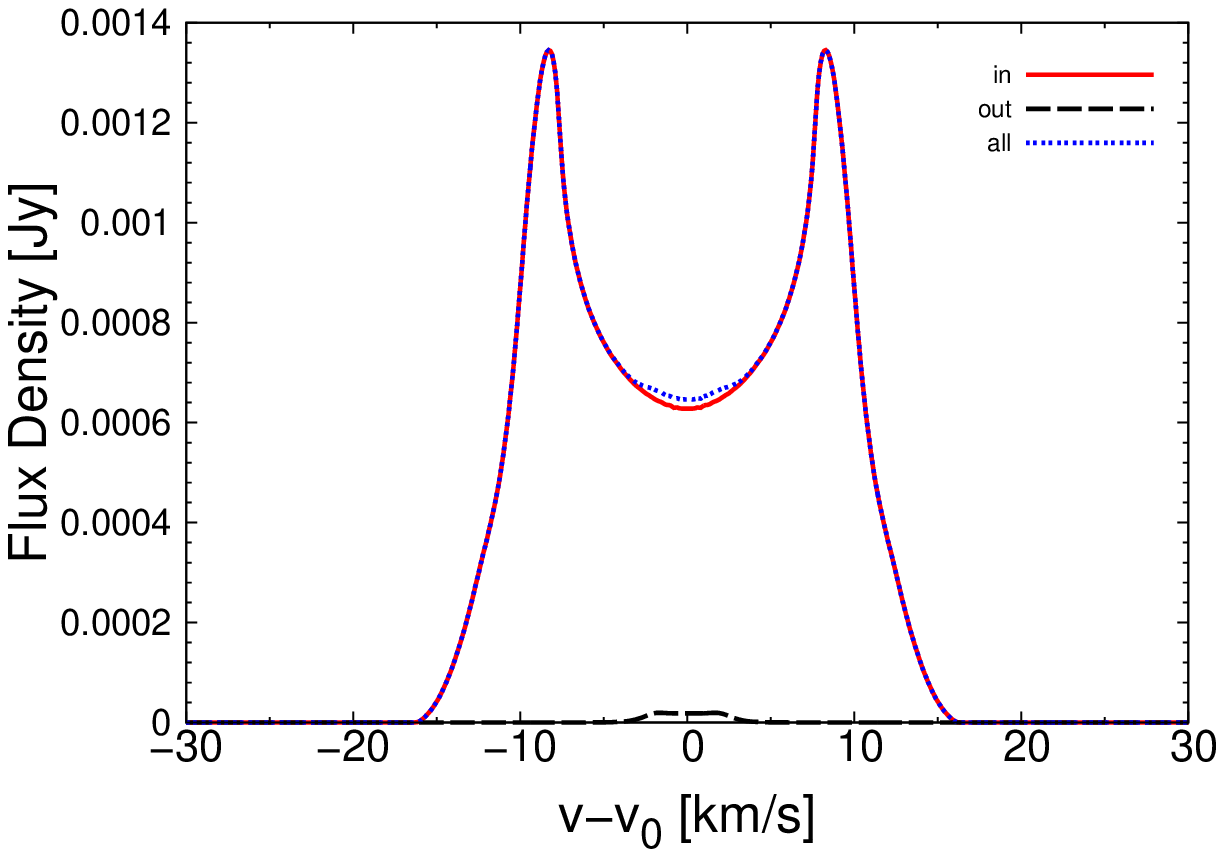}
\end{center}
\vspace{0.3cm}
\caption{\noindent The left two panels: the fractional abundance (relative to total hydrogen nuclei density) distributions of $\mathrm{H_2O}$ gas of a disk around a T Tauri star as a function of disk radius and height (scaled by the radius, $z/r$) up to maximum radii of $r=$100AU. We change the fractional abundance of $\mathrm{H_2O}$ gas in the hot disk surface of the outer disk to a larger value (10$^{-5}$, top left), and to a smaller value (10$^{-8}$, bottom left) compared with the original self-consistently calculated value (see also Figure 2).
The right two panels: The velocity profiles of the pure rotational ortho-$\mathrm{H_2O}$ lines at $\lambda$=682.93$\mu$m ($J_{K_{a}K_{c}}$=6$_{43}$-5$_{50}$). The two panels correspond to the case of the larger value (10$^{-5}$, top right), and to the case of the smaller value (10$^{-8}$, bottom right). 
{\it Red solid lines} are the emission line profiles from inside 2AU ($\sim$inside the $\mathrm{H_2O}$ snowline), {\it black dashed lines} are those from 2-30AU ($\sim$outside the $\mathrm{H_2O}$ snowline), and {\it blue dotted lines} are those from the total area inside 30AU, respectively. 
In calculating these profiles, we assume that the distance to the object $d$ is 140pc ($\sim$ the distance of Taurus molecular cloud), and the inclination angle of the disk $i$ is 30 deg.}\label{FigureA4_add}
\end{figure*} 
\setcounter{figure}{8}
 \begin{figure*}[htbp]
\begin{center}
\includegraphics[scale=0.5]{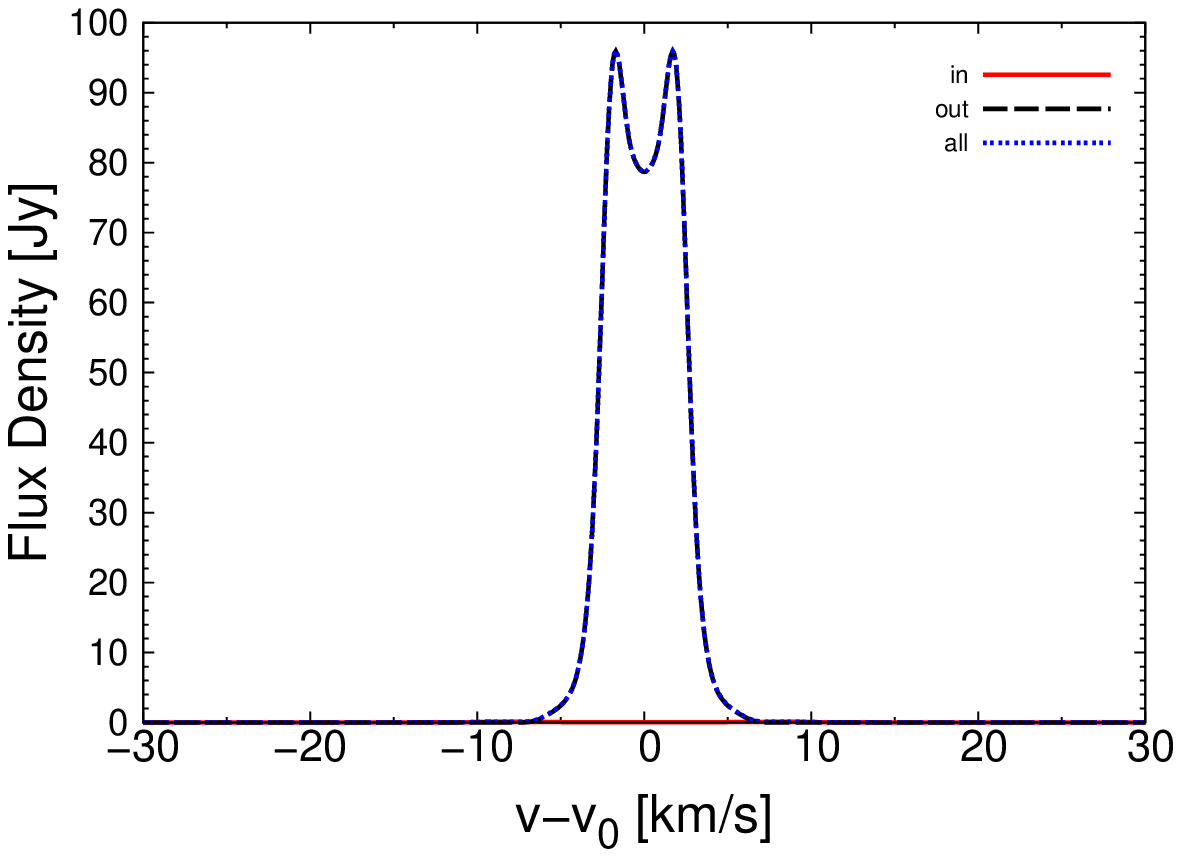}
\includegraphics[scale=0.5]{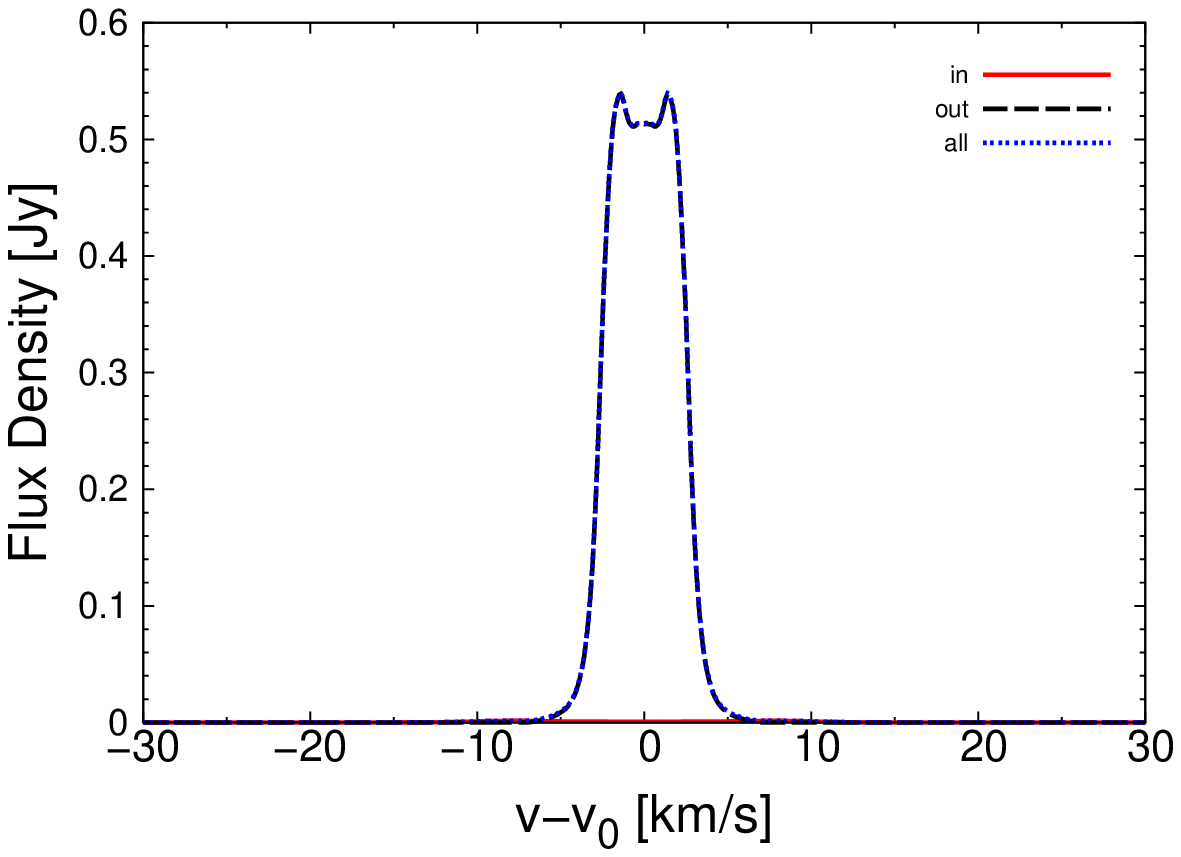}
\includegraphics[scale=0.5]{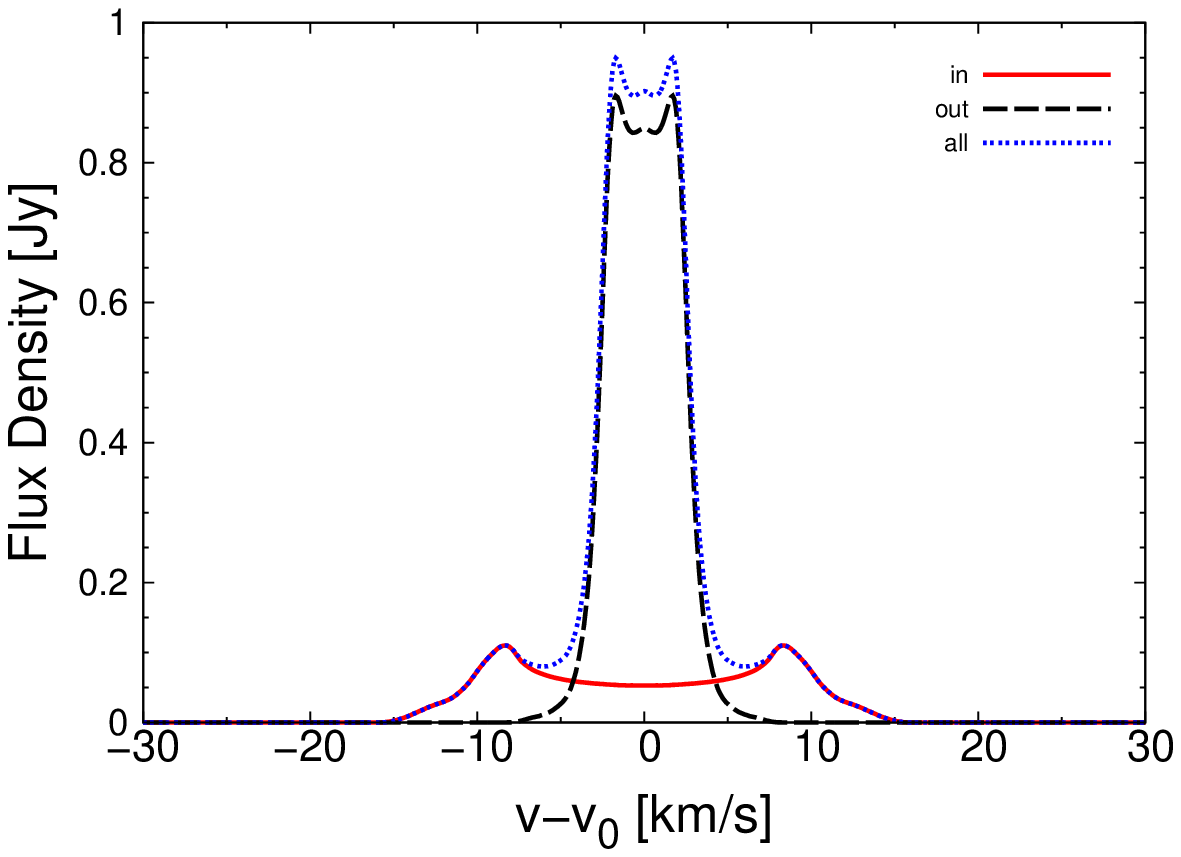}
\includegraphics[scale=0.5]{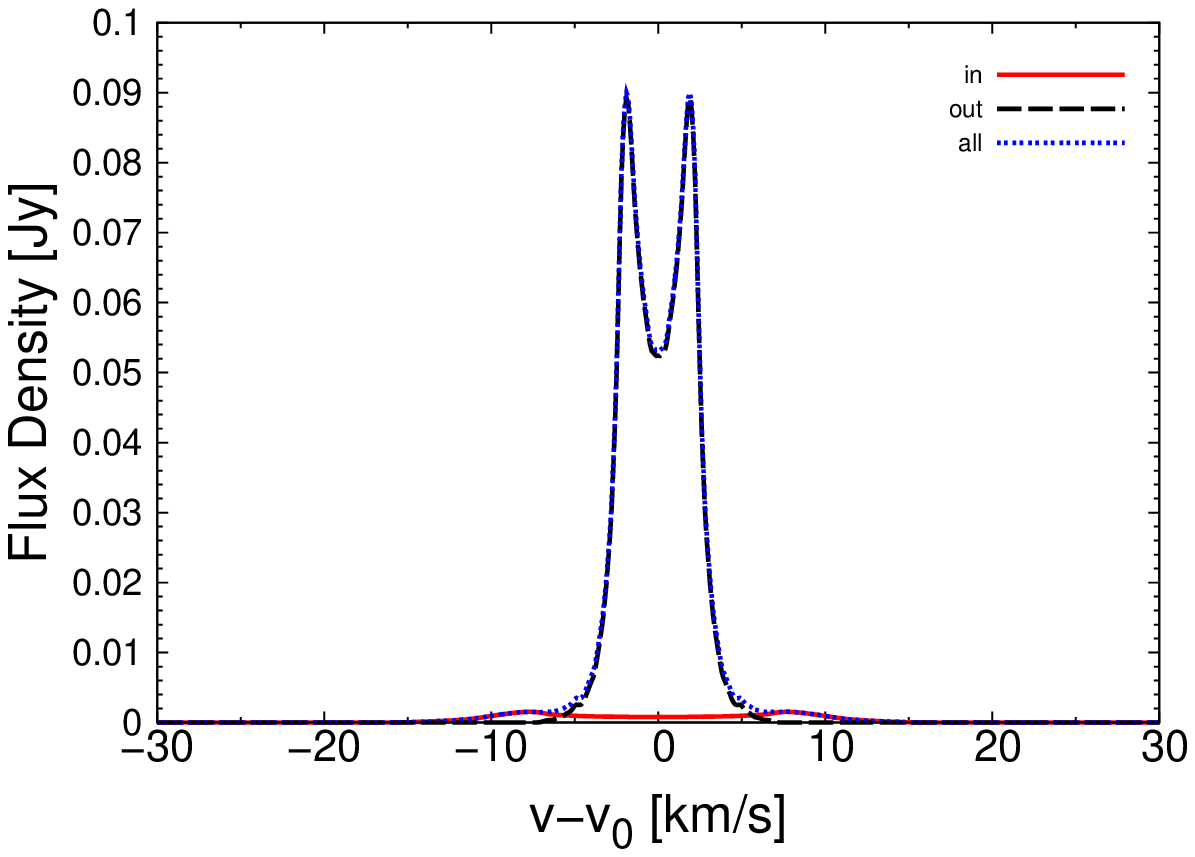}
\end{center}
\vspace{0.3cm}
\caption{
\noindent 
(Continued.) The left two panels: The velocity profiles of the pure rotational ortho-$\mathrm{H_2O}$ lines at $\lambda$=63.37$\mu$m ($J_{K_{a}K_{c}}$=8$_{18}$-7$_{07}$). 
The right two panels: The velocity profiles of the pure rotational ortho-$\mathrm{H_2O}$ lines at $\lambda$=538.66$\mu$m ($J_{K_{a}K_{c}}$=1$_{10}$-1$_{01}$). 
These panels correspond to the case of the larger value (10$^{-5}$, top left and right), and to the case of the smaller value (10$^{-8}$, bottom left and right). 
}\label{Figure11_add}
\end{figure*} 
\\ \\
In Figure \ref{FigureA4_add}, we show the distributions of $\mathrm{H_2O}$ gas and profiles of the 682.93$\mu$m, 63.37$\mu$m, and 538.66 $\mu$m lines when we change the fractional abundance of $\mathrm{H_2O}$ gas by hand in the hot disk surface of the outer disk to a larger value (10$^{-5}$, top panels), and to a smaller value (10$^{-8}$, bottom panels) compared to the original self-consistently calculated value (see also Figure 2), to test the sensitivity of the predictions to the disk surface abundance.
If the fractional abundance of $\mathrm{H_2O}$ gas in the hot disk surface of the outer disk is larger, the flux of the 682.93$\mu$m line from the outer disk becomes larger. Here we note that since the peak velocities of the ``in" and ``out" components are different, we can separate both components with very high sensitivity and high dispersion spectroscopic observations, especially in the very high abundance case (top panels), although the wings of both components are blended.
As the abundance in the hot surface of the outer disk becomes small, the fluxes of the 63.37$\mu$m and 538.66 $\mu$m lines from the outer disk become smaller.
This effect is stronger in the case of the the 63.37$\mu$m line, since this line has a large Einstein $A$ coefficient and high upper state energy compared to those of the 538.66$\mu$m line. 
However, the contributions of the fluxes of these two lines from the outer disk are still larger than that from the inner disk even when the abundance in the hot surface of the outer disk is small.
%
\subsubsection{Critical density and the assumption of LTE}
\noindent As described in section 2.3, the level populations of the water molecule ($n_{u}$ and $n_{l}$) are calculated under the assumption of local thermal equilibrium (LTE).
In this subsection, we discuss the validity of the assumption of LTE within our work.
\\ \\
We calculate the critical density $n_{\mathrm{cr}}=A_{ul}{<\sigma v>}^{-1}$ of the three characteristic lines discussed here (ortho-$\mathrm{H_2O}$ 682.93, 63.37, 538.66$\mu$m lines, see Table 1).
$<\sigma v>$ is the collisional rates for the excitation of $\mathrm{H_2O}$ by H$_{\mathrm{2}}$ and electrons for a an adopted collisional temperature of
200K from \citet{Faure2008}.
The critical density $n_{\mathrm{cr}}$ of these three lines are $1.0\times 10^{6}$, $1.5\times 10^{10}$, $2.9\times 10^{7}$ $\mathrm{cm}^{-3}$, respectively. 
LTE is only realized when collisions dominate the molecular excitation/deexcitation, that is, when the total gas density is larger than $n_{\mathrm{cr}}$.
In contrast, non-LTE allows for the fact that the levels may be sub-thermally excited, when $n_{\mathrm{cr}}$ is higher than the total gas density, or when the emission (deexcitation) dominates collisions, as well as when the levels are super-thermally excited when the radiative excitation dominates the collisions.
When a level is sub-thermally populated in a particular region of the disk, it has a smaller population than in LTE, thus the line flux in non-LTE is smaller than that for LTE (e.g., \citealt{Meijerink2009, Woitke2009b}).
According to \citet{Meijerink2009}, lines with small $A_{ul}$ ($<$10$^{-2}$ $\mathrm{s}^{-1}$) and low $E_{up}$ ($<$2000K at $r=1$AU) are close to LTE, since collisions dominate the radiative excitation/deexcitation in those lines.
\\ \\
As described in Section 2.1 (see also Figure \ref{Figure1_original}), the total gas density decreases as a function of disk radius and disk height.
We found that the densest region of the disk is in the hot disk midplane inside the $\mathrm{H_2O}$ snowline ($\sim10^{12}-10^{14}$ $\mathrm{cm}^{-3}$), where $n_{\mathrm{cr}}$ of the three characteristic lines are much smaller than the total gas density. 
In contrast, the total gas density in the hot surface layer of the outer disk is $\sim10^{7}-10^{8}$ $\mathrm{cm}^{-3}$, and that in the photodesorbed layer of water molecules is $\sim10^{8}-10^{10}$ $\mathrm{cm}^{-3}$. Therefore in these regions, the critical densities of the 63.37 and 538.66$\mu$m lines are similar to and larger than the values of the total gas density, while that of the 682.93$\mu$m line is smaller.
In Section 3.2.1, we showed that the emission flux of the 682.93$\mu$m line which traces the $\mathrm{H_2O}$ snowline mainly comes from the hot disk midplane inside the $\mathrm{H_2O}$ snowline. Since the value of n$_{\mathrm{cr}}$ for this line is much smaller than the total gas density in the line emitting region, it is valid to use the LTE for this region.
\\ \\
On the other hand, in our LTE calculations it remains possible that we have overestimated the emission flux of strong $\mathrm{H_2O}$ lines with large $A_{ul}$ which trace the hot surface layer of the outer disk (e.g., the $\mathrm{H_2O}$ 63.37 $\mu$m line) and lines which trace cold water vapor in the photodesorbed layer (e.g., the $\mathrm{H_2O}$ 538.66 $\mu$m line).
Previous works which model such $\mathrm{H_2O}$ lines (e.g., \citealt{Meijerink2009, Woitke2009b, Banzatti2012, Antonellini2015}) showed that non-LTE calculations are important for these lines. They suggest that non-LTE effects may, however, alter line fluxes by factors of only a few for moderate excitation lines. Moreover, current non-LTE calculations are likely to be inaccurate, due to the incompleteness and uncertainty of collisional rates (e.g., \citealt{Meijerink2009, Banzatti2012, Kamp2013, Zhang2013, Antonellini2015}). 
%
%
%
%
\subsubsection{Requirement for the observations}
\noindent Since the velocity width between the emission peaks is $\sim$20 km s$^{-1}$, high dispersion spectroscopic observations (R=$\lambda$/$\delta \lambda$$>$ tens of thousands) of the identified $\mathrm{H_2O}$ lines are needed to trace emission from the hot water reservoir within the $\mathrm{H_2O}$ snowline.
 Their profiles potentially contain information which can be used to determine the $\mathrm{H_2O}$ snowline position.
Moreover, the lines that are suitable to trace emission from the hot water gas within the $\mathrm{H_2O}$ snowline (e.g., 682.93$\mu$m) tend to have a much smaller $A_{ul}$ than those detected by previous observations (e.g, 63.37$\mu$m, 538.66$\mu$m). Since the area of the emitting regions are small (radii $<$ 2AU for a T Tauri disk) compared with the total disk size, the total flux of each line is very small ($3.12\times 10^{-22}$ W $\mathrm{m}^{-2}$ for the 682.93$\mu$m line). 
In addition, the sensitivity and spectral resolution (of some instruments) used for previous mid-infrared, far-infrared, and sub-millimeter observations (e.g., $Spitzer$/IRS, $Herschel$/PACS, $Herschel$/HIFI) were not sufficient to detect and resolve weak lines. 
\\ \\
Among the various $\mathrm{H_2O}$ lines in ALMA band 8, the $\mathrm{H_2O}$ 682.93$\mu$m line is the most suitable to trace emission from the hot water reservoir within the $\mathrm{H_2O}$ snowline. Several suitable sub-millimeter $\mathrm{H_2O}$ lines exist in ALMA bands 7, 9, and 10 ($\sim$ $300-1000$ $\mu$m), some of which have the same order-of-magnitude fluxes compared with that of the 682.93$\mu$m line.
With ALMA, we can now conduct high sensitivity ($\sim 10^{-21}-10^{-20}$ W $\mathrm{m}^{-2}$ (5$\sigma$, 1 hour)), high dispersion (R$>$ 100,000), and even high spatial resolution ($<$ 100 mas) spectroscopic observations.
Since the total fluxes of the candidate sub-millimeter lines to trace emission from the hot water reservoir within the $\mathrm{H_2O}$ snowline are small in T Tauri disks, they are challenging to detect with current ALMA sensitivity.
However, in hotter Herbig Ae disks and in younger T Tauri disks (e.g., HL Tau), the $\mathrm{H_2O}$ snowline exists at a larger radius and the flux of these lines will be stronger compared with those in our fiducial T Tauri disk ($\sim$1.6AU). Thus the possibility of a successful detection is expected to increase in these sources and could be achieved with current ALMA capabilities.
\\ \\
In addition, suitable lines for detection exist over a wide wavelength range, from mid-infrared (Q band) to sub-millimeter, and there are future mid-infrared instruments including the Q band which will enable high sensitivity and high-dispersion spectroscopic observations: Mid-Infrared Camera High-disperser \& IFU spectrograph on the Thirty Meter Telescope (TMT/MICHI, e.g., \citealt{Packham2012}), and HRS of SPICA\footnote[4]{\url{http://www.ir.isas.jaxa.jp/SPICA/SPICA_HP/research-en.html}} Mid-Infrared Instrument (SPICA/SMI). 
Moreover, since SPICA/SMI has an especially high sensitivity, successful detection is expected even for a T Tauri disk with several hours of observation.
In our companion paper (paper II, \citealt{Notsu2016b}), 
we will discuss in detail the difference in flux between T Tauri and Herbig Ae disks in lines ranging from the mid-infrared to sub-millimeter wavelengths, and their possible detection with future instruments (e.g., ALMA, TMT/MICHI, SPICA/SMI-HRS).
\section{Conclusion}
\noindent In this paper, we identify candidate $\mathrm{H_2O}$ lines to trace emission from the hot water reservoir within the $\mathrm{H_2O}$ snowline through high-dispersion spectroscopic observations in the near future.
First, we calculated the chemical composition of a protoplanetary disk using a self-consistent physical model of a T Tauri disk, and investigated the abundance distributions of $\mathrm{H_2O}$ gas and ice.
We found that the abundance of $\mathrm{H_2O}$ is high ($\sim 10^{-4}$) in the hot inner region within the $\mathrm{H_2O}$ snowline ($\sim$1.6AU) near the equatorial plane, and relatively high $\sim 10^{-7}$ in the hot surface layer of the outer disk, compared to its value in the regions outside the $\mathrm{H_2O}$ snowline near the equatorial plane ($\sim 10^{-12}$).
Second, we calculated the velocity profiles of $\mathrm{H_2O}$ emission lines, and showed that lines (e.g., the ortho-$\mathrm{H_2O}$ 682.93$\mu$m line) with small Einstein $A$ coefficients ($A_{ul}\sim10^{-3}-10^{-6}$ s$^{-1}$) and relatively high upper state energies (E$_{\mathrm{up}}\sim$1000K) are dominated by emission from the disk region inside the $\mathrm{H_2O}$ snowline, and therefore their profiles potentially contain information which can be used to determine the $\mathrm{H_2O}$ snowline position.
This is because the water gas column density of the region inside the $\mathrm{H_2O}$ snowline is sufficiently high that all lines emitting from this region are optically thick
 as long as $A_{ul} > 10^{-6}$ s$^{-1}$.
Instead, the region outside the $\mathrm{H_2O}$ snowline has a lower water gas column density and 
lines with larger Einstein $A$ coefficients have a more significant contribution to their fluxes since the lines are expected to be optically thin there.
Therefore, we argue that the $\mathrm{H_2O}$ lines with small Einstein $A$ coefficients and relatively high upper state energies
are the most suitable to trace emission from the hot water reservoir within the $\mathrm{H_2O}$ snowline in disks through high-dispersion spectroscopic observations in the near future.
The wavelengths of those lines suitable to trace emission from the hot water reservoir within the $\mathrm{H_2O}$ snowline range from mid-infrared (Q band) to sub-millimeter, and they overlap with the capabilities of ALMA and future mid-infrared high dispersion spectrographs (e.g., TMT/MICHI, SPICA/SMI-HRS).
In addition, we calculate the behavior of water lines which have been detected by previous spectroscopic observations 
(e.g., the ortho-$\mathrm{H_2O}$ 63.37$\mu$m line, the ortho-$\mathrm{H_2O}$ 538.66$\mu$m line).
The fluxes calculated for these lines are consistent with those of previous observations and models.
These lines are less suited to trace emission from water vapour within the $\mathrm{H_2O}$ snowline because they are mainly emitted from the region outside the snowline.
In a future paper (paper II, \citealt{Notsu2016b}), we will discuss the differences of fluxes in the suitable lines ranging from mid-infrared (Q band) to sub-millimeter, and the possibility of future observations (e.g., ALMA, TMT/MICHI, SPICA) to locate the position of the $\mathrm{H_2O}$ snowline.
\\
\acknowledgments
\noindent We are grateful to Dr. Itsuki Sakon, Dr. Chris Packham, Dr. Hiroshi Shibai, Dr. Takao Nakagawa, Dr. Satoshi Okuzumi, and Dr. Inga Kamp for their useful comments.
We thank the referee for many important suggestions and comments.
The numerical calculations in this study were carried out on SR16000 at Yukawa Institute for Theoretical Physics (YITP) and computer systems at Kwasan and Hida Observatory (KIPS) in Kyoto University. This work is supported by Grants-in-Aid for Scientific Research, 23103005, 25108004, 25400229, and 16J06887.
S. N. is grateful for the support from the educational program organized by Unit of Synergetic Studies for Space, Kyoto University.
C. W. acknowledges support from the Netherlands Organization for Scientific Research (NWO, program number 639.041.335).
Astrophysics at Queen's University Belfast is supported by a grant from the STFC.
\\ \\

\begin{table}
  \caption{{Molecular Binding Energies}}\label{tab:T2}
 \begin{center}
    \begin{tabular}{lll}
     \hline 
\vspace{-0.2cm}
     Species & Binding Energy & References \\
     &&\\
      & \ \ \ $E_{d}^{\mathrm{K}}(i)$ [K] & \\
     \hline
     CO & 855 & 1 \\
     CO$_{2}$ & 2990 & 2 \\
     $\mathrm{H_2O}$ & 4820 & 3 \\
     CH$_{4}$ & 1080 & 4 \\
     N$_{2}$ & 790 & 1 \\
     NH$_{3}$ & 2790 & 5 \\
     HCN & 4170 & 4 \\
     H$_{2}$CO & 1760 & 6 \\
     C$_{2}$H$_{2}$ & 2400 & 4 \\
\hline
  \multicolumn{3}{l}{\hbox to 0pt{\parbox{70mm}{
   \footnotesize
   \footnotemark[1] \citet{Oberg2005};
   \footnotemark[2] \citet{Edridge2010};
   \footnotemark[3] \citet{Sandford1993};
   \footnotemark[4] \citet{Yamamoto1983};
   \footnotemark[5] \citet{Brown2007};
   \footnotemark[6] \citet{Hasegawa1993} }}}
   \end{tabular}
\end{center}
    \end{table}
\appendix
\twocolumn
\section{The X-ray and UV Radiation Field and The Dust-grain Models}
\noindent We adopt the model X-ray spectrum of a T Tauri star created by fitting the observed $\textit{XMM-Newton}$ spectrum of TW Hya \citep{Kastner2002}, the classical T Tauri star, with a two-temperature thin thermal plasma model (MEKAL model; see, e.g., \citealt{Liedahl1995}). The best-fit parameters are $kT_{\mathrm{1}}=0.8$ keV and $kT_{\mathrm{2}}=0.2$ for the plasma temperatures and $N_{\mathrm{H}}=2.7\times 10^{20}$ $\mathrm{cm}^{-2}$ for the foreground interstellar hydrogen column density. This is the same model that is adopted in \citet{Nomura2007}.
\\ \\
The stellar UV radiation field model we use is based on the observational data of TW Hya, with a stellar UV radiation field that has three components: photospheric blackbody radiation, optically thin hydrogenic bremsstrahlung radiation, and strong Ly$\alpha$ line 
emission (for details, see Appendix C of \citealt{NomuraMillar2005}, \citealt{Walsh2015}).
For the UV extinction, we include absorption and scattering by dust grains.
The interstellar UV radiation field is taken into account, but its contribution is negligible since the UV irradiation of the central star is strong \citep{NomuraMillar2005}.
\\
\\
We adopt the same dust grain model of \citet{NomuraMillar2005}. 
They assume that the dust grains are compact and spherical, and consist of silicate grains, carbonaceous grains, and water ices.
The optical properties of the carbonaceous grains are assumed to have a continuous distribution of graphite-like properties for larger sizes and PAH-like properties in the small size limit 
\citep{LiDraine2001}.
The sublimation temperatures are assumed to be $T_{\mathrm{silicate}}=1500$K, $T_{\mathrm{carbon}}=2300$K, and $T_{\mathrm{ice}}=150$K \citep{AdamsShu1986}. 
The mass fractional abundances are taken to be consistent with the solar elemental abundances: $\zeta_{\mathrm{silicate}}=0.0043$, $\zeta_{\mathrm{carbon}}=0.0030$, and
$\zeta_{\mathrm{ice}}=0.0094$ \citep{Anders1989}.
Their bulk densities are set to be $\rho_{\mathrm{silicate}}=3.5$ g $\mathrm{cm}^{-3}$, $\rho_{\mathrm{graphite}}=2.24$ g $\mathrm{cm}^{-3}$, and $\rho_{\mathrm{ice}}= 0.92$ g $\mathrm{cm}^{-3}$.
They assume that the dust and gas are well mixed.
They use the dust grain size distributions of silicate and carbonaceous grains obtained by \citet{WeingartnerDraine2001}, which reproduces the extinction curve observed in dense clouds with
the ratio of visual extinction to reddening $R_{V}\equiv A(V)/E(B-V)=5.5$.
They adopt the value of $b_{\mathrm{C}}=3.0\times 10^{-5}$, which is the total C abundance per H nucleus in the log-normal size dust-grain distribution, in order to reproduce the distribution of very small hydrocarbon molecules including PAHs (see also Fig. 6 of \citealt{WeingartnerDraine2001}).
It is assumed that the water ice has the simple size distribution of d$n$/d$a$ $\propto a^{-3.5}$, where $a$ is the radius of the dust particles \citep{Mathis1977}.
The maximum radius of dust grains $a_{\mathrm{max}}$ is $\sim$10$\mu$m.
The calculated monochromatic absorption coefficient is shown in Fig. D.1. of \citet{NomuraMillar2005}.
\section{Molecular Binding Energies}
\renewcommand{\thetable}{\Alph{section}.\arabic{table}}
\setcounter{table}{0}
\noindent The binding energy of species $i$ to the dust grain in units of K, $E_{d}^{\mathrm{K}}(i)$, is listed in Table \ref{tab:T2}.

\section{The mechanisms of non-thermal desorption}
\noindent The non-thermal desorption mechanisms we adopt are cosmic-ray-induced desorption \citep{Leger1985, Hasegawa1993} and photodesorption from UV photons \citep{Westley1995, Willacy2000, Oberg2007}, in common with some previous studies (e.g., \citealt{Walsh2010, Walsh2012}).
In this subsection, we explain the details of these non-thermal desorption mechanisms. 
\\
\\
In order to calculate the cosmic-ray-induced desorption rate for each species, $k_{i}^{\mathrm{crd}}$, we assume 
that dust grains with a radius of 0.1 $\mu m$ are impulsively heated by the impact of relativistic Fe nuclei with energies of 20-70 MeV nucleon$^{-1}$
which deposit an energy of 0.4 MeV on average into each dust grain \citep{Leger1985, Hasegawa1993}. 
Assuming that the majority of molecules desorb around 70 K, the cosmic-ray-induced desorption rate can be approximated by
\begin{equation}
k_{i}^{\mathrm{crd}} = f(70\mathrm{K})k_{i}^{d}(70\mathrm{K})\frac{\zeta_{\mathrm{CR}}}{1.36\times10^{-17} \mathrm{s}^{-1}} \ \mathrm{s}^{-1},
\end{equation}
where $\zeta_{\mathrm{CR}}$ is the cosmic-ray ionization rate of H$_{2}$, $k_{i}^{d}(70\mathrm{K})$ is the thermal desorption energy of species $i$ at a dust temperature of 70K computed using Equation. (4).
$f(70\mathrm{K})$ is the fraction of time spent by dust grains around 70K and is defined as the ratio of
the desorption cooling time (10$^{-5}$ s) to the time interval between successive heatings to 70K.
The latter value is estimated to be 3.16$\times 10^{13}$ s from the Fe cosmic-ray flux, then $f(70\mathrm{K})$ is 3.16$\times 10^{-19}$ \citep{Leger1985, Hasegawa1993}. 
We mention that like cosmic-ray particles, X-ray photons can penetrate deep inside the disk and locally heat dust grains. 
However, X-ray desorption is not yet included in our chemical code unlike previous studies (e.g.,  \citealt{Walsh2012, Walsh2014a, Walsh2015}). 
This is because X-ray desorption is the least theoretically or experimentally
constrained of all the non-thermal desorption mechanisms, and thus there remains
significant uncertainties in the reaction rates (e.g., \citealt{Najita2001, Walsh2010}). 
\\
\\
Absorption of a UV photon by a dust-grain surface species can increase the species internal energy enough to induce desorption.
The photodesorption rate of species $i$ is given by 
\begin{equation}
k_{i}^{\mathrm{pd}} = F_{\mathrm{UV}}Y_{\mathrm{UV}}^{i}\sigma_{d}\frac{n_{d}}{n_{act}} \ \mathrm{s}^{-1}, 
\end{equation}
where $F_{\mathrm{UV}}$ is the wavelength integrated UV radiative flux calculated at each $(r,z)$ in units of photons $\mathrm{cm}^{-2}$ s$^{-1}$. $Y_{\mathrm{UV}}^{i}$ is the experimentally determined photodesorption yield in units of molecules photon$^{-1}$. \citet{Woitke2009a} and \citet{Heinzeller2011} used a similar method to calculate $k_{i}^{\mathrm{pd}}$.
$F_{\mathrm{UV}}$ is calculated from the density profile and the dust opacity in our adopted disk physical model \citep{NomuraMillar2005, Nomura2007, Walsh2015}.
We assume that $Y_{\mathrm{UV}}^{i}$ is $3.0\times 10^{-3}$ for all species, which is the same value determined for pure water ice by \citet{Westley1995} and for pure CO ice by \citet{Oberg2007}. \citet{Walsh2010} use the same value of $Y_{\mathrm{UV}}^{i}$. 
$n_{act} = 4\pi a^{2}n_{d}n_{surf}N_{Lay}$ is the number of active surface sites in the ice mantle per unit volume. $N_{Lay}$ is the number of
surface layers to be considered as ``active", and we adopt the value of \citet{Aikawa1996}, $N_{Lay}=2$.
Recent experiments by \citet{Oberg2009a, Oberg2009b, Oberg2009c} suggest that photodesorption rates are dependent on ice composition and the depth of the ice layer on a dust grain.
\\ \\

\end{document}